\PassOptionsToPackage{pdfpagelabels=false}{hyperref} 
\documentclass[fleqn,usenatbib]{mnras}

\usepackage{newtxtext,newtxmath}

\usepackage[T1]{fontenc}
\usepackage{ae,aecompl}

\usepackage{epsfig,graphicx}
\usepackage{amssymb,amsmath}
\usepackage{latexsym}
\usepackage{natbib}
\usepackage[caption=false]{subfig}

\usepackage{enumitem}

\newcommand\nodata{ ~$\cdots$~ }

\title[Irradiated Discs in BH-LMXBs]{Understanding X-ray Irradiation in Low-Mass X-ray Binaries directly from their Light-Curves}

\author[B.E. Tetarenko et al.]{B.E. Tetarenko$^{1}$\thanks{E-mail: btetaren@ualberta.ca} 
G. Dubus$^{2}$
J.-P. Lasota$^{3,4}$
C.O. Heinke$^{1}$
and G.R. Sivakoff$^{1}$
\\
$^{1}$Department of Physics, University of Alberta, CCIS 4-181, Edmonton, AB T6G 2E1, Canada\\
$^{2}$Univ. Grenoble Alpes, CNRS, Institut de Plan\'etologie et d'Astrophysique de Grenoble (IPAG),F-38000, Grenoble, France\\
$^{3}$Institut d'Astrophysique de Paris, CNRS et Sorbonne Universit\'e, UMR 7095, 98bis Bd Arago, 75014 Paris, France\\
$^{4}$Nicolaus Copernicus Astronomical Centre, Polish Academy of Sciences, ul. Bartycka 18, 00-716 Warsaw, Poland
}

\date{Accepted XXX. Received YYY; in original form ZZZ}

\pubyear{2018}

\begin{document}
\label{firstpage}
\pagerange{\pageref{firstpage}--\pageref{lastpage}}
\maketitle

\begin{abstract}
The X-ray light-curves of the recurring outbursts observed in low-mass X-ray binaries provide strong test beds for constraining (still) poorly understood disc-accretion processes. These light-curves act as a powerful diagnostic to probe the physics behind the mechanisms driving mass inflow and outflow in these binary systems. We have thus developed an innovative methodology, combining a foundation of Bayesian statistics, observed X-ray light-curves, and accretion disc theory. With this methodology, we characterize the angular-momentum (and mass) transport processes in an accretion disc, as well as the properties of the X-ray irradiation-heating that regulates the decay from outburst maximum in low-mass X-ray transients. We recently applied our methodology to the Galactic black-hole low-mass X-ray binary population, deriving from their lightcurves the first-ever quantitative measurements of the $\alpha$-viscosity parameter in these systems \citep{tetarenko2018}. In this paper, we continue the study of these binaries, using  Bayesian methods to investigate the X-ray irradiation of their discs during outbursts of strong accretion. We find that the predictions of the disc-instability model, assuming a source of X-ray irradiation proportional to the central accretion rate throughout outburst, do not adequately describe the later stages of BH-LMXB outburst light-curves. We postulate that the complex and varied light-curve morphology observed across the population is evidence for irradiation that varies in time and space within the disc, throughout individual transient outbursts. Lastly, we demonstrate the robustness of our methodology, by accurately reproducing the synthetic model light-curves computed from numerical codes built to simulate accretion flows in binary systems.

\end{abstract}




\section{Introduction}\label{sec:intro}
Throughout their lifetimes, many astrophysical objects (e.g. newborn stars, planets, black holes) grow and evolve by accumulating mass through a disc. For these objects to grow, matter must lose angular momentum to flow inward, and avoid being removed from the system via outflows. Among accreting astrophysical systems, low-mass X-ray binaries (LMXBs), in which compact objects (neutron stars and black holes) accrete from  
nearby, low-mass ($M_2 \lesssim 1 \,  M_{\odot}$) stars, provide us with strong test beds for constraining this poorly understood process of accretion.

So far, 18 confirmed (and $\sim46$ candidate) LMXBs harbouring stellar-mass black holes (BHs) have been identified through their bright X-ray outbursts, indicative of rapid accretion episodes, in our Galaxy \citep[][and refs therein]{mr06,tetarenkob2015,negoro2017,kawamuro2018,kawase2018,barthelmy2018}. All these systems are transient. They display long-term behaviour characterized by extended periods of time (typically years to decades) spent in a quiescent state, where the system is faint ($L_X\sim10^{30} - 10^{33} {\rm \, erg \, s^{-1}}$) as a result of very little accretion occurring onto the compact object \citep[e.g.][]{Garcia01}.
These prolonged quiescent periods are interrupted by occasional bright disc-outbursts, typically lasting hundreds of days,
during which the X-ray luminosity will increase by multiple orders of magnitude ( $L_{\rm X,peak}\sim10^{36} - 10^{39} {\rm \, erg \, s^{-1}}$; \citealt{chen97,tetarenkob2015}).

Although less frequent, the recurring nature of outbursts observed in transient BH-LMXBs is reminiscent of the behaviour observed in dwarf novae (i.e. compact binary systems consisting of a white dwarf accreting from a low-mass companion; \citealt{warner1995}). In dwarf novae, the mechanism behind such outbursts is well understood using the disc-instability model (DIM; \citealt{osaki1974,meyer81,smak83,smak84,cannizzo1985,cannizzo1993,huang89}), which predicts alternating periods of bright disc-outbursts, lasting days, and faint quiescence, lasting weeks. 
According to the DIM, this behaviour results from a thermal-viscous instability developing within the disc, causing it to cycle between a hot, ionized outburst state and a cool, neutral, quiescent state. 
The instability, triggered by the continuous accumulation of matter from the companion star eventually heating and subsequently ionizing the disc, causes a dramatic increase in the viscosity (i.e. the ability of the disc to move angular momentum outwards) of the disc. This increased viscosity results in a rapid in-fall of matter onto the compact object and a bright outburst in the optical and ultraviolet (UV) bands.

X-ray irradiation of the disc must be taken into account when describing transient outbursts of LMXBs.
LMXBs have deeper potential wells and thus undergo brighter X-ray, optical, and UV outbursts that last longer and recur less frequently \citep{tetarenkob2015}, than most dwarf novae\footnote{The notable exceptions here are WZ Sge type dwarf novae, where outbursts typically last months and recur after tens of years, similar to those of BH-LMXBs. See \citealt{kato2015} for a review of WZ Sge dwarf novae.}.
The majority of the UV, optical and infrared (IR) light emitted by the accretion discs in LMXBs comes from reprocessed X-rays. Here the inner regions of the accretion flow heat the outer disc \citep{JvP83,vanp1994,vanp96}. A major contributor to the thermal balance in the accretion flow, this X-ray irradiation keeps the disc in a hot, ionized state controlling most of the outburst decay towards quiescence. Consequently, the light-curve profile for an outburst of an irradiated disc will differ from that of a non-irradiated disc (\citealt{kingrit8,dubus2001}).

Taken as a whole, the multi-wavelength light-curves of the recurring outbursts in LMXBs encode within them key physical parameters describing how (and on what timescale) matter moves through, and is removed from, the discs in these systems.
Thus, LMXB outburst light-curves offer a means in which to understand the mechanism behind the X-ray irradiation affecting these discs which still remains poorly understood (see \citealt{dubus1999} and references therein).
Accordingly, we have  developed an innovative methodology,
combining a foundation of Bayesian statistics, the observed X-ray light-curves, and accretion disc
theory. With this methodology, we characterize the angular-momentum (and mass) transport
processes in an accretion disc, as well as the properties of the X-ray irradiation-heating that
the discs are subject too. 

\defcitealias{tetarenko2018}{Paper~I}
In \cite{tetarenko2018} (hereafter \citetalias{tetarenko2018}), we presented the details of this methodology. By applying this approach to the BH-LMXB population, we were able to derive the first-ever measurements of the efficiency of the angular-momentum (and mass) transport process (parametrized via $\alpha$-viscosity; \citealt{ss73}) in the X-ray irradiated discs of LMXBs, directly from observations. 
In this paper, we continue our analysis of Galactic BH-LMXB discs with our methodology, studying the physical properties of the X-ray irradiation heating these discs.

The outline of this paper is as follows. In Section 2 we describe how our model of the X-ray irradiation affects the accretion discs in LMXB systems and the Bayesian methodology we employ. 
Section 3 describes the application of our methodology to the BH-LMXB population of the Galaxy, 
including details behind the selection of our BH-LMXB source and outburst sample, and X-ray data collection, reduction, and analysis procedures.
In Section 4, we present the results of fitting the X-ray light-curve profiles of our BH-LMXB outburst sample and the observational constraints that can be derived
using these characterized light-curve profiles. In Section 5 we discuss what LMXB light-curve profiles can tell us about the structure and geometry of the irradiation source heating LMXB discs and 
how our observationally based methodology compares to the output of numerical disc codes. Lastly, Section 6 provides a summary of this work.

\section{Modelling the X-ray Irradiation affecting low-mass X-ray binary discs}\label{sec:model_irr}

\subsection{The Irradiation Prescription}

X-ray irradiation from the inner accretion region is the dominant factor that determines the temperature over most of the accretion disc during outbursts of BH-LMXBs. The fraction of the X-ray flux that is intercepted and reprocessed in the outer disc is not well understood. Simple prescriptions based on the radial profile of the disc height lead to shadowing of the outer disc, suggesting part of the irradiation process may occur via a larger-sized scattering corona \citep{kim1999,dubus1999}. 
We make use of the prescription used by \cite{dubus2001} to model the lightcurves of X-ray irradiated BH-LMXB accretion discs,
\begin{equation}
T_{\rm irr}^4=\frac{C_{\rm irr} L_{\rm bol}}{4 \pi \sigma_{\rm SB}  R^2}.
\label{eq:irr}
\end{equation}
Here, $C_{\rm irr}$ is a constant encapsulating the information about the fraction of the bolometric 
accretion (mostly X-ray) 
luminosity ($L_{\rm bol}=\eta c^2 \dot{M_c}$ for radiative efficiency $\eta$) that is intercepted and reprocessed by the disc (i.e. it encapsulates the irradiation geometry, the X-ray albedo, the X-ray spectrum, etc.).
Since the effective temperature of the disc is defined through
\begin{equation}
\label{eq:flacc}
T_{\rm eff}^4 = \frac{3GM\dot M}{8\pi\sigma_{\rm SB} R^3},
\end{equation}
the ratio of the irradiation to effective temperatures is
\begin{equation}
\label{eq:flratio}
\frac{T^4_{\rm irr}}{T^4_{\rm eff}}=\frac{4}{3}{C_{\rm irr}} \eta \frac{R}{R_S},
\end{equation}
where $R_S=2GM/c^2$ is the Schwarzschild radius, disc irradiation is important only in the outer disc regions ($R\>>10^3\,R_S$ for $\eta=0.1$ and $C_{\rm irr}\sim 5\times 10^{-3}$, see below).

Physically, $C_{\rm irr}$ controls the overall outburst duration and sets a limit on the amount of mass that the black hole can accrete during the outburst. A larger value of $C_{\rm irr}$, corresponding to stronger irradiation in the outer disc, will increase the duration of the outburst and thus, the relative amount of matter that can be accreted during a given outburst. A larger $C_{\rm irr}$ during outburst will also result in a more lengthy quiescent period following the outburst, as the disc will require more time to build up again.

The actual value $C_{\rm irr}$ takes in accretion discs has been a matter of debate for decades. $C_{\rm irr}$ (in the outer disc) has been previously measured  in five BH-LMXBs (by modelling a combination of X-ray and optical data) 
and two persistently accreting (non-transient) neutron star LMXBs. In these cases, the authors assumed a vertically-isothermal disc and derived a disc opening angle and albedo from optical observations. For the BHs: \citet{Hynes02} found $C_{\rm irr}\sim7.4\times10^{-3}$ for XTE J1859$+$226; \citet{suleimanov2008} estimated $\sim7\times10^{-4}$ and $3\times10^{-4}$ for A0620$-$00 and GRS 1124$-$68, respectively; for XTE J1817$-$330, \citet{gierlinski2009} found $C_{\rm irr}\sim1\times10^{-3}$ in the soft state and $\sim6\times10^{-3}$ in the hard state (consistent with predictions of increased absorption of hard X-ray photons); and finally 
\citet{lipunova2017} constrain $C_{\rm irr}<6\times10^{-4}$ for 4U 1543$-$47.
\citet{vrtilek1990} and \citet{dejong1996} model two persistent neutron star LMXBs, 
leading to the so-called ``standard'' value of $C_{\rm irr}\sim 5 \times 10^{-3}$ typically assumed in theoretical work. This value of $C_{\rm irr}$ has also been shown to be consistent with the amount of X-ray heating required to stabilize persistent neutron star and transient BH LMXB systems against the thermal-viscous instability \citep{vanp96, coriat2012}.  
However, as we only have this limited sample of BH-LMXBs where $C_{\rm irr}$ has actually been estimated, it remains unclear whether the same value would describe the outburst properties of discs across the Galactic BH-LMXB population. Moreover, it is also unknown how (or if) the value of $C_{\rm irr}$ varies from source to source (e.g., with changing $P_{\rm orb}$ or component masses \citep{esin2000b}) or even between outbursts of the same source (e.g. with changing peak outburst luminosity or outburst duration; \citealt{esin2000}).

\subsection{The Light-Curve Model}\label{sec:model}

The outburst light-curve of an LMXB, as predicted by DIM+irradiation, involves a characteristic three-stage decay profile after the outburst peak (see \citealt{kingrit8,dubus2001}). The outburst decay begins with a viscous phase during which the X-ray irradiation from the inner accretion flow can keep the whole disc in a hot (ionized) state, preventing the onset and propagation of a cooling front. Since the accretion rate is larger than the mass-transfer rate, and the mass of the hot disc can only change through central accretion onto the black hole, the light curve will show an exponential-shaped decline on the viscous timescale. Over time, the mass in the disc and central mass-accretion rate will decrease. When the dimming X-ray irradiation can no longer keep the outer regions of the disc in the hot (ionized) state (i.e. above the hydrogen ionization temperature $T_{\rm irr}(R_{\rm disc})>10^4$ K), a cooling front forms and propagates down the disc, bringing the disc to a cold state. 

At this point, the second phase of the decay begins, during which the speed of the propagation of this cooling front, and thus the timescale of the phase itself, is controlled by the temperature of the decaying irradiating X-ray flux. Here, the cooling-front inward propagation is hindered by irradiation. The farthest it can move inward is set by the radius at which $T=10^4$ K. While the hot (ionized) zone of the disc will continue to flow and accrete, it must now gradually shrink in size as the central mass-accretion rate decreases ($R_{\rm hot}\sim\dot{M}_{1}^{1/2}$), leading to a linear-shaped decline in the light-curve. 

Eventually, the central mass accretion rate will become small enough that X-ray irradiation will no longer play a role and the system will enter the final (thermal) decay stage. X-ray irradiation may also decline faster than the mass accretion rate when the inner disc switches to a radiatively-inefficient accretion flow with a smaller radiative efficiency $\eta$. At this point the cooling front will be allowed to propagate inward through the thin disc on a thermal-viscous timescale (where the speed of the front can be written as $v_f\sim \alpha c_S$), where $c_S$ is the sound speed for $T_{\rm eff}\sim 10^4$K. Ultimately resulting in a steeper final dwarf-nova type decline in the light-curve down to the quiescent accretion level.

As detailed in \citetalias{tetarenko2018}, we have built an improved analytical version of this ``classic'' irradiated disc-instability model. Our version builds on the simple model of irradiated discs by \cite{kingrit8}, using the irradiation flux as set by \autoref{eq:irr}.
This analytical model effectively characterizes the light curve profile of a transient LMXB using five parameters as follows,
\[ 
f_X =
\begin{cases} 
      (f_t-f_2)\exp \left( {-(t-t_{\rm break})}/{\tau_e} \right)+f_2 & t\leq t_{\rm break} \\
      f_t\left(1-{(t-t_{\rm break})}/{\tau_l}\right) & t>t_{\rm break} 
   \end{cases}
\]
where $\tau_e$ is the viscous timescale in the hot (ionized) zone of the disc, $\tau_l$ is the timescale of the irradiation-controlled stage of the decay, $t_{\rm break}$ defines the transition time between viscous and irradiation-controlled stages, $f_t$ is the corresponding bolometric X-ray flux of the system at time $t_{\rm break}$, and $f_2$ represents the bolometric flux limit of the viscous stage of the decay, dependent upon the mass-transfer rate from the companion ($-\dot{M_2}$) and source distance ($d$). See \cite{powell2007} and \cite{heinke2015} for full derivation of this analytical form and \citetalias{tetarenko2018} for a more detailed discussion on the development of this model.

While the formalism developed by \citet{kingrit8} is simplified compared to other formalisms \citep[e.g.,][where the kinematic viscosity is allowed to vary with surface density and time]{lipunova2000}, it remains unclear if the additional layers of complexity in a more detailed semi-analytical model provide a correspondingly clearer physical insight. In addition, we continue using the \citet{kingrit8} formalism for continuity with \citetalias{tetarenko2018}.

\subsection{The Bayesian Hierarchical Methodology}\label{sec:bays_model}

As detailed in \citetalias{tetarenko2018}, the viscous timescale $\tau_e$ in the disc can be written in terms of the $\alpha$-viscosity parameter ($\alpha_h$), which describes the efficiency of angular-momentum and mass transport through the hot zone of the disc, compact object mass ($M_1$) and accretion disc radius ($R_{\rm disc}$) such that,
\begin{equation}
\left(\frac{\tau_e}{1s}\right)=(1\times10^6)\left(\frac{G^{0.5} m_H M_{\odot}^{0.5}}{3 \gamma k_b T_c}\right) \left( \frac{\alpha_h}{0.1} \right)^{-1}  \left( \frac{M_1}{M_{\odot}} \right)^{0.5}  \left( \frac{R_{\rm disc}}{10^{10} \rm{cm}} \right)^{0.5}.
\label{eq:exp}
\end{equation}
This expression was used to constrain $\alpha$ in \citetalias{tetarenko2018}. Note that, as discussed in Paper I and shown in \citet[][see their Fig. 6]{dubus2001}, the central midplane temperature of the disc ($T_c$) is only weakly dependent on viscosity and X-ray irradiation in irradiated discs, thus we can approximate its value as a constant.

The transition from the viscous to the irradiation-controlled (linear) phase of the outburst's decay occurs when the irradiation temperature at the outer radius is $T_{\rm irr}\approx 10^4$ K, the temperature at which hydrogen starts to recombine, since it is the outermost region of the disc that starts the transition to the quasi-neutral, cold state. Therefore we make use of the irradiation law of \autoref{eq:irr} to obtain the value of $C_{\rm irr}$ by assuming $T_{\rm irr}=10^4$ K in
\begin{equation}
\left( \frac{C_{\rm irr}}{T_{\rm irr}^4} \right)=(5.4\times10^{6})\left( \frac{f_t}{10^{-12} {\rm erg s^{-1} cm^{-2}}}\right)^{-1}\left( \frac{d}{{\rm kpc}} \right)^{-2} \left( \frac{R_{\rm disc}}{10^{10} {\rm cm}} \right)^2 .
\label{eq:irr_params}
\end{equation}
Here $C_{\rm irr}$ depends only on the transition luminosity between these two stages of the outburst decay and the known measurements of compact object mass ($M_1$), binary mass ratio ($q$), and orbital period ($P_{\rm orb}$). As these quantities are readily obtained from a combination of fitting X-ray outburst light curves with the analytical decay model (described in \citetalias{tetarenko2018} and \autoref{sec:model}) and a literature search (see \autoref{tab:binaryinfo}), it is possible to
derive observational constraints on the strength of the X-ray irradiation heating the outer regions of LMXB discs using a multi-level Bayesian statistical sampling technique.

\citetalias{tetarenko2018} describes in detail the development and (python) implementation of the Bayesian methodology we use to sample $C_{\rm irr}$ effectively.
In short, we have built a hierarchical model, a multi-level statistical model that makes use of a combination of known prior distributions and observational data to estimate a posterior distribution of a physical quantity effectively. 

Together with \autoref{eq:irr_params}, we sample $C_{\rm irr}$ using only the established binary orbital parameters ($M_1$, $q$, $P_{\rm orb}$) for a system as known priors and the observed X-ray light-curve data. From the light-curves we are able to measure the posterior distribution of the observed flux of the system at the transition between viscous and irradiation-controlled decay stages ($f_t$). This quantity acts as the observational data in our hierarchical model.

\section{Application to the BH-LMXB Population of the Galaxy}\label{sec:data}

\begin{table}
{
\setlength{\tabcolsep}{4pt}
\centering
\caption{ The binary orbital parameters of our Galactic BH-LMXB sample. }
\medskip
\label{tab:binaryinfo}
\begin{tabular}{lcccl} 
\hline
		Source Name & distance & $M_1$ & $q$& $P_{\rm orb}$   \\
		 & (kpc) & $(M_{\odot})$ & $(M_2/M_1$) & (hrs)  \\
\hline
XTE J1118+480 & $1.72\pm0.1$ & $7.2\pm0.72$ & $0.024\pm0.009$ & $4.1$\\
	MAXI J1305$-$704& \nodata &-& \nodata &9.74\\
	Swift J1357.2$-$0933&2.3--6.3&$12.4\pm3.6$&$0.04^{+0.005}_{-0.003}$&2.8\\
	GS 1354$-$64& \nodata &-&0.12$\pm$0.04&61.1\\

	4U 1543$-$475&7.5$\pm$0.5&9.4$\pm$2.0&0.25--0.31&26.8\\
	XTE J1550$-$564&4.4$\pm$0.5&10.39$\pm$2.3&0.031--0.037&37.0\\

	XTE J1650$-$500&2.6$\pm$0.7&4.7$\pm$2.2& \nodata &7.7\\

	GRO J1655$-$40&3.2$\pm$0.5&5.4$\pm$0.3&0.38$\pm$0.05&62.9\\
	MAXI J1659$-$152&1.6--8.0& \nodata &-&2.414\\
	GX 339$-$4&8.0$\pm$2.0& \nodata &-&42.1\\

	Swift J1745$-$26& \nodata &-& \nodata &$\leq$21\\

	MAXI J1836$-$194& \nodata &-& \nodata &$<$4.9\\

	XTE J1859+226&8$\pm$3&10.83$\pm$4.67& \nodata &6.6\\
\hline
 \multicolumn{5}{p{0.95\columnwidth}}{\hangindent=1ex NOTE. -- All binary parameters taken from the WATCHDOG catalogue \citep{tetarenkob2015}, with the exception of SwiftJ1357.2$-$0933 \citep{casares2016}. The observed Galactic BH distributions from \cite{tetarenkob2015} and \cite{oz10} are used when no acceptable estimates of BH mass $M_1$ or binary mass ratio $q$ are available in the literature. A distance of 8kpc is assumed when no distance estimates exist in the literature.} 
\end{tabular}\\
}

\end{table}

\begin{table*}
	\centering
	\caption{Outburst History for our Galactic BH-LMXB Source Sample}
	\medskip
	\label{tab:outinfo}
	\begin{tabular}{lccccl} 
		\hline
		Source Name & Outburst & $t_b$ & $t_e$ & Data & Refs.\\
		& Year & (mjd) & (mjd) &Available & \\
		\hline

XTE J1118+480 & 1999/2000 & 51538.0 & 51770.0 &PCA&-\\[0.045cm]
& 2005 & 53380.0 & 53420.0&PCA&-\\[0.2cm]
MAXI J1305$-$704&2012& 56009.5& 56190.0&GSC,XRT&-\\[0.2cm]
Swift J1357.2$-$0933&2011& 55576.5 & 55653.0&EPIC,PCA,XRT&1\\[0.045cm]
&2017& 57874.0 & 57977.0&XRT&-\\[0.2cm]
GS 1354$-$64&1997/1998& 50714.0& 50870.0&PCA&-\\[0.045cm]
&2015 &57153.0& 57315.0&GSC,XRT&-\\[0.2cm]

4U 1543$-$475&2002& 52435.0& 52503.0&EPIC,PCA&2\\[0.2cm]

XTE J1550$-$564&1998/1999*& 51062.0& 51316.0&PCA&-\\[0.045cm]
&2000& 51597.0& 51719.0&ACIS-S,PCA&3--5\\[0.045cm]
&2001 &51934.0 &51986.0&PCA&-\\[0.045cm]
&2001/2002& 52261.0& 52312.0&ACIS-S,PCA&5\\[0.045cm]
&2003& 52725.0& 52775.0&PCA&-\\[0.2cm]

XTE J1650$-$500&2001/2002 &52149.0 &52366.0&ACIS-S,PCA&7\\[0.2cm]

GRO J1655$-$40&1996/1997*& 50184.0 &50690.0&PCA&-\\[0.045cm]
&2005*& 53415.0& 53654.0&PCA&-\\[0.2cm]

MAXI J1659$-$152&2010/2011& 55456.5 &55685.0&ACIS-S,GSC,PCA,XRT&8\\[0.2cm]

GX 339$-$4& 1996-1999 &50259.0 &51298.0 &PCA&-\\[0.045cm]
& 2002/2003* &52350.0& 52750.0 &PCA&-\\[0.045cm]
& 2004/2005*& 53054.0 &53515.0 &PCA&-\\[0.045cm]
& 2006& 53751.0 &53876.0 &PCA&-\\[0.045cm]
& 2006/2007*& 54053.0 &54391.0 &PCA,XRT&-\\[0.045cm]
&2008 &54624.0 &54748.0 &PCA,XRT&-\\[0.045cm]
&2009 &54875.0 &55024.0 &EPIC,PCA,XRT&9,10\\[0.045cm]
& 2009-2011*& 55182.5 &55665.0 &ACIS-S,GSC,PCA,XRT&11\\[0.045cm]
&2013 &56505.5 &56608.0 &GSC,XRT&-\\[0.045cm]
&2014/2015 &56936.0& 57311.0&GSC,XRT&-\\[0.2cm]

Swift J1745$-$26&2012/2013& 56178.0& 56463.0&XRT&-\\[0.2cm]

MAXI J1836$-$194&2011/2012& 55793.5 &56154.5&GSC,PCA,XRT&-\\[0.2cm]

XTE J1859+226&1999/2000 &51437.0& 51661.0&PCA&-\\[0.2cm]

\hline
\multicolumn{6}{p{1.3\columnwidth}}{\hangindent=1ex NOTE. -- The outburst year and start ($t_b$) and end ($t_e$) times of the outburst are taken from the WATCHDOG catalogue \citep{tetarenkob2015}. A ``*'' in the outburst year indicates that the outburst in question displays complex variability, and thus is not included in the analysis of this paper. References for Chandra and XMM-Newton data used -- [1] \citet{armaspadilla2014}, [2] \citet{lapalombara2005}, [3] \citet{tom01a}, [4] \citet{tomsick3}, [5] \citet{corbel2006}, [7] \citet{tom4}, [8] \citet{jonk12}, [9] \citet{basak2016}, [10] \citet{plant2014}, and [11] \citet{co13}.} 
	\end{tabular}
\end{table*}

\begin{table*}
{
\setlength{\tabcolsep}{0.8pt}
\centering
\caption{ Results of our Bayesian Methodology applied to Outbursts of BH-LMXBs}
\medskip
\label{tab:outfits}
\begin{tabular}{lcccccccccc} 
\hline
Source Name & Outburst & Function&Outburst &$f_{\rm t} \, (\times 10^{-12})$ & $t_{\rm break}$ &$\tau_{l}$ & $f_{\rm 2} \, (\times 10^{-12})$ &$\tau_{e}$&$\alpha_h$$^b$&$C_{\rm irr}$$^{c}$ \\
  & Year & Type&Class$^a$&($\rm{ergs^{-1}cm^{-2}s^{-1}}$)  & (mjd) &(days)& ($\rm{ergs^{-1}cm^{-2}s^{-1}}$)& (days)&&\\[0.2cm]
  \hline
XTEJ1118+480 & 1999/2000 & exp+lin&A & ${2843}^{+15}_{-16}$ & ${51726.23}^{+0.36}_{-0.37}$ & ${34.18}^{+0.47}_{-0.45}$ & ${2831}^{+16}_{-17}$ & ${85.96}^{+0.55}_{-0.56}$&${0.279}^{+0.017}_{-0.018}$&
$\left(1.20_{-0.40}^{+0.44}\right) \times 10^{-1}$\\[0.045cm]
 & 2005 & exp &C& ${0.00002}^{+0.0045}_{-0.00000044}$ & ${53465.37}^{+3.37}_{-0.02}$ & \nodata & ${0.0000079}^{+0.0019}_{-0.00000049}$ & ${79.0}^{+1.3}_{-1.0}$&${0.303}^{+0.019}_{-0.021}$&$>4.9\times 10^{0}$\\[0.2cm]
MAXIJ1305-704 & 2012 & exp+lin&A & ${896}^{+10}_{-12}$ & ${56128.28}^{+0.19}_{-0.10}$ & ${97.7}^{+4.2}_{-4.0}$ & ${0.78}^{+0.21}_{-0.20}$ & ${52.90}^{+0.11}_{-0.12}$&${0.49}^{+0.11}_{-0.11}$&
$\left(2.31_{-1.40}^{+1.96}\right) \times 10^{-2}$\\[0.2cm]
SWIFTJ1357.2-0933 & 2011 & exp+lin&A & ${425}^{+32}_{-33}$ & ${55647.0}^{+3.3}_{-2.8}$ & ${57.6}^{+2.4}_{-2.8}$ & ${173}^{+28}_{-30}$ & ${68.3}^{+2.2}_{-2.0}$&${0.346}^{+0.067}_{-0.065}$&
$\left(4.5_{-2.8}^{+6.8}\right) \times 10^{-2}$\\[0.045cm]
 & 2017 & exp+lin &A& ${142}^{+20}_{-18}$ & ${57909.9}^{+5.2}_{-4.9}$ & ${63.4}^{+3.6}_{-3.7}$ & ${7.0}^{+1.0}_{-1.0}$ & ${64.9}^{+3.5}_{-3.7}$&${0.366}_{-0.070}^{+0.066}$&
 $\left(1.31_{-0.81}^{+2.05}\right) \times 10^{-1}$\\[0.2cm]
 GS1354-64 & 1997/1998 & lin&C & ${7266}^{+77}_{-77}$ & ${50774.25}^{+0.86}_{-0.88}$ & ${90.6}^{+1.5}_{-1.5}$ & \nodata & \nodata&\nodata&$<2.8\times 10^{-3}$\\[0.045cm]
 & 2015 & exp&B & ${57.8}^{+2.8}_{-2.7}$ & ${57358.49}^{+0.94}_{-0.91}$ & \nodata & ${0.0060}^{+0.0010}_{-0.0011}$ & ${139.69}^{+0.63}_{-0.65}$&${0.362}^{+0.070}_{-0.066}$&$>3.7\times 10^{-3}$\\[0.2cm]
4U1543-475 & 2002 & exp+lin&A & ${79.0}^{+6.1}_{-5.6}$ & ${52501.36}^{+0.60}_{-0.59}$ & ${3.43}^{+0.69}_{-0.72}$ & $5.6^{+1.0}_{-1.0}$ & $58.94^{+0.42}_{-0.42}$&${0.66}^{+0.16}_{-0.14}$&
$\left(1.16_{-0.69}^{+0.99}\right) \times 10^{0}$\\[0.2cm]
XTEJ1550-564 & 2000 & exp+lin &A& ${50.9}^{+2.9}_{-3.3}$ & ${51715.25}^{+0.57}_{-0.48}$ & ${34.2}^{+3.0}_{-2.6}$ & $0.37^{+0.10}_{-0.10}$ & ${61.78}^{+0.38}_{-0.37}$&${0.96}^{+0.15}_{-0.16}$&
$\left(19.8_{-6.7}^{+8.3}\right) \times 10^{0}$\\[0.045cm]
 & 2001 & exp &B& ${52.4}^{+6.0}_{-3.9}$ & ${52014.5}^{+9.5}_{-6.5}$ & \nodata & ${47.4}^{+2.0}_{-2.0}$ & ${61.9}^{+5.0}_{-5.8}$&${0.962}^{+0.101}_{-0.089}$&$>9.8\times 10^{0}$\\[0.045cm]
 & 2001/2002 & exp+lin&A & ${37.0}^{+3.4}_{-3.4}$ & $52339.91^{+0.94}_{-0.94}$ & ${5.18}^{+0.96}_{-0.99}$ & ${30.6}^{+3.6}_{-3.6}$ & ${60.38}^{+0.64}_{-0.63}$&${0.99}^{+0.15}_{-0.15}$&
 $\left(27.2_{-9.5}^{+12.1}\right) \times 10^{0}$\\[0.045cm]
 & 2003 & exp+lin&A & ${1000}^{+10}_{-10}$ & ${52776.93}^{+0.74}_{-0.73}$ & ${4.61}^{+0.83}_{-0.84}$ & ${4.5}^{+2.1}_{-2.0}$ & ${61.89}^{+0.55}_{-0.52}$&${0.96}^{+0.15}_{-0.14}$&
 $\left(1.00_{-0.32}^{+0.39}\right) \times 10^{0}$\\[0.2cm]
XTEJ1650-500 & 2001/2002 & exp+lin&B & ${1267}^{+47}_{-59}$ & ${52230.90}^{+2.1}_{-1.5}$ & ${45.8}^{+1.7}_{-2.1}$ & ${533}^{+16}_{-16}$ & $93.1^{+1.3}_{-1.3}$&${0.185}^{+0.034}_{-0.052}$&
$\left(7.3_{-4.6}^{+7.8}\right) \times 10^{-2}$\\[0.2cm]
MAXIJ1659-152 & 2010/2011 & exp+lin &A& ${3000}^{+350}_{-380}$ & ${55522.6}^{+1.9}_{-1.6}$ & ${30.0}^{+3.2}_{-2.8}$ & ${5.8}^{+2.1}_{-2.1}$ & ${60.7}^{+1.2}_{-1.2}$&${0.265}^{+0.059}_{-0.064}$&
$\left(2.9_{-2.0}^{+8.2}\right) \times 10^{-3}$\\[0.2cm]
GX339-4 & 1996--1999 & exp+lin&B & ${2700}^{+10}_{-10}$ & ${51254.8}^{+1.3}_{-1.3}$ & ${75.6}^{+1.7}_{-1.6}$ & ${10.0}^{+2.1}_{-2.1}$ & ${167.2}^{+2.1}_{-2.3}$&${0.250}^{+0.059}_{-0.056}$&
$\left(4.7_{-3.1}^{+7.0}\right) \times 10^{-2}$\\[0.045cm]
 & 2006 & lin &A& ${2456}^{+10}_{-10}$ & $53742.7^{+1.1}_{-1.1}$ & ${160.0}^{+1.0}_{-1.0}$ & \nodata & \nodata & \nodata &$<1.2\times 10^{-1}$\\[0.045cm]
 & 2008 & exp &B& ${16.7}^{+3.2}_{-2.9}$ & ${54802.3}^{+8.5}_{-8.2}$ & \nodata & ${6.9}^{+2.0}_{-2.0}$ & ${168.2}^{+5.9}_{-5.8}$&${0.247}^{+0.061}_{-0.056}$&$>6.8\times 10^{-1}$\\[0.045cm]
 & 2009 & exp&B & ${22.8}^{+6.0}_{-3.5}$ & ${55048.3}^{+5.6}_{-7.6}$ & \nodata & ${1.31}^{+0.49}_{-0.52}$ & ${166.9}^{+5.0}_{-4.5}$&${0.249}^{+0.060}_{-0.057}$&$>6.9\times 10^{-1}$\\[0.045cm]
 & 2013 & exp &B& ${0.0310}^{+0.0084}_{-0.0069}$ & ${56716.0}^{+4.8}_{-4.5}$ & \nodata & ${0.0084}^{+0.0048}_{-0.0046}$ & ${172.4}^{+3.1}_{-3.5}$&${0.242}^{+0.058}_{-0.054}$&$>3.3\times 10^{-1}$\\[0.045cm]
 & 2014/2015 & exp+lin &B& ${2218}^{+16}_{-15}$ & ${57233.70}^{+0.34}_{-0.34}$ & ${56.77}^{+0.34}_{-0.33}$ & ${0.14}^{+0.23}_{-0.11}$ & ${188.90}^{+0.25}_{-0.23}$&${0.222}^{+0.049}_{-0.052}$&
 $\left(5.7_{-3.7}^{+8.5}\right) \times 10^{-2}$\\[0.2cm]
SWIFTJ1745-26 & 2012/2013 & exp+lin&B & ${13280}^{+100}_{-100}$ & ${56266.5}^{+2.8}_{-2.6}$ & ${104.0}^{+4.2}_{-4.4}$ & ${3070}^{+100}_{-100}$ &
${81.5}^{+1.9}_{-1.9}$&${0.410}^{+0.097}_{-0.091}$&
$\left(4.4_{-2.7}^{+3.7}\right) \times 10^{-3}$\\[0.2cm]
MAXIJ1836-194 & 2011/2012 & exp+lin&B & ${1132}^{+25}_{-22}$ & ${55894.4}^{+2.7}_{-2.6}$ & ${212.8}^{+2.6}_{-2.7}$ & ${1027}^{+16}_{-15}$
& ${93.1}^{+1.8}_{-2.0}$&${0.220}^{+0.049}_{-0.053}$&
$\left(7.3_{-4.5}^{+6.3}\right) \times 10^{-3}$\\[0.2cm]
XTEJ1859+226 & 1999/2000 & exp+lin &A& ${2648}^{+13}_{-13}$ & ${51507.12}^{+0.12}_{-0.11}$ & ${111.55}^{+0.52}_{-0.50}$ & ${152}^{+10}_{-10}$ & ${56.61}^{+0.066}_{-0.084}$&${0.505}^{+0.142}_{-0.093}$&
$\left(5.0_{-3.2}^{+6.8}\right) \times 10^{-3}$\\[0.2cm]
\hline
\multicolumn{11}{p{2.2\columnwidth}}{\hangindent=1ex $^a$Class of the outburst describing how confident we are in the fit given the available data. See \autoref{sec:fitsdiscuss} for a detailed explanation for each individual outburst. }\\
\multicolumn{11}{p{1.3\columnwidth}}{\hangindent=1ex $^b$from \citetalias{tetarenko2018}.}\\
\multicolumn{11}{p{2.1\columnwidth}}{\hangindent=1ex $^{c}$Upper and lower limits on $C_{\rm irr}$ are calculated in the cases of pure linear decays by assuming $f_t$ is the maximum observed flux and pure exponential decays by using the minimum observed flux, respectively.}\\
\end{tabular}\\
}
\end{table*}

\subsection{Source and Outburst Selection}

We have used the WATCHDOG catalogue \citep{tetarenkob2015} to compile a representative sample of BH (and BH candidate) LMXBs in our Galaxy. This sample, consisting of 13 BH-LMXBs and 30 individual outbursts undergone by these sources,
includes only those systems with a known $P_{\rm orb}$ that have underwent at least one outburst since 1996. Tables \ref{tab:binaryinfo} and \ref{tab:outinfo} display binary parameter information, outburst information, and data availability for our source/outburst sample.

\subsection{Mining X-ray Light-Curves of the Galactic Population}

We have collected X-ray data available during outbursts occurring in our source sample from the following instruments: (i) Proportional Counter Array (PCA) aboard the Rossi X-ray Timing Explorer (RXTE), (ii) X-ray Telescope (XRT) aboard the Neil Gehrels Swift Observatory, (iii) Gas-Slit Camera (GSC) aboard the Monitor of All-sky Image (MAXI) Telescope, (iv) Advanced CCD Imaging Spectrometer (ACIS-S) and High Resolution Camera (HRC-S) aboard the Chandra X-ray Observatory, and (v) European Photon Imaging Camera (EPIC) aboard XMM-Newton.

We used the RXTE/PCA and MAXI/GSC data obtained with the WATCHDOG project \citep{tetarenkob2015}. This compilation includes all (i) good pointed PCA observations (i.e. no scans or slews) available (over the 16-year RXTE mission) in the HEASARC archive and (ii) publicly available MAXI/GSC data from the MAXI archive\footnote{http://maxi.riken.jp/top/}.
We obtained Swift/XRT data, including all available windowed-timing and photon-counting mode pointed observations, from the Swift/XRT online product builder\footnote{http://www.swift.ac.uk/user\_objects/index.php}\citep{evans2009}.
Finally, we collected select pointed observations with Chandra/ACIS-S, Chandra/HRC-S, and XMM-Newton/EPIC, occurring during the decay phase of outbursts in our sample, from the literature. See \autoref{tab:outinfo} for details.

All RXTE/PCA, Swift/XRT, and MAXI/GSC light-curves were extracted in the 2--10 keV band. Following \citet{tetarenkob2015}, individual instrument count-rates were then converted to flux by using crabs as a baseline unit and calculating approximate count rate equivalences. Count-rates from Chandra/ACIS-S, Chandra/HRC-S, and XMM-Newton/EPIC were converted to flux in the 2--10 keV band using PIMMS v4.8c\footnote{http://cxc.harvard.edu/toolkit/pimms.jsp} and spectral information available in the literature. 
Lastly, all 2--10 keV band flux light-curves were converted to bolometric flux light curves using a combination of the bolometric corrections estimated for each BH-LMXB accretion state by \cite{migliari2006} and WATCHDOG project's online Accretion-State-By-Day tool\footnote{http://astro.physics.ualberta.ca/WATCHDOG}, the latter of which provides accretion state information on daily timescales during outbursts of BH-LMXBs.
For a detailed account of the complete data reduction and analysis procedures used refer to \citetalias{tetarenko2018}.

\section{Results}\label{sec:results}

\subsection{X-ray Light Curve Fitting}\label{sec:fits}

By fitting the decay profiles found in our sample of BH-LMXB X-ray light curves with the 
analytic irradiated disc instability model described in \citetalias{tetarenko2018} and \autoref{sec:model}, we can derive 
the flux level at which the transition occurs between the viscous and irradiation-controlled decay stages in a light-curve. We find this transition flux found in BH-LMXB light-curves to occur between $\sim3.6\times10^{-11}-1.3\times10^{-8} {\rm\,erg\,cm^{-2}\,s^{-1}}$ (for models whose fits we classified as trusted -- Class A; see \autoref{tab:outfits} and \autoref{sec:fitsdiscuss}).

All fitting was performed in logarithmic bolometric flux space, as opposed to luminosity space, to avoid the possibility of correlated errors resulting from uncertain distance estimates. Uncertainties in the distance (as well as other binary parameters) are incorporated within the Bayesian Hierarchical model itself.
Secondary maxima and other rebrightening events
that can contaminate BH-LMXB decay profiles are removed by hand before fitting occurs. Removing such events has been found to have no effect on either of the characteristic timescales derived from the X-ray light-curves. 

All 23 fitted X-ray light-curves are 
presented in panels of \autoref{fig:real_lcs}.
Each light-curve has been plotted in logarithmic space on the main axis. In addition, a small zoomed-in inset, displaying the outburst in linear space, is also included. Data in each figure has been colour-coded by instrument: RXTE/PCA (purple), Swift/XRT (blue), MAXI/GSC (green), Chandra/ACIS-S and Chandra/HRC-S (pink), and XMM-Newton/EPIC (orange). All data not included in the fits (including the outburst rise and rebrightening events) are displayed in translucent versions of these colors. Shaded background colours show accretion state information of the source, computed with the WATCHDOG project's Accretion State-by-Day tool \citep{tetarenkob2015}, throughout the outburst on a daily timescale.

A sizeable fraction of BH-LMXB outburst light-curves in our sample do not display simple ``clean'' decays. In fact, of the 30 outbursts in our sample, 23\% (7/30) exhibit complex variability, in the form of multiple intermediate flares and decays, throughout the individual outbursts themselves. While 50\% (15/30) show a combination of exponential plus linear decays, 20\% (6/30) show pure exponential decays and 7\% (2/30) show pure linear decays. 
We reiterate that one should by no means assume that  the standard disc-instability picture governs the complex variability observed in the form of intermediate flares/decays. 
As our analytical decay model is  too simple to draw any conclusions about the cause of this complex variability, we do not fit or include these outbursts that exhibit ``complex variability'' (marked by a ``*'' in \autoref{tab:outinfo}) in any further analysis presented in this paper. Instead, we review possible causes of this behaviour in the discussion.

\subsection{The Outburst Light-Curve Sample}\label{sec:fitsdiscuss}

In \autoref{tab:outfits}, each outburst in our sample has been assigned a class (A, B, or C) to indicate how confident we are that the best fit preferred by our algorithm accurately describes and constrains the outburst light-curve behaviour. We define these three classes as follows: (A) 
the data clearly constrain the shape of both the viscous (exponential) and  irradiation-controlled (linear) stages of the decay, as well as the transition point between these two stages;
(B) While the data clearly indicate an exponential or linear decay type,
missing data in the early (near the outburst peak) or late (in the irradiation-controlled decay) stages of the outburst introduce uncertainty in the fitted transition flux or irradiation-controlled decay timescale; (C) Due to insufficient data available, 
we cannot be confident in our identification of the decay type, or other fit parameters.
In the following paragraphs, we explain our reasoning behind our classifying individual outbursts as Class B or C.

\begin{itemize}[label={},leftmargin=1em, labelwidth=0em, labelsep=0em, itemsep=0.5\baselineskip]
\item \noindent\textit{GS1354-64 (1997/1998): (Class C)} While the algorithm prefers a pure linear fit, the limited data for this outburst does not clearly discriminate between a linear or exponential fit. 
The 2015 outburst of this source (for which we have relatively complete coverage of both the rise and viscous decay stage) peaks at a similar flux level to the first available data of the 1997/1998 outburst. 
 Stochastic variability in an exponential decay may have led our algorithm to select a pure linear decay instead.

\item\noindent \textit{GS1354-64 (2015), GX339-4 (2013), and XTEJ1550-564 (2001): (Class B)} We have good coverage of the rise and viscous portion of the decay in these outbursts. While this is sufficient to derive a viscous timescale (see \citetalias{tetarenko2018}), we do not observe the transition to the irradiation-controlled decay. Thus, our transition flux estimates cannot be considered reliable.

\item\noindent \textit{GX339-4 (1996-1999): (Class B)} While we have no coverage of the outburst peak, sufficient data is available from the later stages of the viscous decay through to quiescence. Thus, we are confident in the fitted transition flux and irradiation-controlled decay timescale. We note that even though we are missing the outburst peak, comparison to other outbursts of the same source with more complete data coverage validates the fitted viscous timescale and value of $\alpha$-viscosity derived from it (see \citetalias{tetarenko2018}).

\item\noindent \textit{GX339-4 (2008 and 2009): (Class B)} In both of these outbursts we have good data coverage of both the rise and a significant portion of the viscous decay, allowing for an accurate fitted viscous timescale. However, both light-curves display a significant data gap later in the viscous decay stage. It is possible that the source could have decayed to quiescence and exhibited a reflare during these gaps, bringing the validity of the fitted transition flux calculated by our algorithm into question.

\item\noindent \textit{GX339-4 (2014/2015): (Class B)} 
 We have good coverage of the rise and viscous portion of the decay in this outburst, and thus an accurate fitted viscous timescale. However, stochastic variability (e.g. secondary maxima) occurring around the transition between viscous and irradiation-controlled decay stages introduces uncertainty in the transition flux found by our algorithm. Further, clear structure is seen in the residuals during the late stages of the decay. 
Fitting synthetic model light-curves, which include the effects of disc evaporation (see \autoref{sec:discuss_codes}), with our analytical algorithm, we encounter similar residual behavior. We postulate that the steeper decline seen in the data may be the result of the inner disc transitioning to a radiatively inefficient accretion flow, an effect not taken into account in our analytical algorithm.

\item\noindent \textit{MAXIJ1836-194 (2011/2012): (Class B)}
We have good coverage of the rise and viscous decay, then a data gap, after which the source is brighter than before the gap. It is unclear whether the transition to quiescence at the end of our data can be associated with the initial viscous decay, or whether the source would have transitioned to quiescence during the data gap, in the absence of the rebrightening episode.

\item\noindent \textit{SWIFTJ1745-26 (2012/2013) and XTEJ1650-500 (2001/2002): (Class B)} We have sufficient data coverage during the rise and initial portion of the viscous decay stage, allowing for our algorithm to  determine a viscous timescale from these light-curves. 
However, the irregular flaring behaviour seen in these outbursts (e.g. \citealt{yan2017}) requires the removal of much of the later data to fit an appropriate decay curve. The choice of which data to include is subjective and affects the final fitted parameters (transition flux and irradiation-controlled decay timescales) of these outbursts.

\item\noindent \textit{XTEJ1118+480 (2005): (Class C)} 
We have only 11 data points in this decay. Although these are best-fit by an exponential decay, this conclusion is very uncertain. Furthermore, the best-fit decay from our algorithm generates an extremely low transition flux. These lead us to suggest that this decay is actually an irradiation-controlled decay and that this outburst completely lacks a viscous decay.

\end{itemize}

\begin{figure}
  \center
\includegraphics[width=\columnwidth]{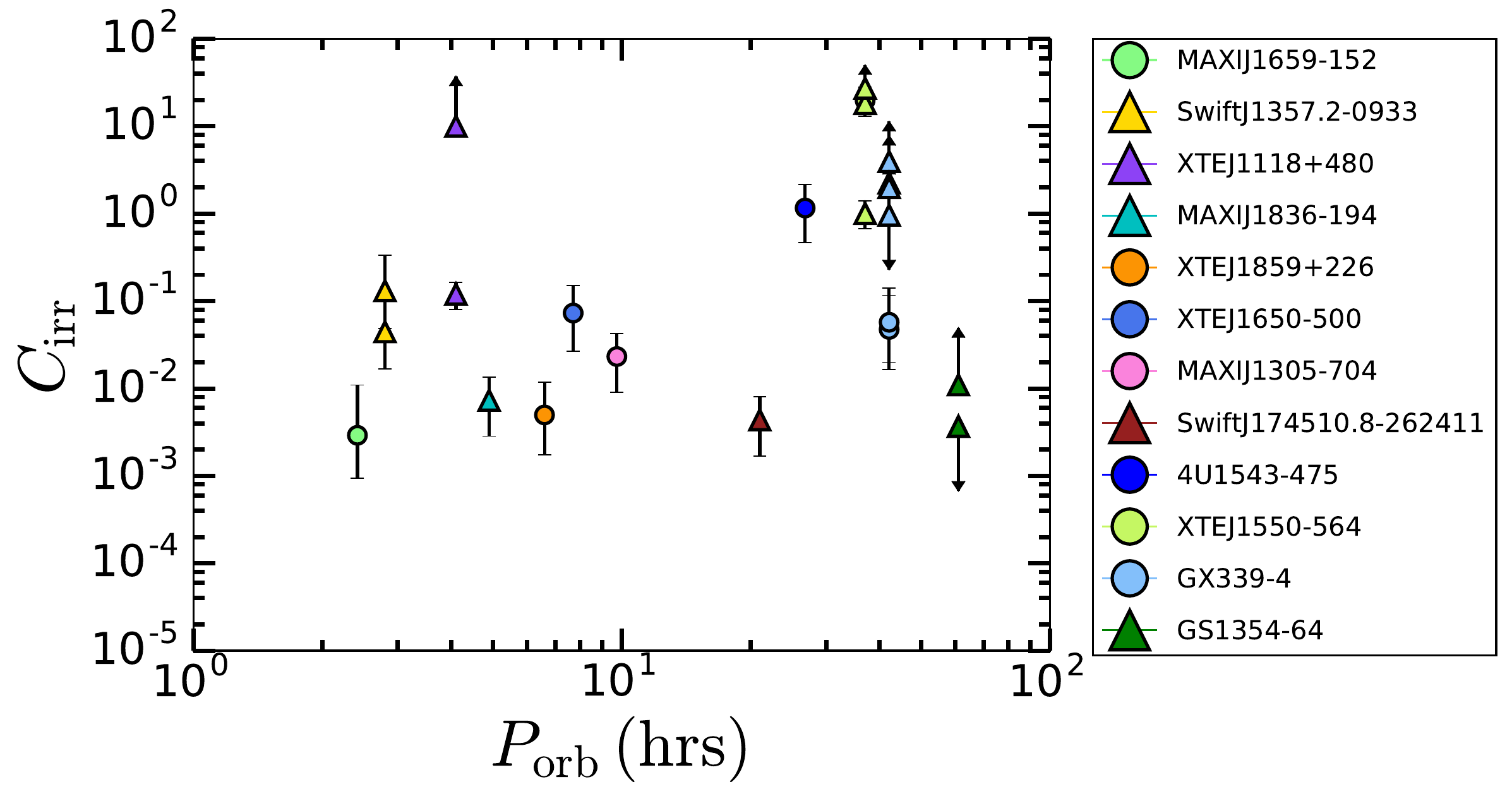}
\caption{
$C_{\rm irr}$, the parameter that encompasses the strength of the X-ray irradiation heating the surface of the outer regions of BH-LMXB accretion discs (derived by our Bayesian methodology) is plotted vs.\ binary orbital period ($P_{\rm orb}$). We include the 23 individual outbursts 
in our sample of 12 Galactic BH-LMXBs with measured orbital periods. Marker colours represent individual sources and marker shape indicates accretion state(s) reached during outburst: (circles) hard/intermediate/soft states and (triangles) only hard state. The error bars show the 68\% confidence interval on $C_{\rm irr}$. $C_{\rm irr}$ is derived during both outbursts where the source cycles through all the accretion states (canonical) and those where the source remains only in the hard state (failed).
}
\label{fig:irr_porb_plot}
\end{figure}

\subsection{The Irradiation Constant (\texorpdfstring{$C_{\rm irr}$}{C\_irr})}
\label{sec:cirr}

\begin{figure}
  \center
\includegraphics[width=\columnwidth]{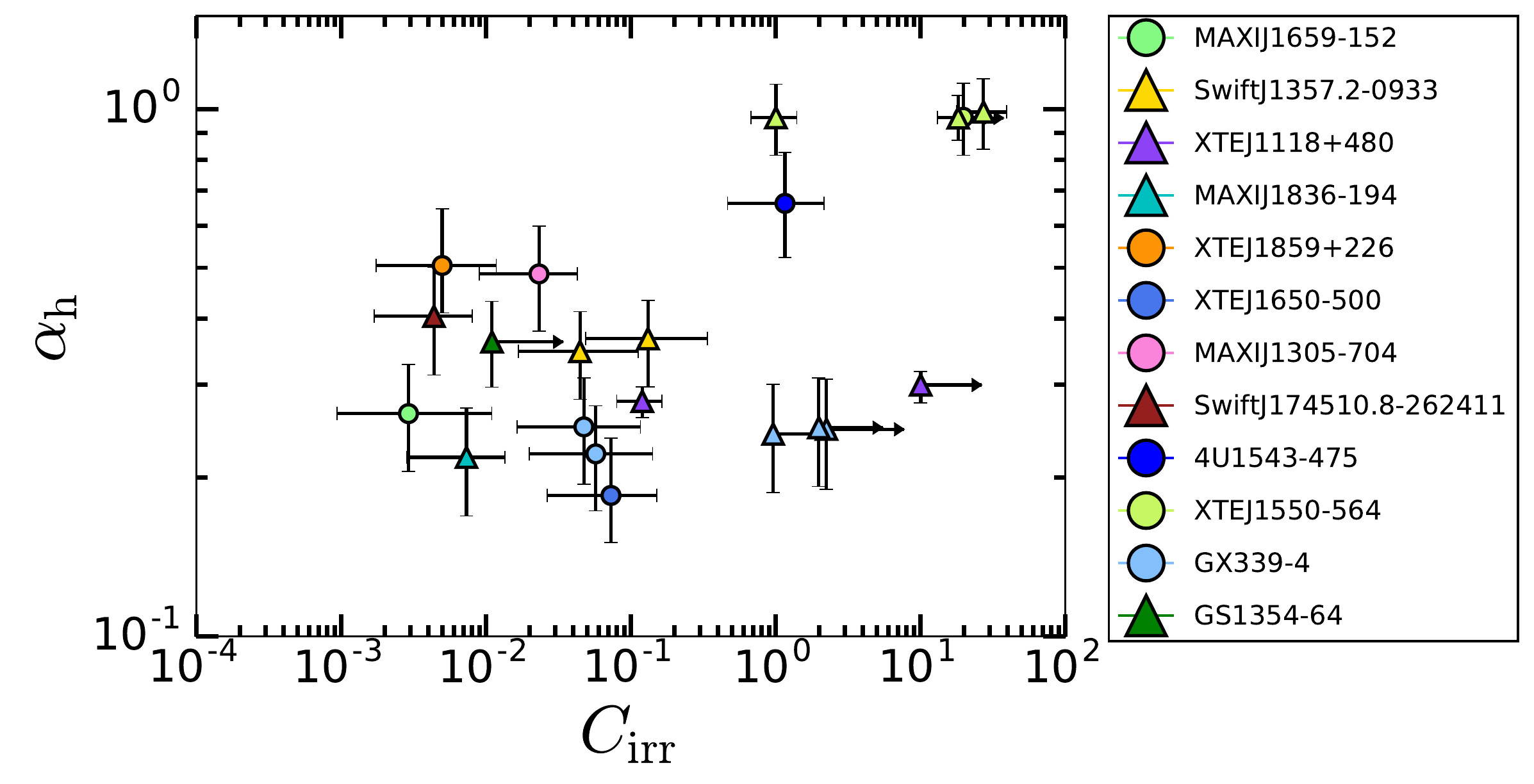}
\caption{
$\alpha$-viscosity ($\alpha_h$) plotted vs.\ $C_{\rm irr}$, derived from our Bayesian methodology. We include the 21 individual outbursts that exhibit exp+lin or pure exp decays. For the pure exp decays, only lower limits on $C_{\rm irr}$ are available. Marker colours represent individual sources and marker shape indicates accretion state(s) reached during outburst: (circles) hard/intermediate/soft states and (triangles) only hard state. The error bars show the 68\% confidence intervals on $C_{\rm irr}$ and $\alpha_h$.
}
\label{fig:C_alpha_plot}
\end{figure}

Using our Bayesian hierarchical methodology (as described in \citetalias{tetarenko2018} and  \autoref{sec:bays_model}), we have sampled the strength of the X-ray irradiation heating the outer regions of BH-LMXB discs, parametrized with the irradiation constant $C_{\rm irr}$. For the 15 outbursts in our sample that display the full exp+lin decay profile, we derive $3\times10^{-3}<C_{\rm irr}<30$.
See \autoref{fig:irr_porb_plot}, 
\autoref{fig:C_alpha_plot},
and  \autoref{tab:outfits}. 

In \autoref{fig:irr_porb_plot}, we see that most, but not all of the systems with $C_{\rm irr} >1$ (i.e., the most unphysically high $C_{\rm irr}$) are associated with long-period systems. Similarly, most, but not all, systems with $C_{\rm irr} >1$ underwent failed outbursts. However, there are at least two long-period, failed outburst systems that do not have unphysical $C_{\rm irr}$. On the other hand, in \autoref{fig:C_alpha_plot}, we see that systems with $C_{\rm irr} >1$ can occur in systems that are more strongly ($\alpha\sim1$) and less strongly ($\alpha\sim0.2$) transferring angular momentum, regardless of the accretion state transitions made during the outburst.
Future work on larger samples will be needed to test if long-period, failed outburst systems continue to dominate the systems where our Bayesian methodology predicts unphysically high $C_{\rm irr}$.

\section{Discussion}\label{sec:discuss}

\subsection{The light-curve profiles of BH-LMXB systems}\label{sec:disc_discuss}

We have found 15 outbursts that display the full exponential+linear shaped decay profile and thus allow us to determine $C_{\rm irr}$ from the transition luminosity. We find values that are typically a factor $\sim 5 $ higher than the expected value, $C_{\rm irr,expected}\sim 5\times 10^{-3}$. Such values can arise if the albedo of the disc is low and if the intercepted fraction is high, both of which might result from an irradiation source that is large and causes X-rays to impinge on the disc vertically (e.g. via a corona). A value of $C_{\rm irr}\sim 3\times 10^{-2}$ would still be compatible with the stability limits between transient and persistent LMXBs \citep{coriat2012}. 

However, we also find unphysical values of $C_{\rm irr}>1$ in 4 outbursts and values $>$0.1 in 2 outbursts.
The latter stretch credibility as they require an unrealistically high fraction of the X-ray flux to be reprocessed. In three cases (panels h, k, l of Figure A1) the transition luminosity is essentially set by the last flux measurement in the lightcurve. 
At worst, we have an upper limit on the transition luminosity, hence a lower limit on $C_{\rm irr}$. This issue is not the case for the others, where the transition can be traced very well in the data. It is interesting to note that the most physically unrealistic values of $C_{\rm irr}$ occour in the largest-orbital-period systems.

We could overestimate $C_{\rm irr}$ for a variety of reasons in the context of the model that we applied to the data: because we underestimate the distance (hence $L_X$); because we underestimate the irradiation flux (e.g. if there is a large FUV contribution that is not accounted for); and because we overestimate the disc radius. We consider that these issues may lead to corrections of $O(1)$ but are unlikely to explain values of $C_{\rm irr}$ reaching 20, more than 1000 times the expected value.

We also find that the linear decay timescale $\tau_l$ and the exponential decay timescale $\tau_e$ differ significantly in some cases, whereas both should be comparable according to the model of \citet{kingrit8}. This model implicitly assumes that the viscosity does not depend on the radius within the hot region of the disc. This assumption is unlikely to be realized since the disc is close to a steady-state \citet{ss73} disc in this region, for which $\nu\propto R^{3/4}$ (see e.g., the physical model for non-stationary viscous $\alpha$-discs from \cite{lipunova2000} and its application to observations in \citealt{suleimanov2008} and \citealt{lipunova2017}). In addition, $\alpha_h$ might be a function of radius \citep{Coleman2016}. In this case, $\tau_l$ will change slightly compared to $\tau_e$ and lead to a more complex relationship of $\dot{M}$ with time. The effect of mass loss via a wind in the hot region is also likely to change $\tau_l$. However, in these cases, toy-model calculations lead us to expect differences of O(1) between $\tau_e$ and $\tau_l$, whereas differences of O(10) are found in \autoref{tab:outfits}, notably when $C_{\rm irr}$ is high.

The standard DIM interpretation of the linear-shaped portion of the decay profile is the result of a cooling front propagating inward through the disc at a speed controlled purely by the decaying X-ray irradiating flux. This model is most likely an oversimplification. 
Realistically, the ways in which these discs are irradiated are complicated by a number of factors. Some possible explanations to explain the light-curve profiles we observe are as follows:
\begin{itemize}[label={},leftmargin=1em, labelwidth=0em, labelsep=0em, itemsep=0.5\baselineskip]
    \item At some point in the outburst decay, the inner disc switches from a radiatively efficient thin disc to a radiatively inefficient corona (i.e. an advection dominated accretion flow; ADAF). The radiative efficiency $\eta$ will decrease with time, whereas it is assumed constant in the model. The transition radius (between these accretion flows) will also move outward as the inner thin disc evaporates \citep{liu1999,menou2000}, on a timescale that may be comparable to the cooling front propagation timescale. This change differs from the model we use where the inner radius of the thin disc is assumed constant. A transition radius that propagates outward will terminate the decay prematurely and result in a small $\tau_l$ \citep{dubus2001}.
    \item The irradiation geometry may not be constant during outbursts, due to changes in a disc warp or to properties of the X-ray corona (leading to $C_{\rm irr}$ changing during outburst; \citealt{esin2000}). A major change would be if the source of irradiating X-rays is not at a distance $R$ but is much closer to the reprocessing site (for instance if the X-rays are produced in the corona directly above the disc),  leading to a measured $C_{\rm irr}>1$ given our definition. This implies the size of the X-ray emitting region would have to be comparable to the size of the optical emission region. We consider this unlikely since most of the energy dissipation in the accretion flow naturally occurs close to the compact object. Furthermore, the irradiation flux would not be decoupled from the local conditions, contrary to what is assumed in the model of the linear decay. 
    \item Spectral state transitions observed during outbursts \citep[e.g.][]{mr06} may change the amount of X-rays absorbed by the outer disc, either because the geometry changes (for instance, because the inner thin disc gives way to a geometrically thick disc, or the size of the corona changes, or an X-ray emitting jet structure appears)  or because harder X-rays deposit heat deeper in the disc, 
    thus leading to a temporally varying $C_{\rm irr}$. However, there is no clear relation between the X-ray state and the value of $C_{\rm irr}$ in the systems investigated here (\S4.3).
    \item Heating of the outer disc by tidal heating of the expanding disc or by the stream impact of incoming material may keep the disc hot longer \citep{buat2001}, especially if the mass transfer rate from the companion is enhanced during outburst \citep{augusteijn1993,esin2000b}.
    \item A disc wind with the ability to remove a significant portion of disc mass throughout the outburst decay could affect how $\dot{M}$ changes with disc radius and, therefore, how $L_X$ evolves with time \citep[see e.g.][]{cannizzo2000}. In our model here, we assumed $\dot{M}$  was constant with radius in the hot region. In \citetalias{tetarenko2018}, we found strong evidence for disc winds throughout the outbursts, due to the unusually short viscous timescales (high $\alpha$) we observe in the light-curves. There is some evidence that high values of $C_{\rm irr}$ are correlated with high values of $\alpha$  (\autoref{fig:C_alpha_plot}).
\end{itemize}

 More generically, the transition luminosity that we are fitting might not be produced by a transition between an exponential and a linear (irradiation-controlled) decay. Instead, the transition luminosity may be produced by some other physical process going on in the disc or X-ray emission region, unrelated to the DIM (e.g. a change in how large-scale magnetic fields diffuse, or in rotational energy-extraction from the black hole, etc). In (at least) a quarter of our systems, some other physics must be altering the later parts of the lightcurves.
However, 
the observed exponential decay is a robust feature of a fully-irradiated disc accreting on a viscous timescale. Hence the results presented in \citetalias{tetarenko2018} are unaffected by the issues raised above.

The 2002 outburst of 4U 1543$-$564 is the only outburst (of the 15 in this paper) where $C_{\rm irr}$ has been estimated previously. Comparing the optical/near-infrared and X-ray lightcurves, \citet{lipunova2017} find $C_{\rm irr} <  6\times10^{-4}$, which is in conflict with our measurement of  $C_{\rm irr} = 1.16_{-0.69}^{+0.99}$. This is one of the sources where we find a linear decay timescale $\tau_l$ and the exponential decay timescale $\tau_e$ that differ significantly.
While this may point to an issue with the simplifying assumptions in \citet{kingrit8}, this outburst remains difficult to fit when the formalism of \citet{lipunova2000} and \citet{lipunova2017} is adopted. \citet{lipunova2017} attempted to fit theoretical lightcurves to the outburst: either  $L_{\rm bol}\propto t^{-10/3}$ for a viscous decay (our ``exponential'' decay) or $L_{\rm bol}\propto (t-t_{\rm end})^{40/13}$ for an irradiation-controlled decay (our ``linear'' decay), where $t_{\rm end}$ is the time the source returns to quiescence. Although they found an acceptable solution for the latter, they only fit to X-ray data taken within $\sim 30$ days of the peak X-ray flux. The X-ray data continue another $\sim 30$ days. While we can reproduce their fit when only $\sim 30$ days of X-ray data are included, a pure irradiation-controlled decay cannot fit the entire lightcurve decay. This demonstrates the need for a more-detailed comparison (which is beyond the scope of this paper) of how different formalisms fit existing data, as well as how different formalisms can or cannot constrain $\alpha$ and $C_{\rm irr}$ based on X-ray lightcurves alone.

\subsection{Comparison of our Bayesian Methodology with Numerical Disc Codes}\label{sec:discuss_codes}

Given the occasional high values of  $C_{\rm irr}$ that we measured in \autoref{sec:cirr} and the potential issues regarding the simplifying assumptions that we discussed in \autoref{sec:disc_discuss}, we investigate here how our Bayesian statistical methodology compares to numerical disc codes that were built to simulate accretion flows in binary systems. We have applied our method to a set of synthetic light-curves computed with the code described in \citet{dubus2001}, which uses the same description of the irradiation flux that we used.
This code is developed from the numerical scheme of \citet{hameury1998}, adapted to include irradiation heating from \citet{dubus1999} and inner disc evaporation \citep{menou2000}.
Using this code we have run 46 individual disc models that cover the large BH-LMXB parameter space well. These models vary from $4\,M_{\odot}<M_1<15 \,M_{\odot}$, $3\times10^{10}<R_{\rm disc}<1\times10^{12}$ cm, $0.1<\alpha_h<1.0$, and $0.005<C_{\rm irr}<0.1$.

By reversing the direction of our Bayesian hierarchical methodology, we gain the ability to predict a light-curve profile.
In this case, the known priors used are $M_1$ and $R_{\rm circ}$ (specified for each code run) and $q$, 
taken as a uniform distribution between the minimum and maximum of the known values of $q$ for all dynamically confirmed BHs in the Galaxy. The ``backwards'' Bayesian hierarchical methodology then uses these known priors in combination with known disc/system properties ($\alpha_h$, $C_{\rm irr}$, $-\dot{M}_2$) specified for each code run, to sample the remaining parameters ($\tau_e$, $\tau_l$, and $L_t$) that our analytical irradiated disc instability model needs to describe a LMXB light-curve profile. For a detailed description of the implementation and use of our Bayesian hierarchical model, see \citetalias{tetarenko2018}.

In 34 of the 46 runs, the heating fronts reach the outer edge of the discs. At the peak of each outburst in these runs,
the entire disc is in the hot, ionized state (i.e. $R_h=R_{\rm disc}$). Thus, (as expected) we observe the characteristic exp+lin shaped decay profile. In the remaining 12 runs the heating fronts do not reach the outer edges of the discs due to weaker irradiation. As $R_h<R_{\rm disc}$ in these cases, the synthetic light-curves exhibit only a pure linear-shaped decay. Unfortunately, in these cases, where the heating front does not reach the outer edge of the disc, we are not able to predict the light-curve profile with the ``backwards'' Bayesian hierarchical methodology.

Taking into account only the runs in which the characteristic exp+lin shaped decay profile is observed, we find that the $1 \sigma$ confidence intervals for the lightcurves generated by the``backwards'' Bayesian methodology include the synthetic model light-curve output by the numerical code 
in 74\% (25/34) of the runs. 

\autoref{fig:lc_compare1}, in the Appendix, display light-curve comparison plots for a representative sample of disc models we have run, demonstrating how our Bayesian hierarchical methodology matches the light-curve profile predicted by the numerical code.

\begin{figure}
  \center
  \includegraphics[width=\columnwidth,scale=0.5]{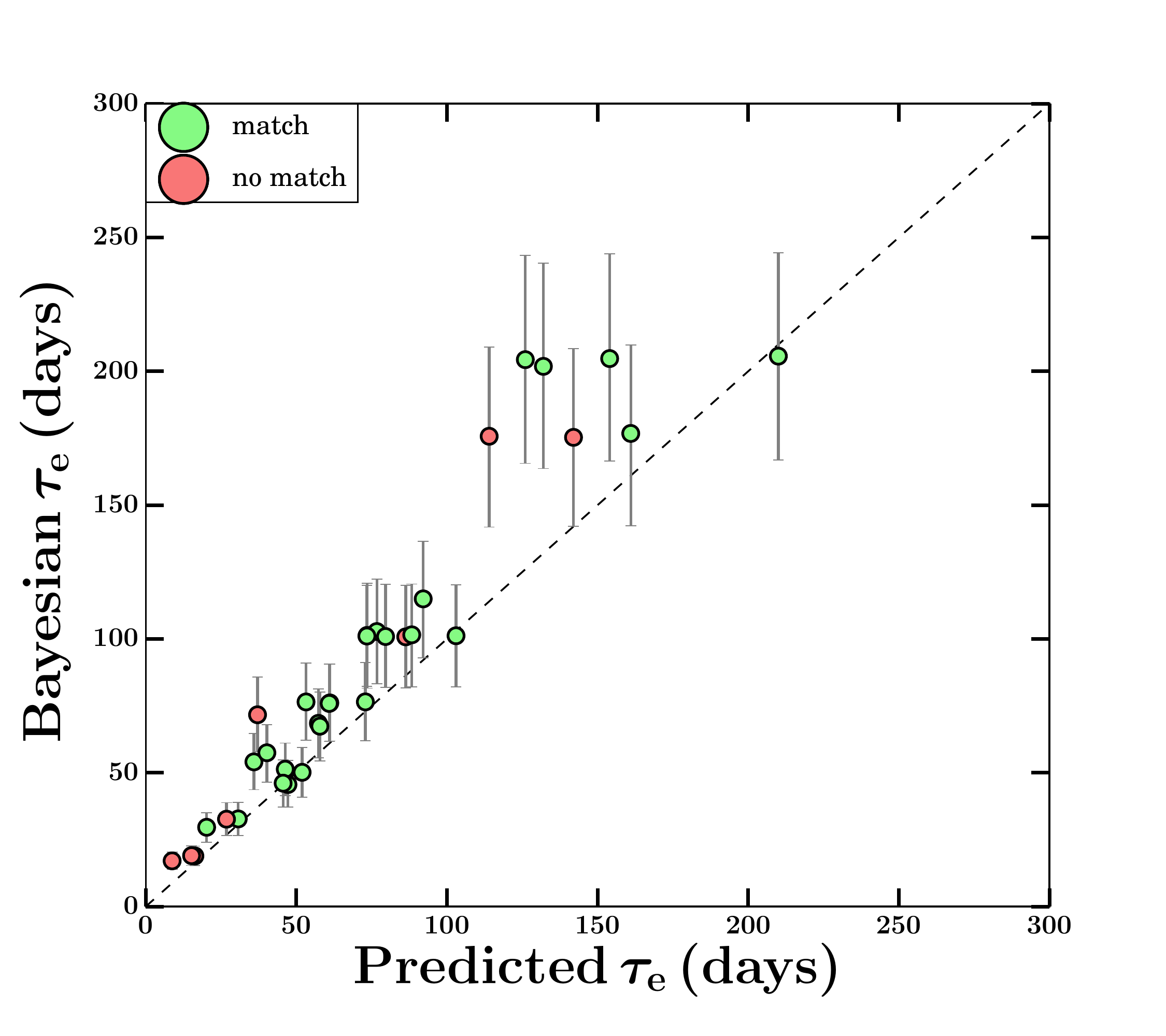}
  \caption{Correlation plot for the viscous timescale in the hot disc ($\tau_e$), comparing the predicted value (from the numerical code) to the Bayesian value (from our methodology). Error bars show the $1\sigma$ confidence interval from our Bayesian methodology. Data is colour coded to show whether or not we can reproduce the entire model light-curve decay with our Bayesian method. The black dotted line represents the 1-to-1 line on the plot.}%
  \label{fig:taue_compare}%
\end{figure}

\begin{figure}

  \center
  \includegraphics[width=\columnwidth,scale=0.5]{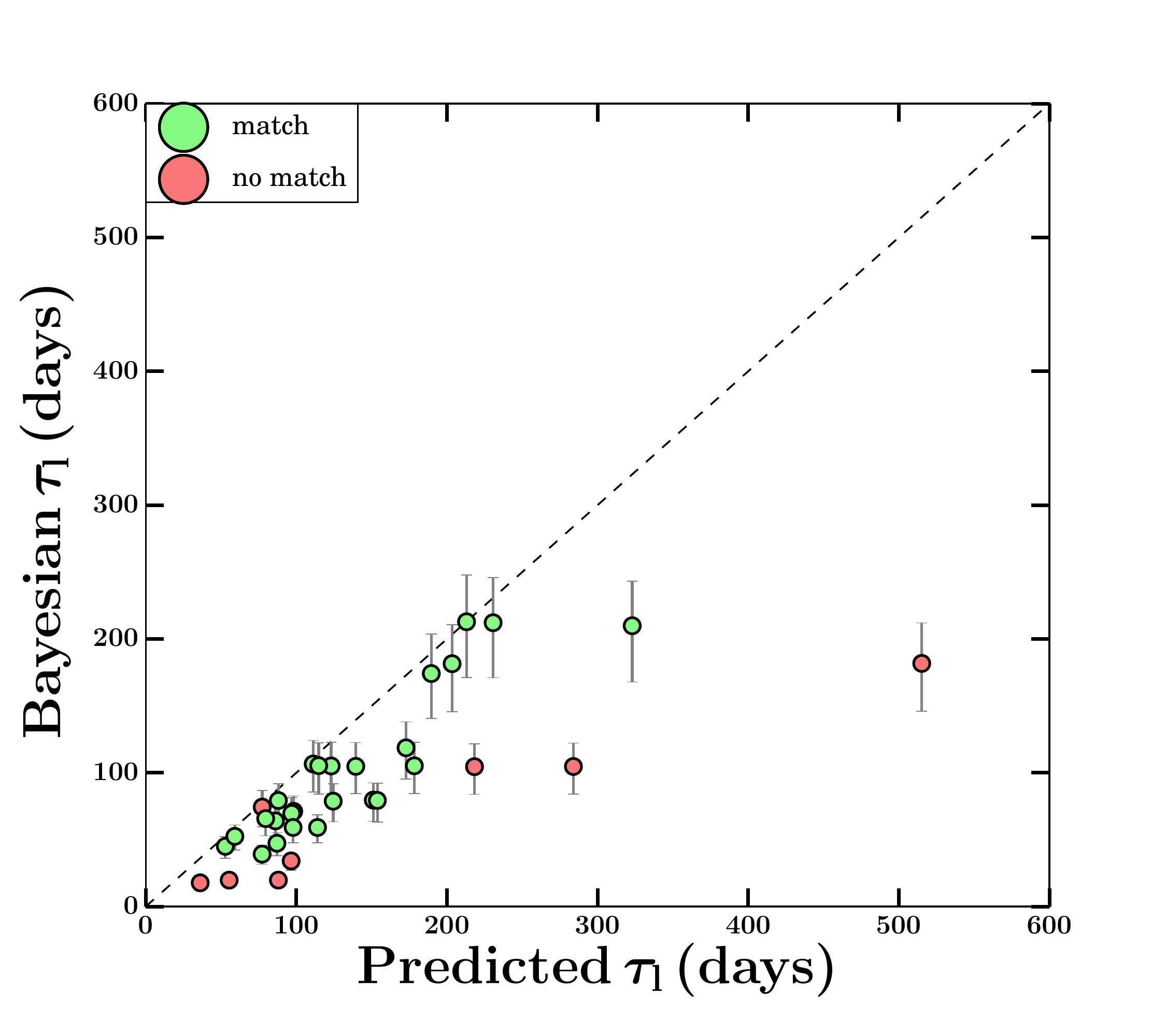}

  \caption{Correlation plot for the linear decay timescale in the disc ($\tau_l$), comparing the predicted value (from the numerical code) to the Bayesian value (from our methodology). Error bars show the $1\sigma$ confidence interval from our Bayesian methodology. Colours are the same as in \autoref{fig:taue_compare}. The black dotted line represents the 1-to-1 line on the plot.}%
  \label{fig:taul_compare}%
\end{figure}

\begin{figure}
  \center
  \includegraphics[width=\columnwidth,scale=0.5]{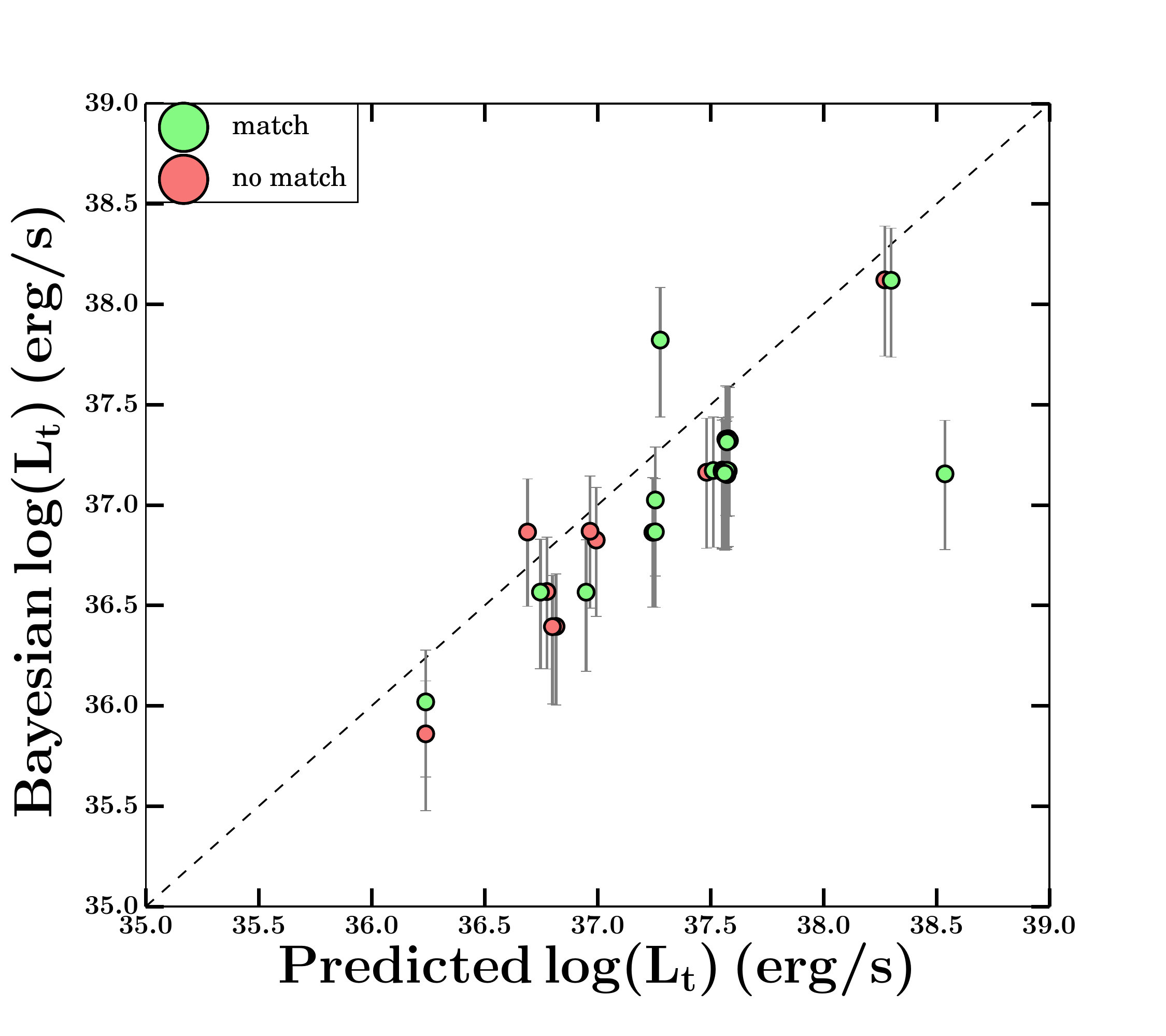}

  \caption{Correlation plot for the transition luminosity in the disc ($L_t$), comparing the predicted value (from the numerical code) to the Bayesian value (from our methodology). Error bars show the $1\sigma$ confidence interval from our Bayesian methodology. Colours are the same as in \autoref{fig:taue_compare}. The black dotted line represents the 1-to-1 line on the plot.}%
  \label{fig:Lt_compare}%
\end{figure}

For each model, the ``backwards'' hierarchical methodology samples $\tau_e$, $\tau_l$, and $L_t$. These parameters can then be used to estimate $\alpha$, and $C_{\rm irr}$ using the same method we used on the observed data.
In Figures \ref{fig:taue_compare}-\ref{fig:Lt_compare}, we display correlation plots, comparing the three light-curve parameters ($\tau_e$, $\tau_l$, and $L_t$, where the latter corresponds to $f_t$ at a known distance) derived from our Bayesian methodology to the same set of parameters predicted by the disc code. Here, each disc model run has been colour coded, with green and red representing those runs in which we effectively match and cannot match the model light-curves to within $1 \sigma$ confidence intervals, respectively. For the well-matched light curves,  individual values of $L_t$ are within 1 (9/25) -- 2 (24/25) $\sigma$ of the model values; we typically underpredict $L_t$ by a factor of $\sim2$.
Similarly, individual values of $\tau_e$ are within 1 (12/25) -- 2 (24/25) $\sigma$ of the model values; we typically overpredict $\tau_e$ by a factor of $\sim1.2$.
We have more difficulties reproducing values of $\tau_l$:
8, 12, and 16 out of 25 models 
are within 1, 2, and 3 $\sigma$ of the model $\tau_l$ values, respectively. Here, if we correct for our underpredicting $\tau_l$ by a factor of $\sim1.5$, we get much stronger agreement: 10 and 24 models are within 1 and 2 $\sigma$ of the model $\tau_l$, respectively. 

Our slight overprediction of $\tau_e$ might suggest that the intrinisic $\alpha$ may be slightly higher than that we measured in \citetalias{tetarenko2018}. This highlights that we were conservative there, even when claiming high values of $\alpha$. We also note that the values of the $\alpha$-viscosity in the hot disc ($\alpha_h$) used to create the synthetic light-curves in each of the well-matched code runs are enclosed within the one-sigma confidence interval of the value of these parameters implied by the ``backwards'' Bayesian methodology in 24/25 cases (the other is within the 2 $\sigma$ confidence interval).

While we underpredict $L_t$ by a factor of $\sim2$, this does not strictly transfer to our having overpredicted $C_{\rm irr}$ by a factor of $\sim 2$, as might be implied from \autoref{eq:irr_params}. In our Bayesian approach, we do not have a strong constraint on $R_{\rm disc}$. And in fact, our Bayesian values of $C_{\rm irr}$ are a factor of $\sim2$ lower than the model's input value. 
Since $R_{\rm disc}$ is sampled from a uniform distribution between the circularization radius $R_{\rm circ}$ and outer disc radius $R_{\rm max}$, $R_{\rm disc, median} \approx R_{\rm max}/2$). 
Given \autoref{eq:irr_params}, this explains how we can both underpredict $L_t$ and $C_{\rm irr}$.
Because of the large range in the $R_{\rm disc}$ prior, all but one of the 1 $\sigma$ confidence intervals for $C_{\rm irr}$ from the Bayesian approach include the model value of $C_{\rm irr}$.

We note that correcting for our underprediction of $C_{\rm irr}$ exacerbates the issue of too-high $C_{\rm irr}$ values we report on in this paper.  The large (and sometimes unphysical) values of $C_{\rm irr}$ that we are deriving via our Bayesian methodology are likely caused by a physical mechanism in the binary systems themselves.

Analyzing the 26\% (9/34) of the runs that are unable to reproduce the model light-curves from the code, we find that our Bayesian methodology has trouble dealing with strong irradiation ($0.01<C_{\rm irr}<0.1$), when combined with large discs ($R_{\rm circ}>1\times10^{11}$ cm) and  large viscosities ($\alpha_h>0.7$). We postulate that a possible explanation for this could stem from the fact that our Bayesian method is underestimating the increase in outburst duration that should happen, as a result of the delay in cooling-front propagation allowing more mass to be accreted, when irradiation is stronger. It remains unclear why our Bayesian method underestimates the timescale of the linear-shaped portion of the decay in these cases.

\section{Summary}\label{sec:conclusion}

The X-ray light-curves of the recurring transient outbursts occurring in LMXBs 
encode within them the physics behind the mechanisms driving mass inflow and outflow in these binary systems.
We have developed an 
algorithm that 
effectively links the disc-instability picture (including irradiation) to observations of real accretion discs. 
This algorithm characterizes a light-curve profile into definitive stages based on observable properties (i.e. timescales, flux) 
describing how matter moves through LMXB discs throughout an outburst.

We have tested this method against model light-curves calculated under the assumptions of the disc instability model, including irradiation and evaporation. We reproduce (to within 1$\sigma$ confidence) the model light-curves derived from the numerical code for 74\% of the disc models we ran, only having trouble reproducing specific models involving a combination of very strong irradiation, large discs, and large values of the $\alpha$-viscosity parameter. We note that, with only the knowledge of the peak outburst flux/luminosity required, our Bayesian methodology can predict the outburst decay profile, and thus, may prove a tool to aid ongoing observational monitoring campaigns of X-ray binaries at optical through X-ray wavelengths.

Applying this Bayesian methodology to a representative sample of X-ray light-curves from outbursts occurring in BH-LMXBs, we have derived observational constraints on the efficiency of the angular-momentum transport process  ($\alpha$-viscosity; presented in \citetalias{tetarenko2018}), and  the strength of the X-ray irradiation heating (parametrized by $C_{\rm irr}$) , 
in the outbursts of LMXBs according to the DIM (this paper). 
We find that the strength of the X-ray irradiation parameter, describing the heating of the outer regions of the discs in these systems, lies in the range $3\times10^{-3}<C_{\rm irr}<30$. Values of $C_{\rm irr}\geq 1$ are clearly unphysical. 
The outburst decay profile is predicted to show a final, linear-shaped stage, due to a cooling front propagating inward through the disc, at a rate controlled by the amount of irradiation heating.
We conclude that our modeling of this stage inadequately describes part of our sample of BH-LMXB outburst light-curves. 
We suggest that the varied light-curve morphology we observe proves that the late-time evolution of the disc is more complex than linear (a dependence that has been obtained using strong simplifying assumptions). It also provides indirect evidence for the existence of a temporal and spatially varying X-ray irradiation source heating the discs in these systems. More likely, given the high values of $C_{\rm irr}$, it suggests that the lightcurve morphology, beyond the exponential decays that are well-accounted for by a viscously-accreting fully-irradiated disc, involve a variety of physical mechanisms of which irradiation is only one. In particular, mass loss through inner disc evaporation to a radiatively-inefficient structure or through a magnetized disc wind may play a prominent role in shaping the outburst lightcurves, a significant change in paradigm.

To begin to understand the evolution of accretion disc structure and the geometry of the X-ray irradiating source heating the discs through the course of a LMXB outburst, it is clear that we require a method that is (i) not limited by the complexity of light-curve morphology observed (e.g. can deal with variability on a range of timescales), or is (ii) tied directly to the simplifying assumptions of the DIM. Possible future avenues of investigation to effectively tackle this complex, multi-scale problem include: making use of simultaneous, multi-wavelength, time-series data sets and phase-resolved spectroscopic data.
For example, one could use the observed UVOIR spectral energy distribution (SED), at different times during an outburst, to model the irradiated disc in the binary system, with the goal of trying to understand the time-series evolution of the X-ray irradiation heating the disc in the system (e.g. \citealt{hynes2005}, \citealt{russell2006}, \citealt{gierlinski2009}, \citealt{meshcheryakov2018}). Another possibility is to make use of a combination of optical and X-ray light-curves of these systems. Here, constraints on $C_{\rm irr}$ can be derived by computing the fraction of X-ray emission needed to be reprocessed to explain the observed optical luminosity (e.g., see \citealt{suleimanov2008,lipunova2017}). A third possibility involves using the correlation between X-ray and optical variability often observed in LMXBs to understand physical properties of the different components (i.e. disc vs. corona) that make up the accretion flow in LMXBs. These properties include the size of the emitting regions, and the characteristic timescales at which matter moves through different regions of the accretion flow (e.g. \citealt{malzac2003,hynes2004,veledina2017,gandhi2017}).

\section*{Acknowledgements}\label{sec:acknowledge}
B.E.T. is grateful to the participants of the ``disks17: Confronting MHD Theories of Accretion Disks with Observations'' program, held at the Kavli Institute for Theoretical Physics (KITP) for their feedback during the early stages of this project. B.E.T. would also like to thank Diego Altamirano, Omer Blaes, and Alexandra Veledina for insightful discussions on the subject matter.  B.E.T., G.R.S., and C.O.H. acknowledge support by NSERC Discovery Grants, and C.O.H. by a Discovery Accelerator Supplement. This research was supported in part by the National Science Foundation under Grant No. NSF PHY-1125915, via support for KITP. J.-P.L acknowledges support by the Polish National Science Centre OPUS grant 2015/19/B/ST9/01099. J.-P.L and G.D. also acknowledge support from the French Space Agency CNES. This research has made use of data, software, and/or web tools obtained from the High Energy Astrophysics Science Archive Research Center (HEASARC), a service of the Astrophysics Science Division at NASA/GSFC and of the Smithsonian Astrophysical Observatory's High Energy Astrophysics Division, data supplied by the UK Swift Science Data Centre at the University of Leicester, and data provided by RIKEN, JAXA, and the MAXI team. This work has also made extensive use of NASA's Astrophysics Data System (ADS).




\bibliographystyle{mnras}
\bibliography{mnras_refs_new}



\appendix

\section{X-ray Light Curves and Light Curve Comparison Plots}

\begin{figure*}
\subfloat[1999/2000 outburst of XTEJ1118+480\label{fig:xte11181}]
  {\includegraphics[width=.3\linewidth]{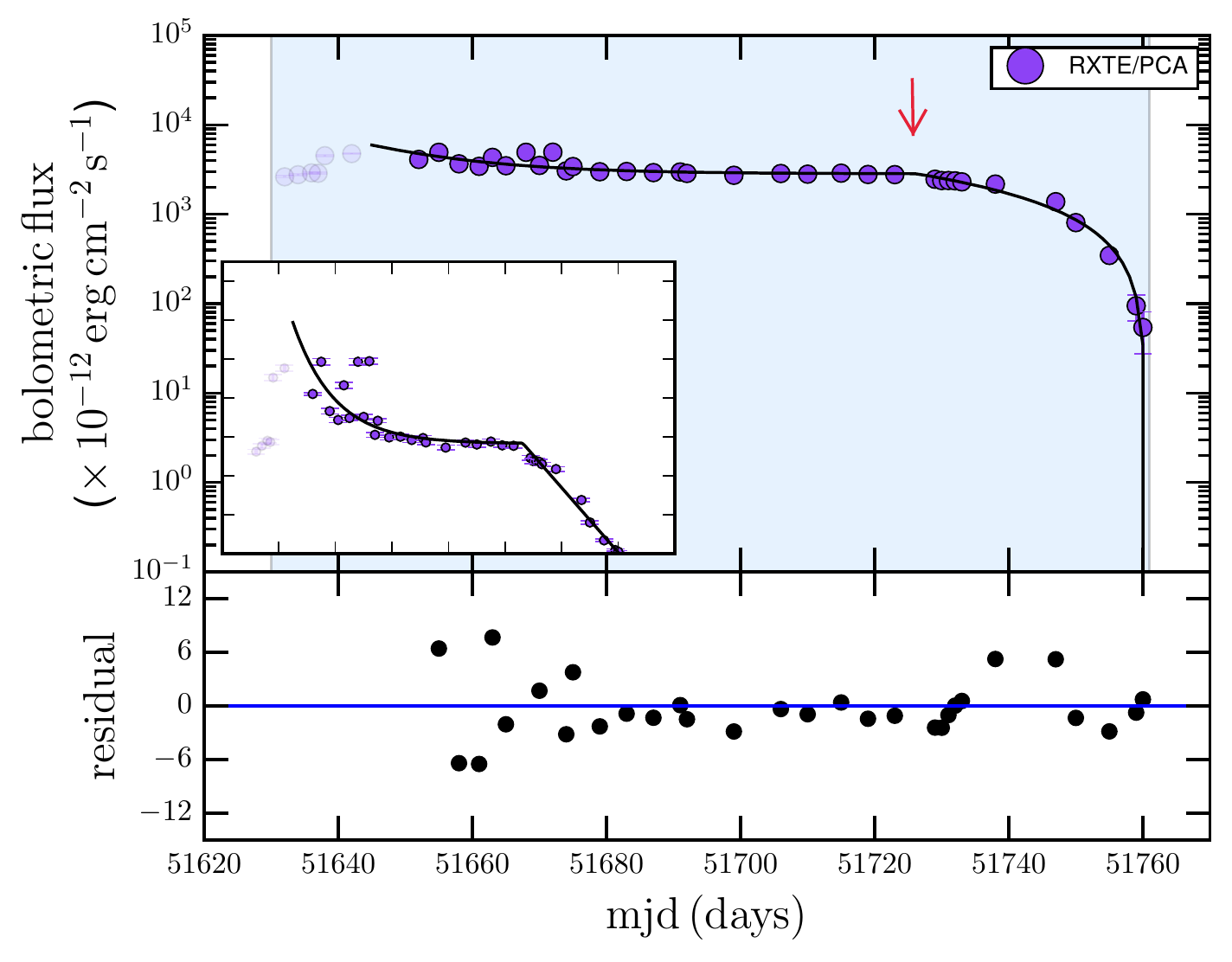}}\hfill
\subfloat[2005 outburst of XTEJ1118+480\label{fig:xte11182}]
  {\includegraphics[width=.3\linewidth]{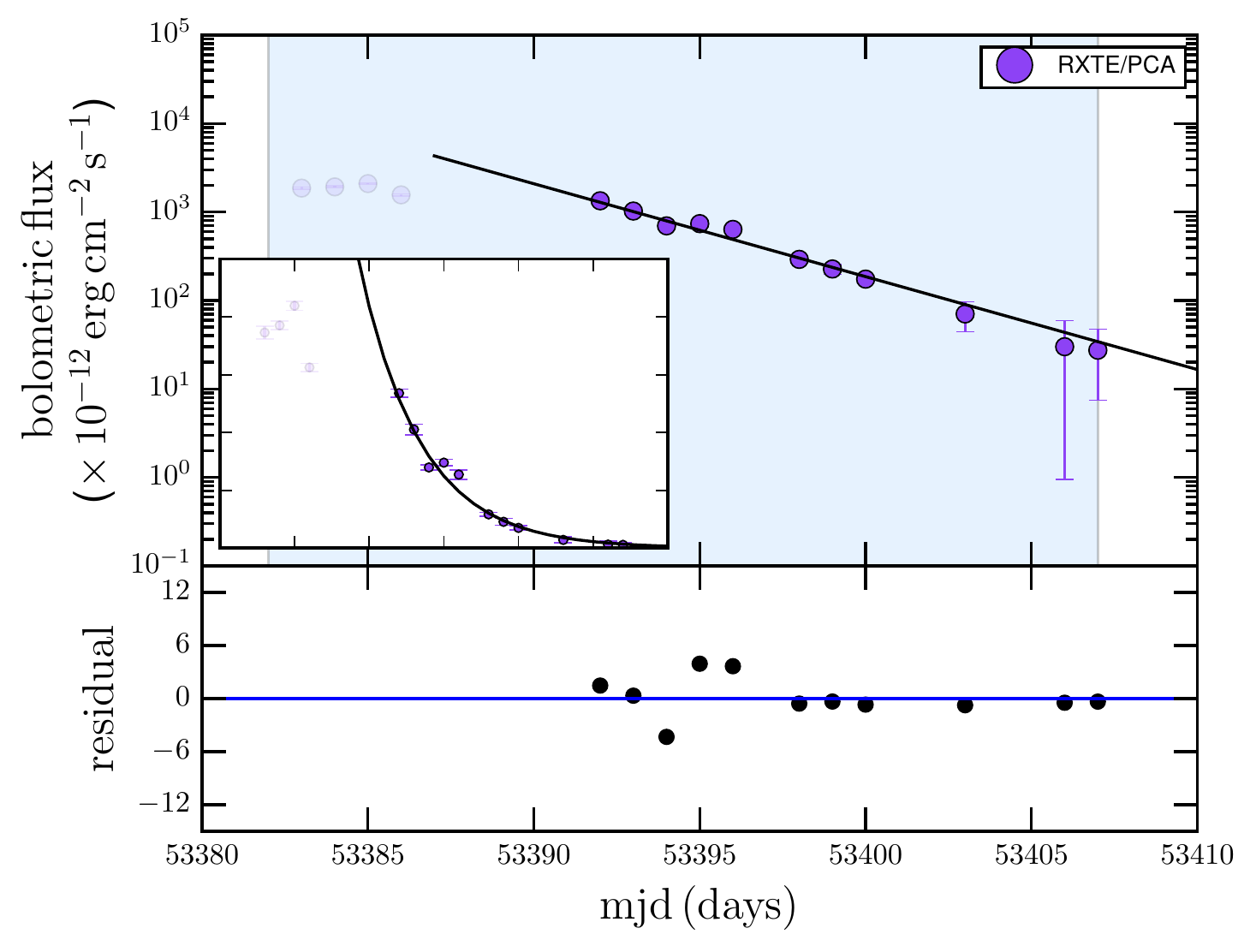}}\hfill
\subfloat[2012 outburst of MAXIJ1305$-$704\label{fig:maxi1305}]
  {\includegraphics[width=.29\linewidth]{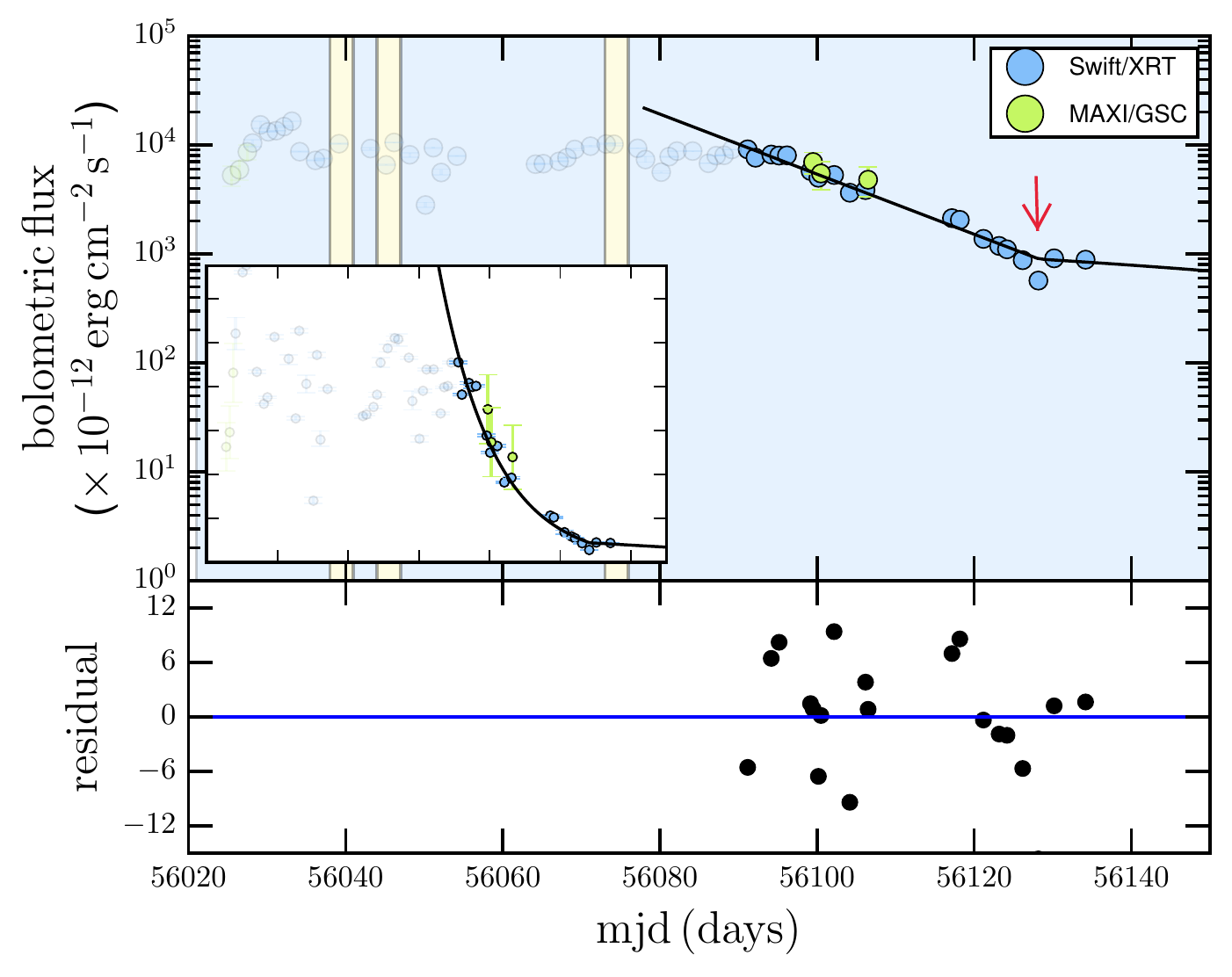}}\hfill
  
\subfloat[2011 outburst of SwiftJ1357.2$-$0933\label{fig:swift13571}]
  {\includegraphics[width=.3\linewidth]{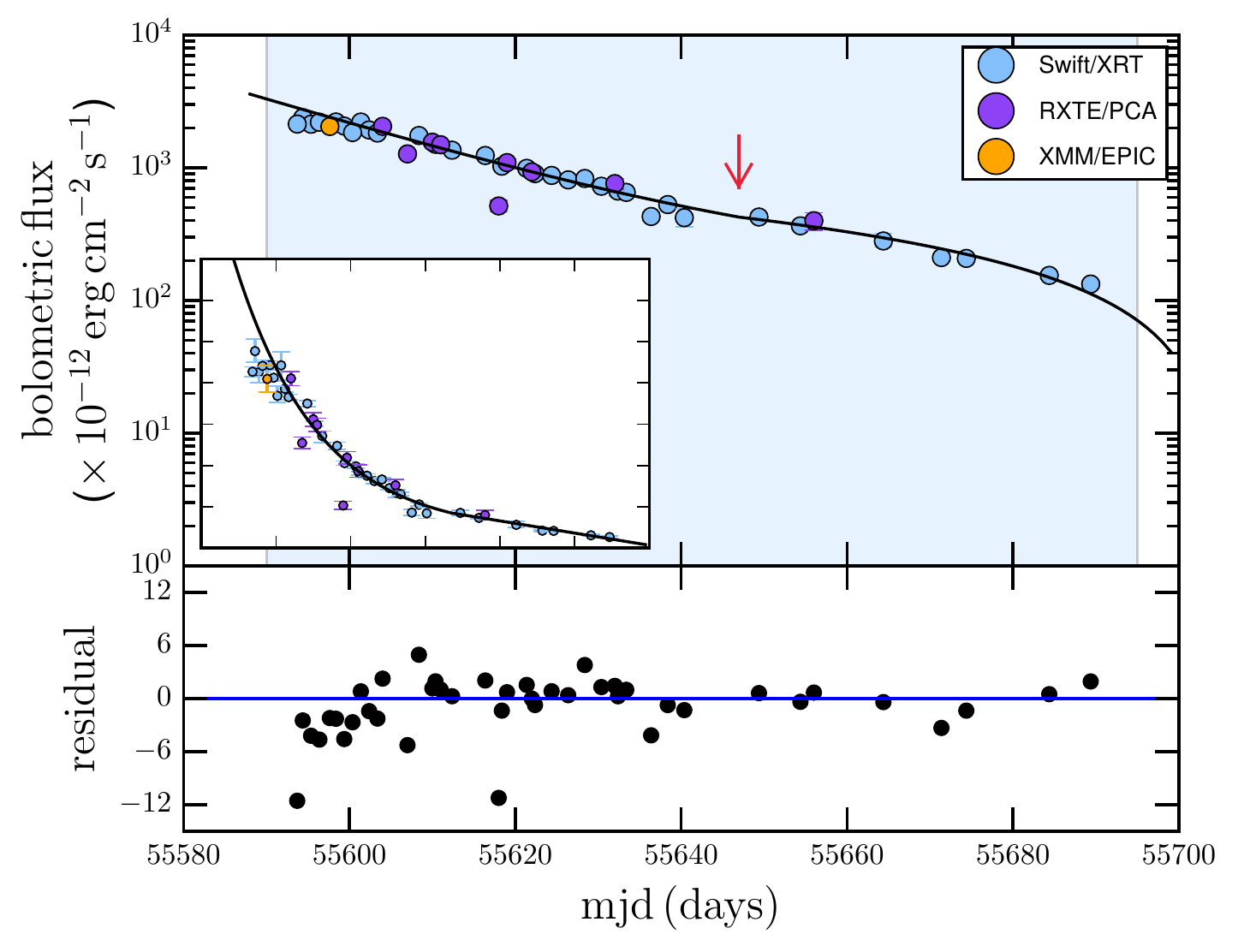}}\hfill
\subfloat[2017 outburst of SwiftJ1357.2$-$0933\label{fig:swift13572}]
  {\includegraphics[width=.3\linewidth]{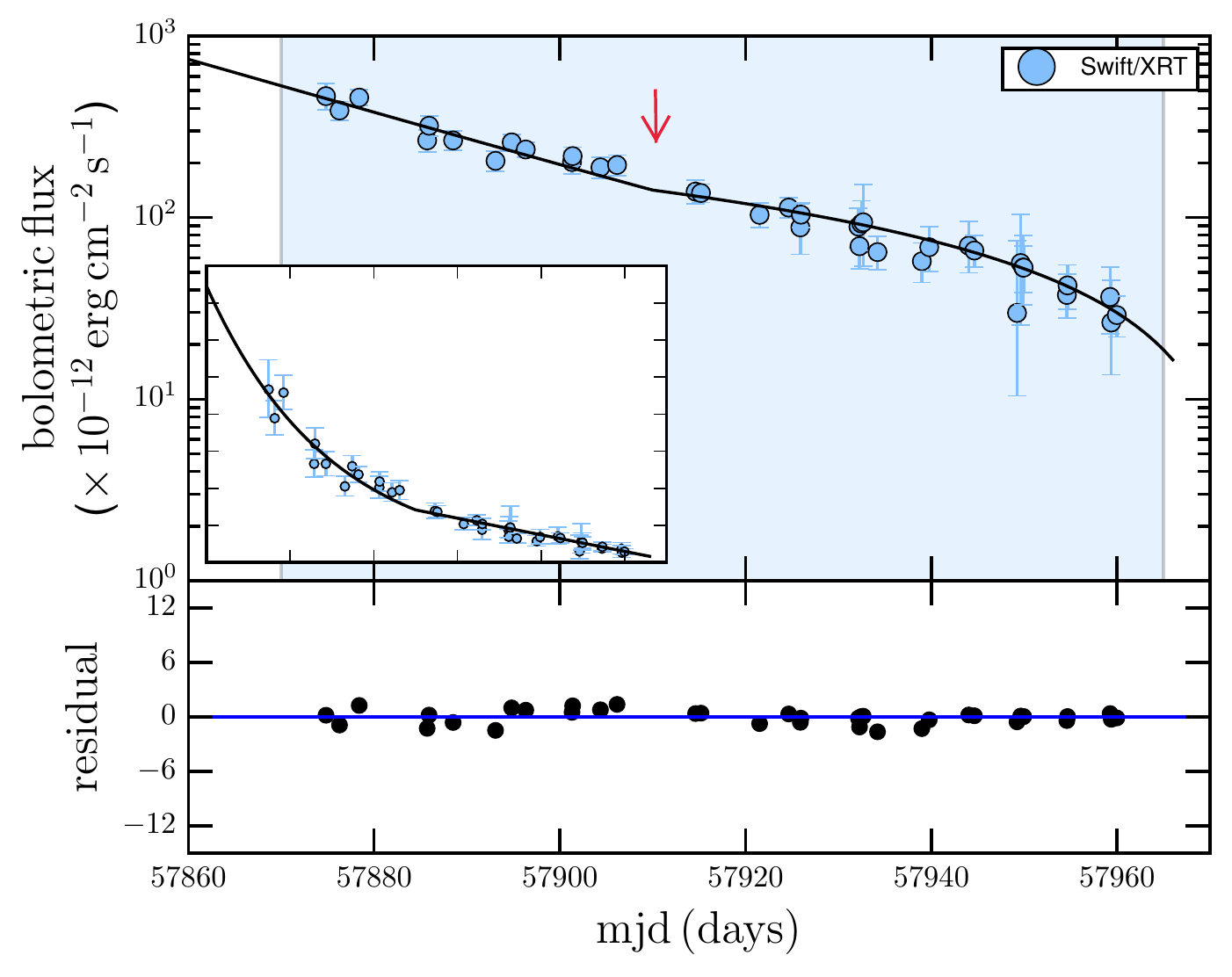}}\hfill
\subfloat[1997/1998 outburst of GS1354$-$64\label{fig:gs13541}]
  {\includegraphics[width=.3\linewidth]{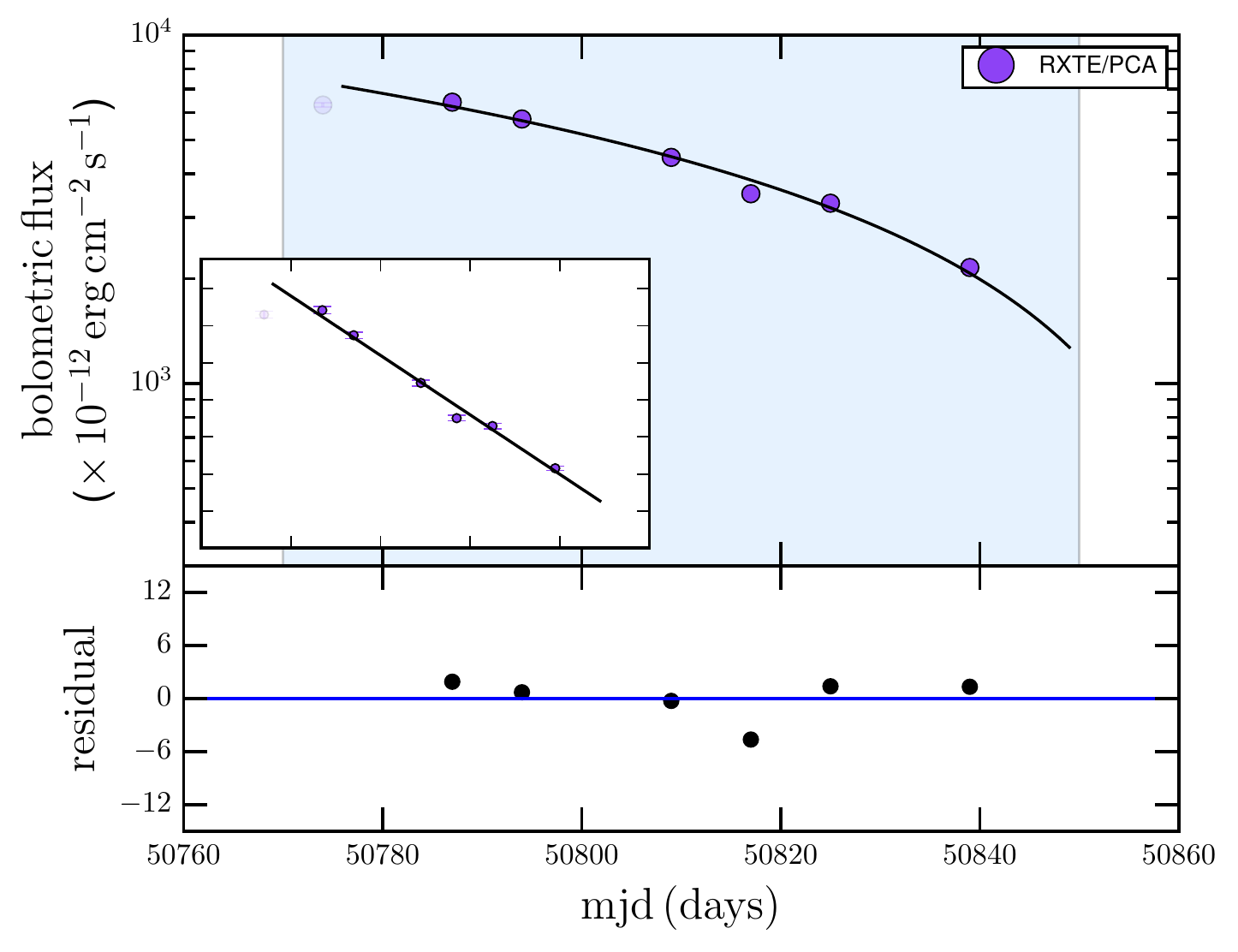}}\hfill
  
\subfloat[2015 outburst of GS1354$-$64\label{fig:gs13542}]
  {\includegraphics[width=.3\linewidth]{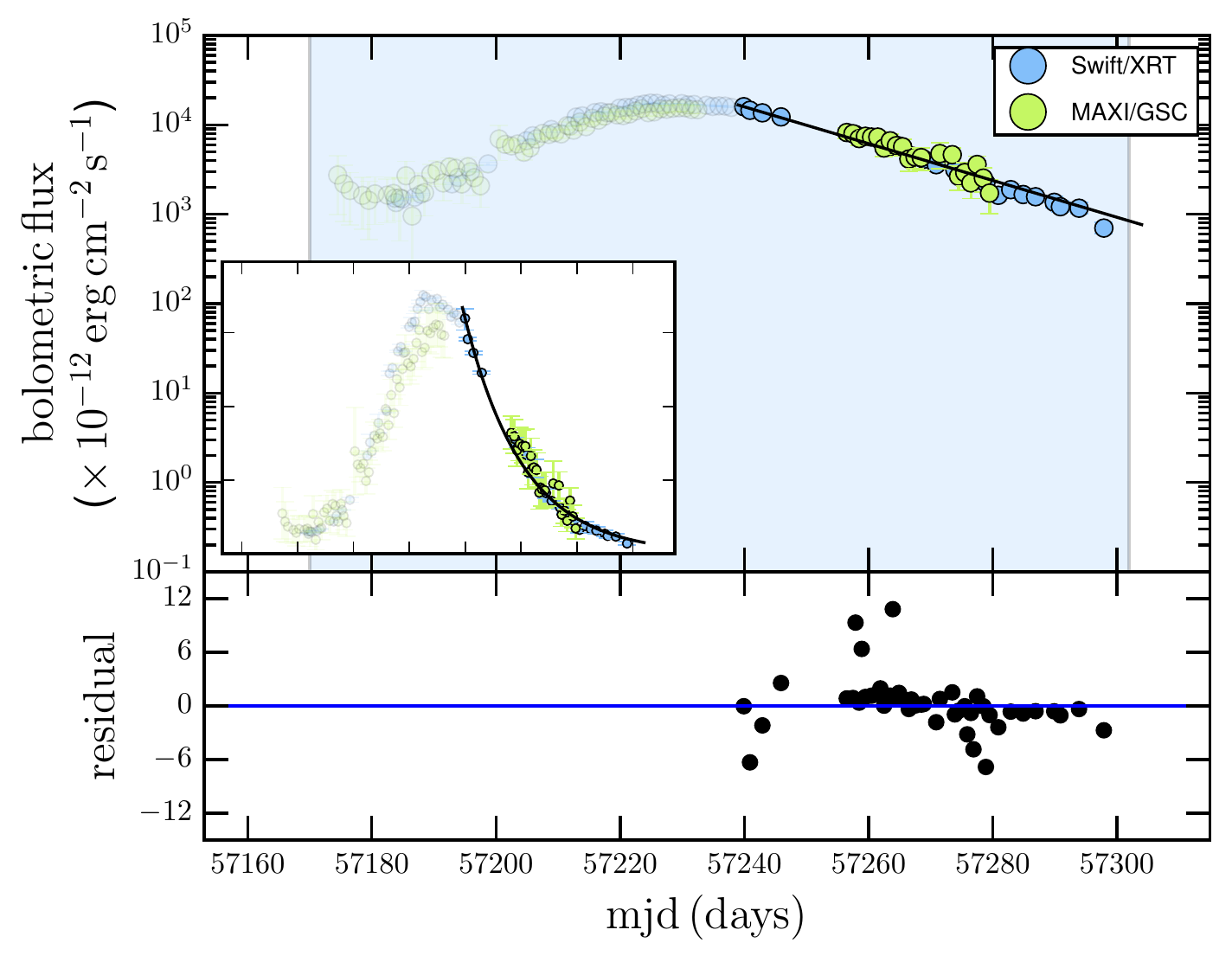}}\hfill
\subfloat[2002 outburst of 4U1543$-$475\label{fig:4u1543}]
  {\includegraphics[width=.3\linewidth]{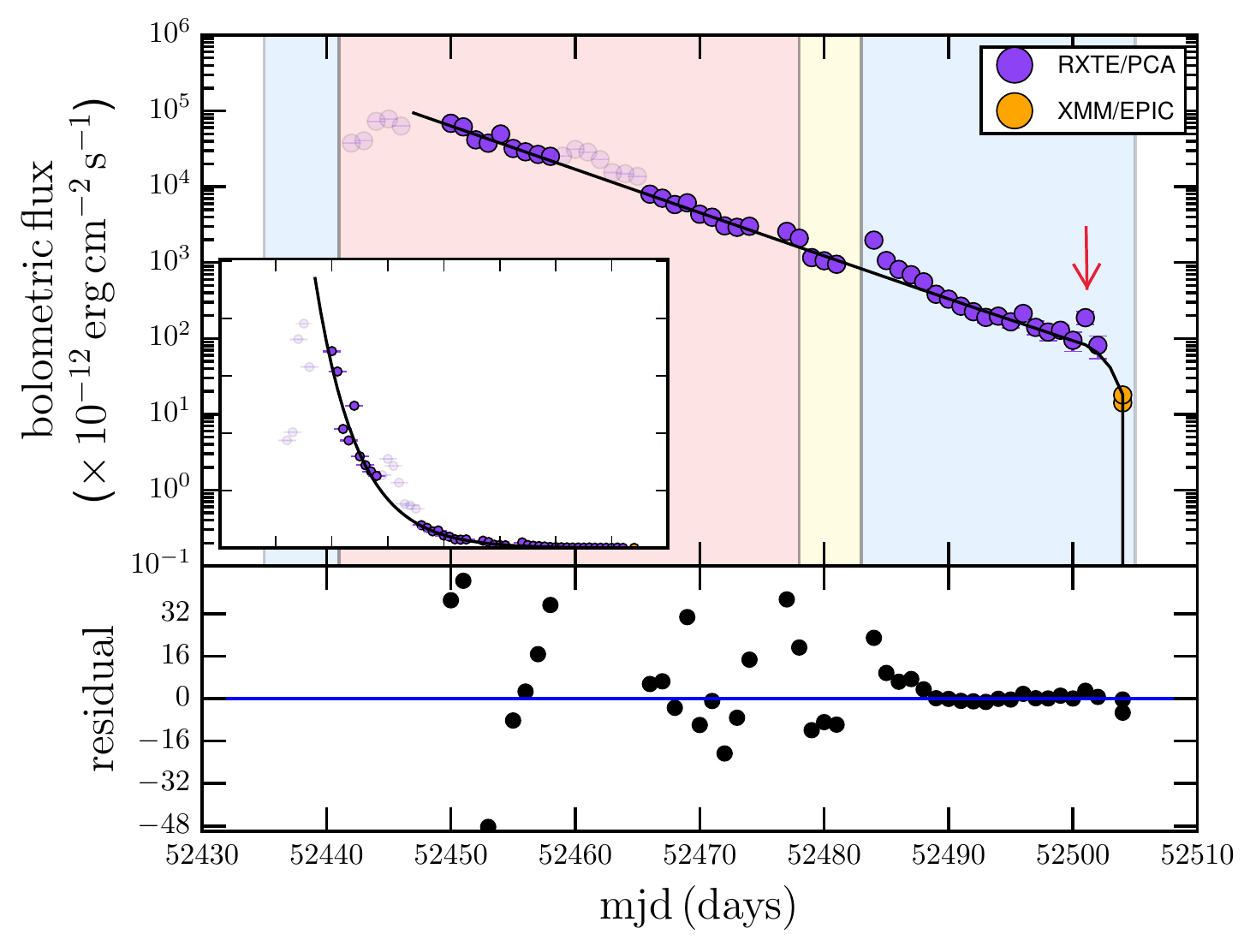}}\hfill
\subfloat[2000 outburst of XTEJ1550$-$564\label{fig:xte15501}]
  {\includegraphics[width=.3\linewidth]{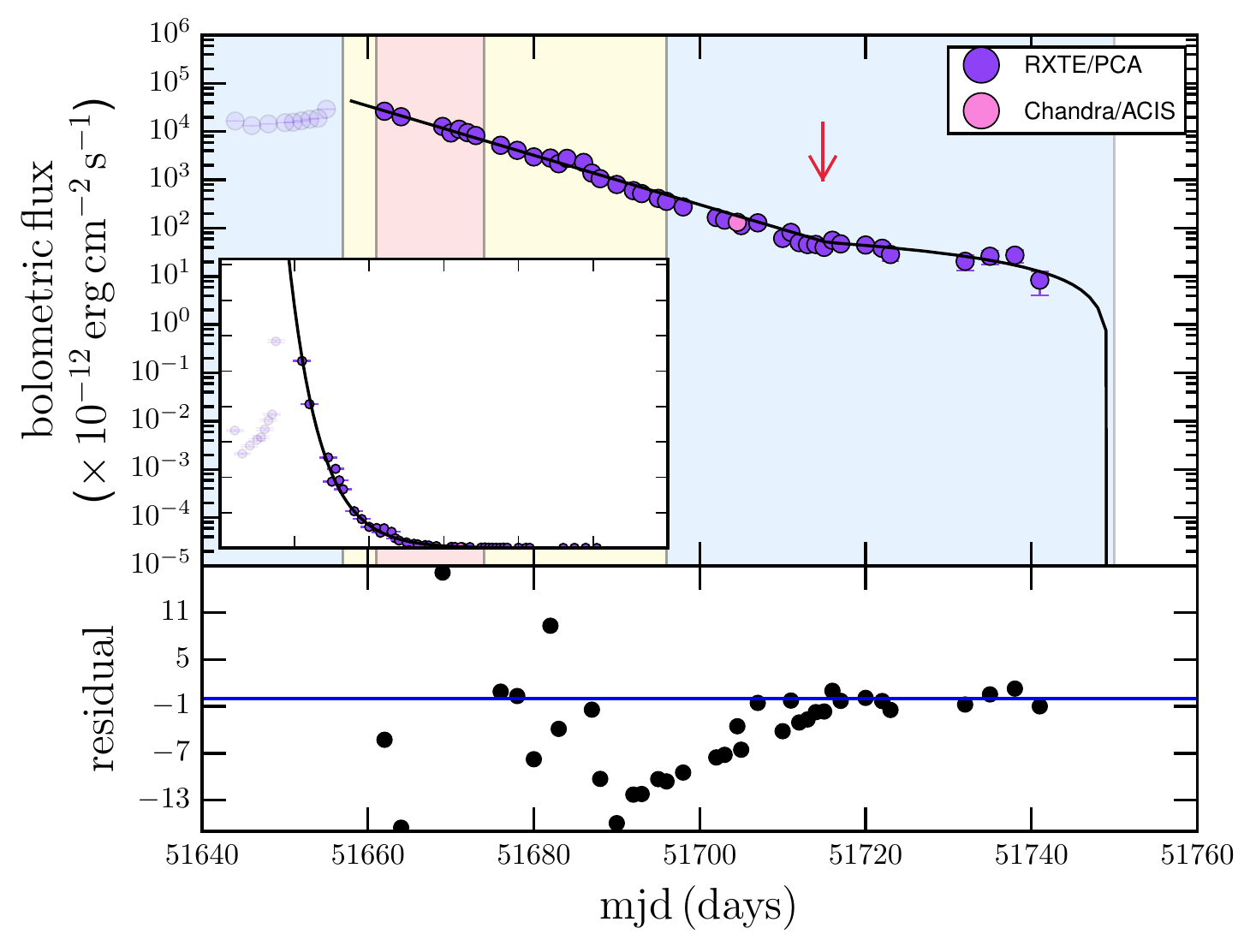}}\hfill
  
\subfloat[2001 outburst of XTEJ1550$-$564\label{fig:xte15502}]
  {\includegraphics[width=.3\linewidth]{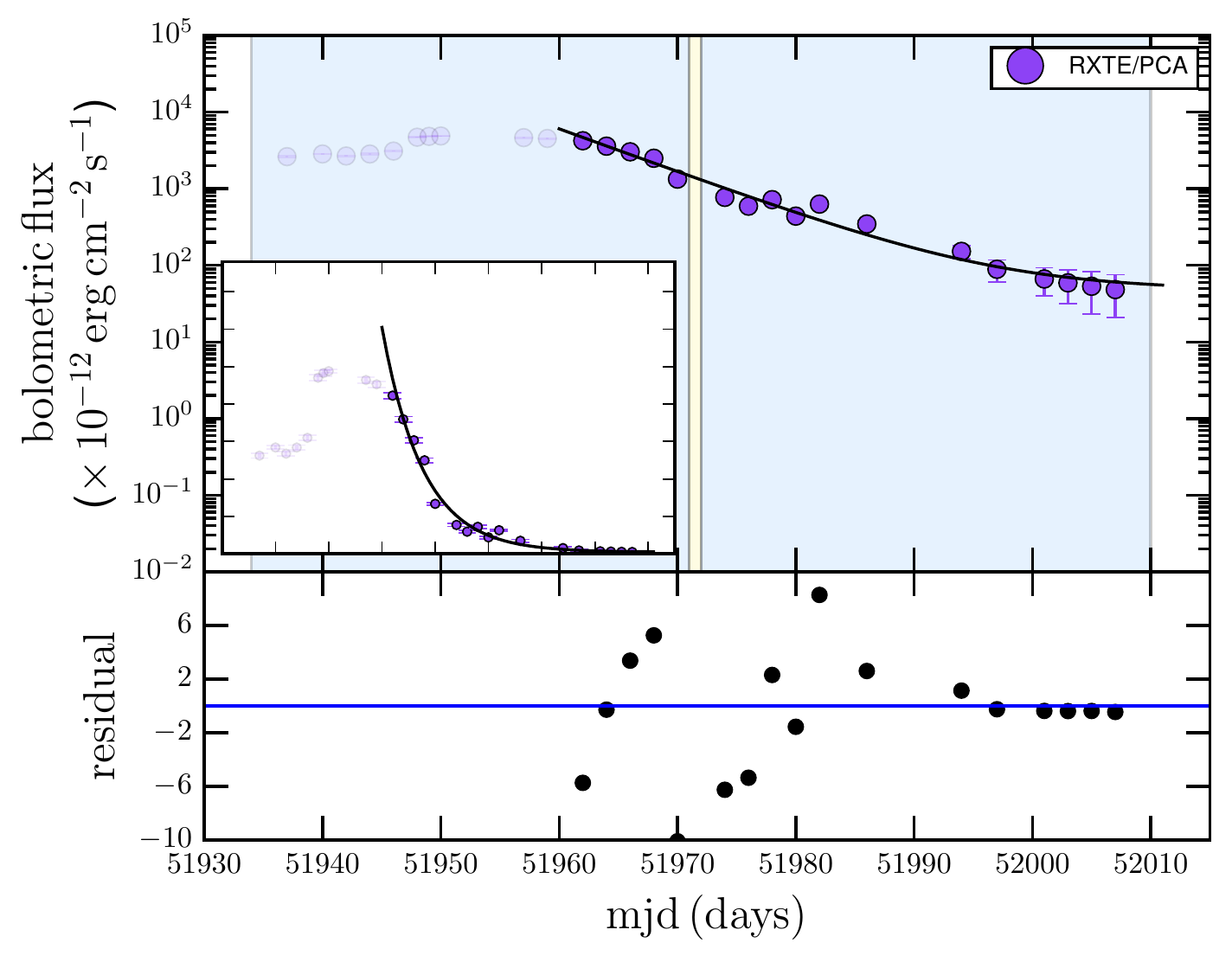}}\hfill
\subfloat[2001/2002 outburst of XTEJ1550$-$564\label{fig:xte15503}]
  {\includegraphics[width=.3\linewidth]{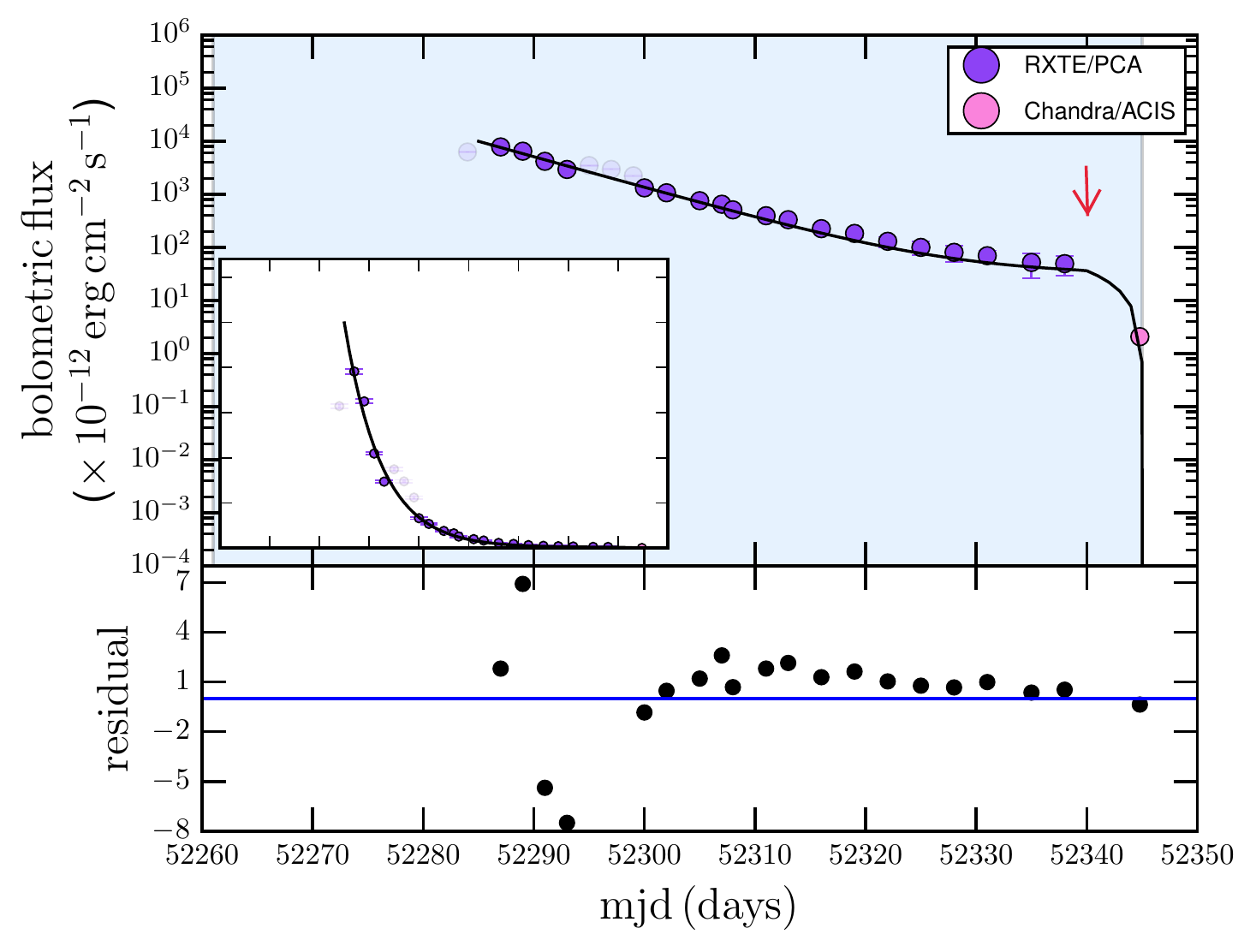}}\hfill
\subfloat[2003 outburst of XTEJ1550$-$564\label{fig:xte15504}]
  {\includegraphics[width=.3\linewidth]{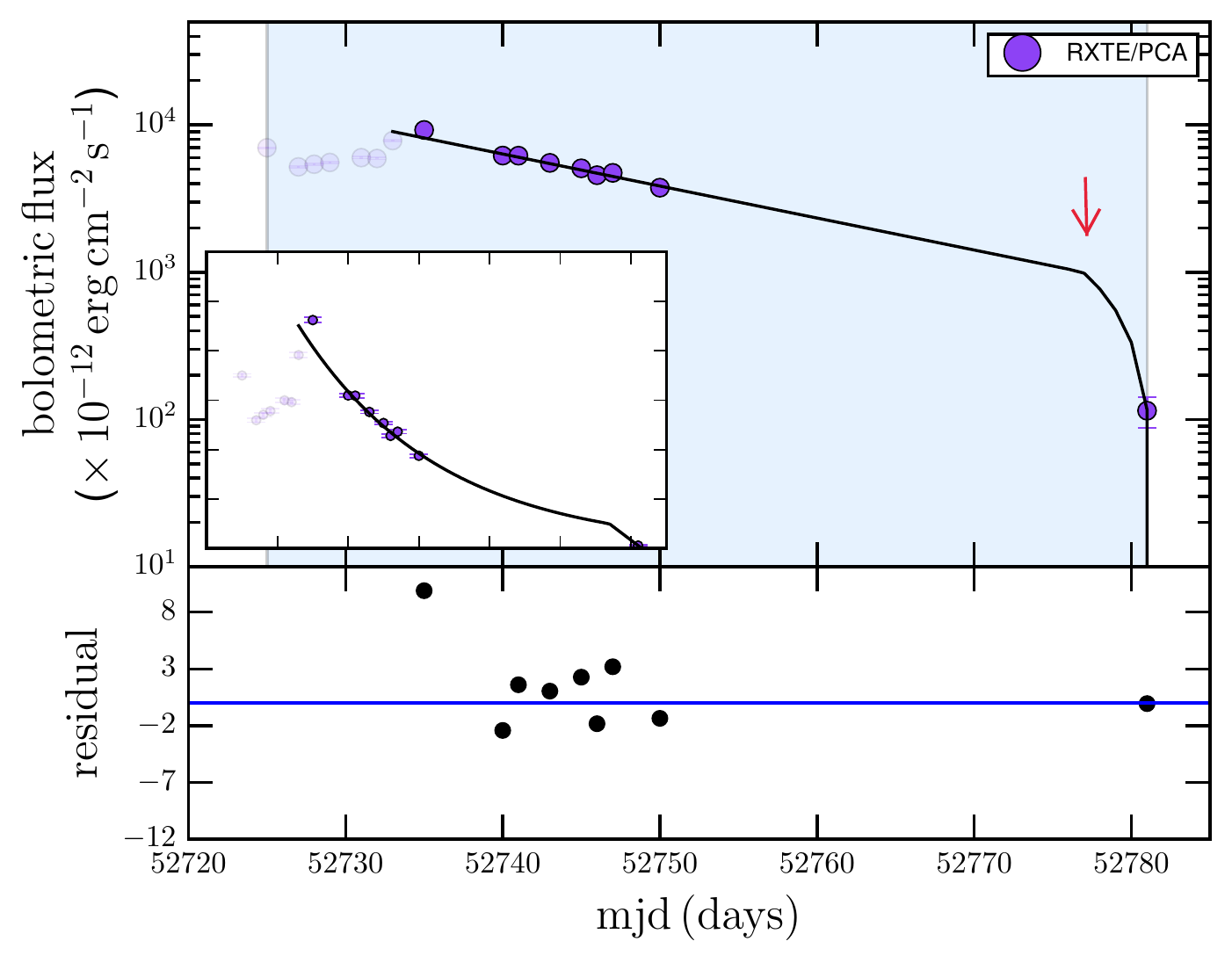}}\hfill
\caption{X-ray outburst light-curves for our BH-LMXB sample. Error bars are individual instrument statistical uncertainties only. The inset axes shows the data on a linear scale. Red arrows indicate where the transition between viscous and irradiation-controlled decay stages occurs (where applicable). Background shaded colours show the accretion state(s) of the source, computed from the WATCHDOG project \citep{tetarenkob2015}, throughout the outburst: blue = hard, yellow = intermediate, red = soft. The best fit analytical model is represented by the solid black line and residuals are presented in the lower panel of each figure.
Coloured circular markers represent data from individual X-ray instruments: XTE/PCA (purple), Swift/XRT (blue), MAXI/GSC (green), Chandra/ACIS-S and Chandra/HRC-S (pink), and XMM-Newton/EPIC (orange). Translucent data markers indicate portions of the outburst not included in the fit (e.g. the rise of the outburst, flares and re-brightening events). }
\label{fig:real_lcs}
\end{figure*}

\begin{figure*}
 \setcounter{subfigure}{12}
\subfloat[2001/2002 outburst of XTEJ1650$-$500\label{fig:xte1650}]
  {\includegraphics[width=.3\linewidth]{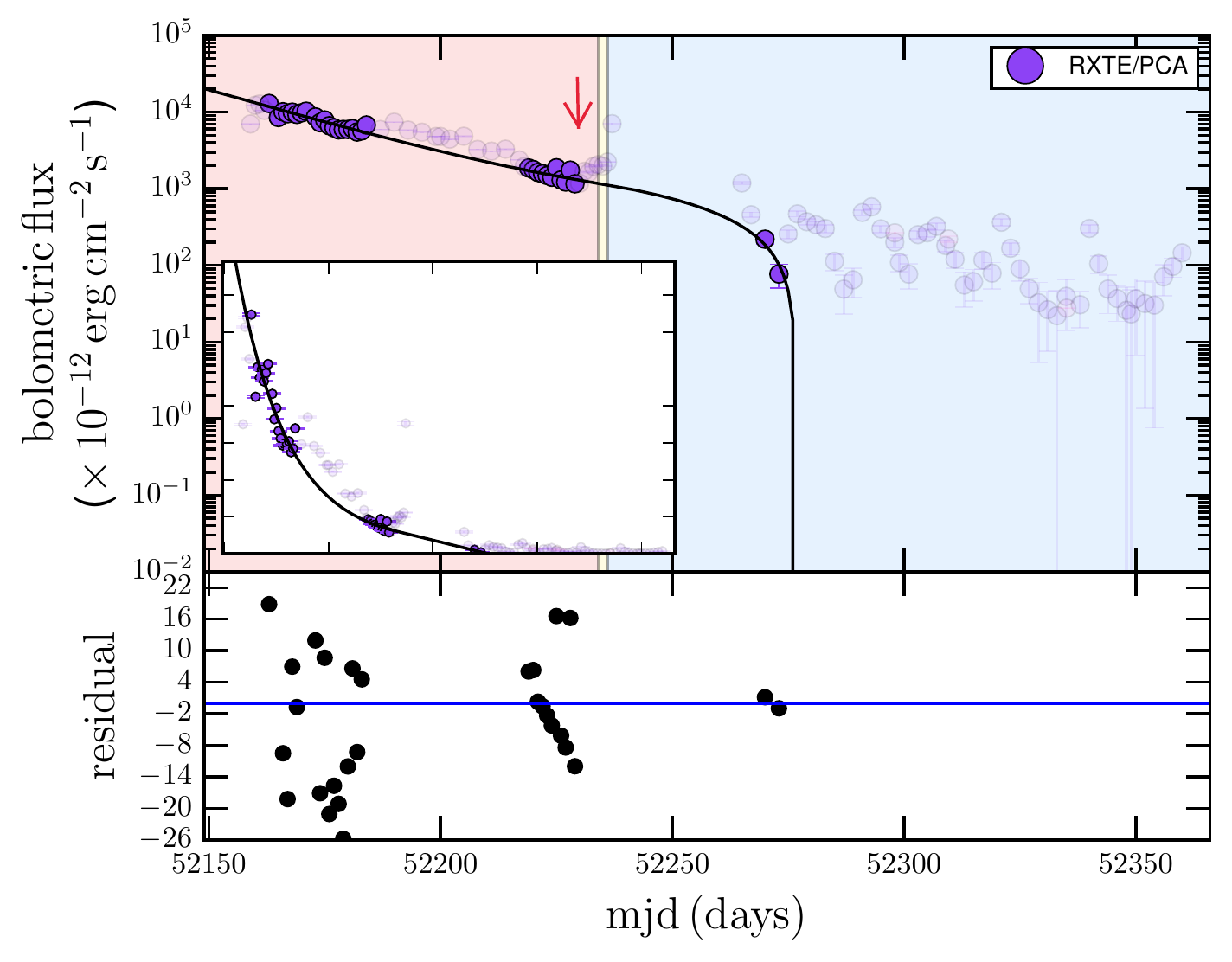}}\hfill
\subfloat[2010/2011 outburst of MAXIJ1659$-$152\label{fig:maxi1659}]
  {\includegraphics[width=.3\linewidth]{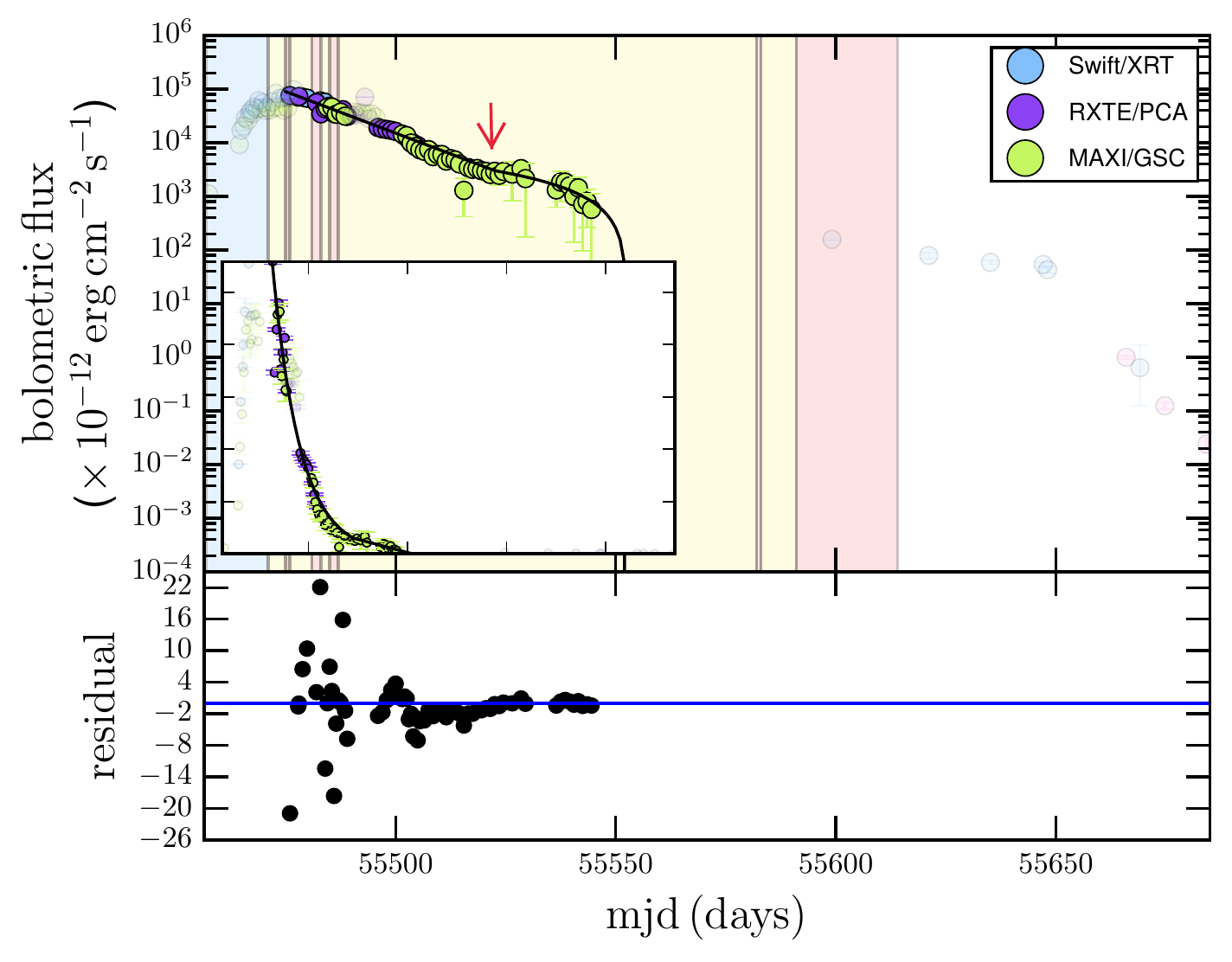}}\hfill
\subfloat[1996-1999 outburst of GX339$-$4\label{fig:gx3391}]
  {\includegraphics[width=.3\linewidth]{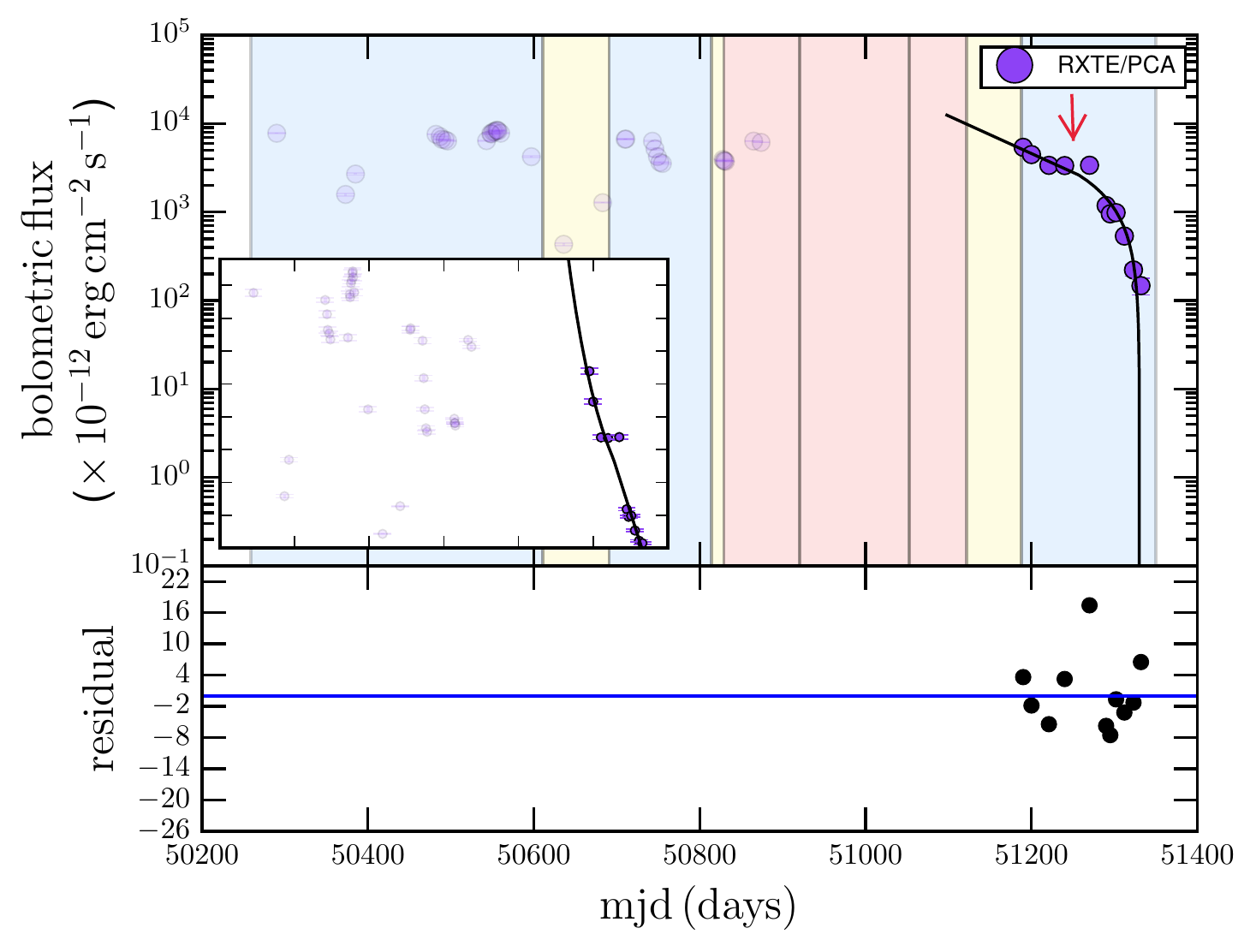}}\hfill

\subfloat[2006 outburst of GX339$-$4\label{fig:gx3392}]
  {\includegraphics[width=.3\linewidth]{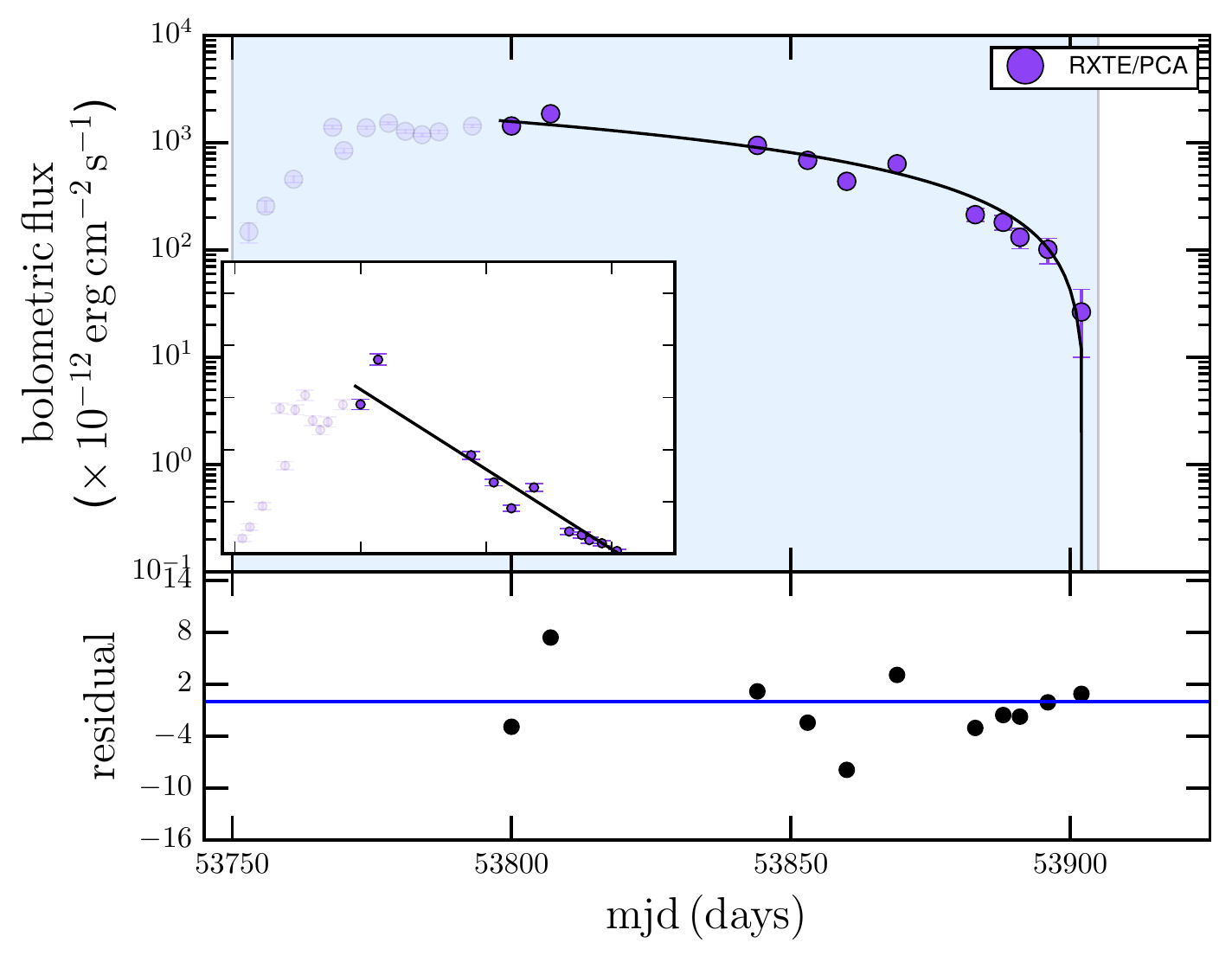}}\hfill
\subfloat[2008 outburst of GX339$-$4\label{fig:gx3393}]
  {\includegraphics[width=.3\linewidth]{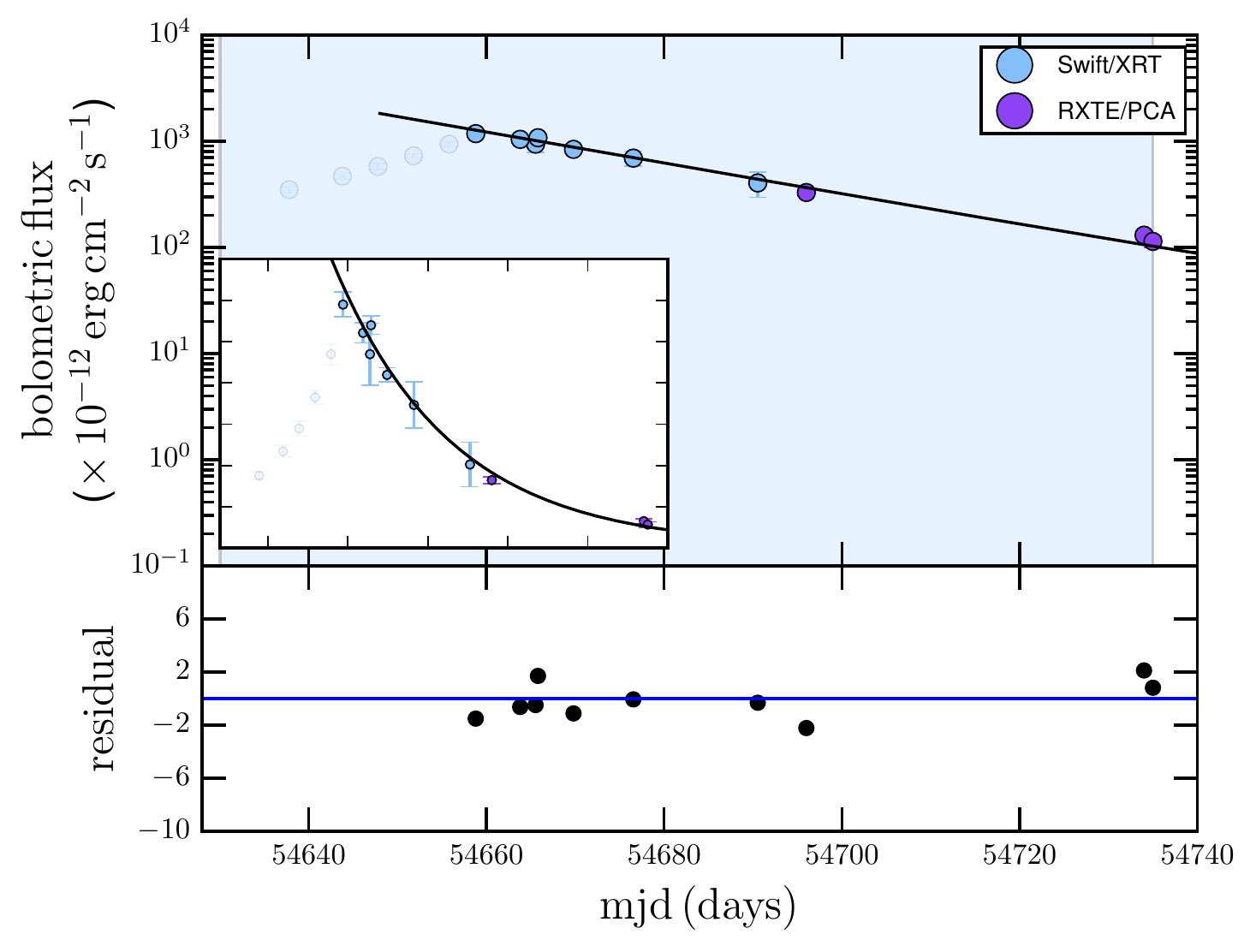}}\hfill
\subfloat[2009 outburst of GX339$-$4\label{fig:gx3394}]
  {\includegraphics[width=.3\linewidth]{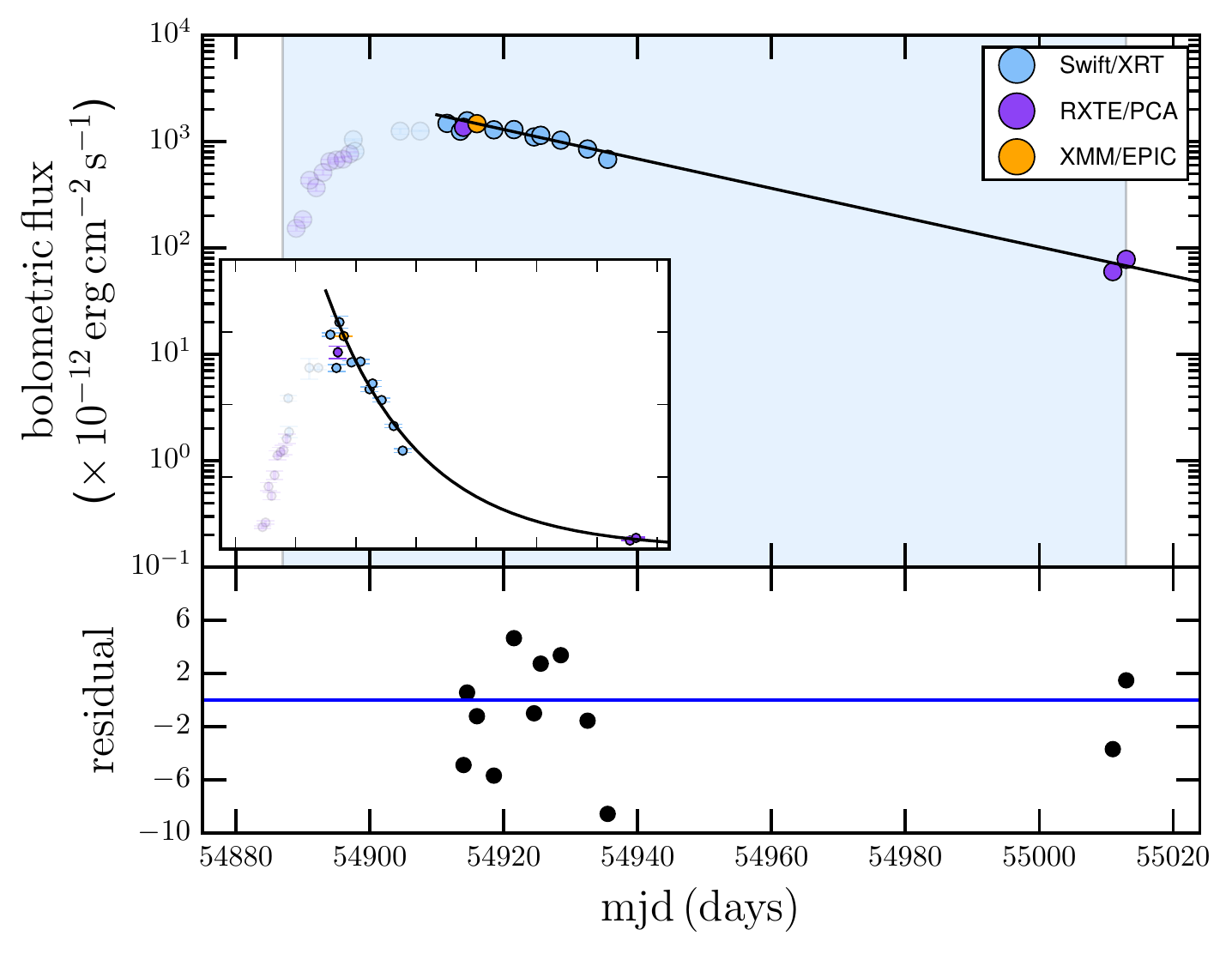}}\hfill

\subfloat[2013 outburst of GX339$-$4\label{fig:gx3395}]
  {\includegraphics[width=.3\linewidth]{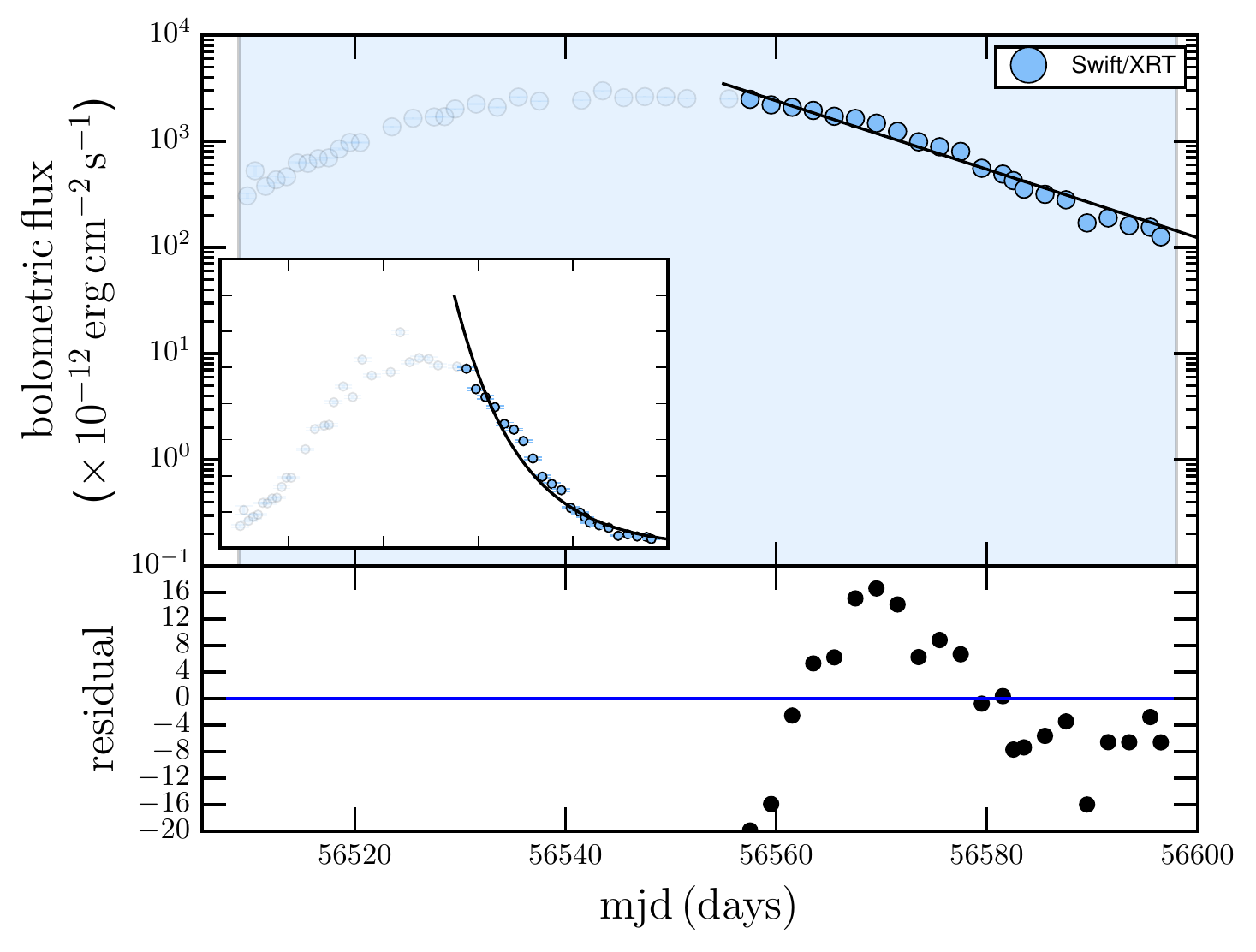}}\hfill
\subfloat[2014/2015 outburst of GX339$-$4\label{fig:gx3396}]
  {\includegraphics[width=.3\linewidth]{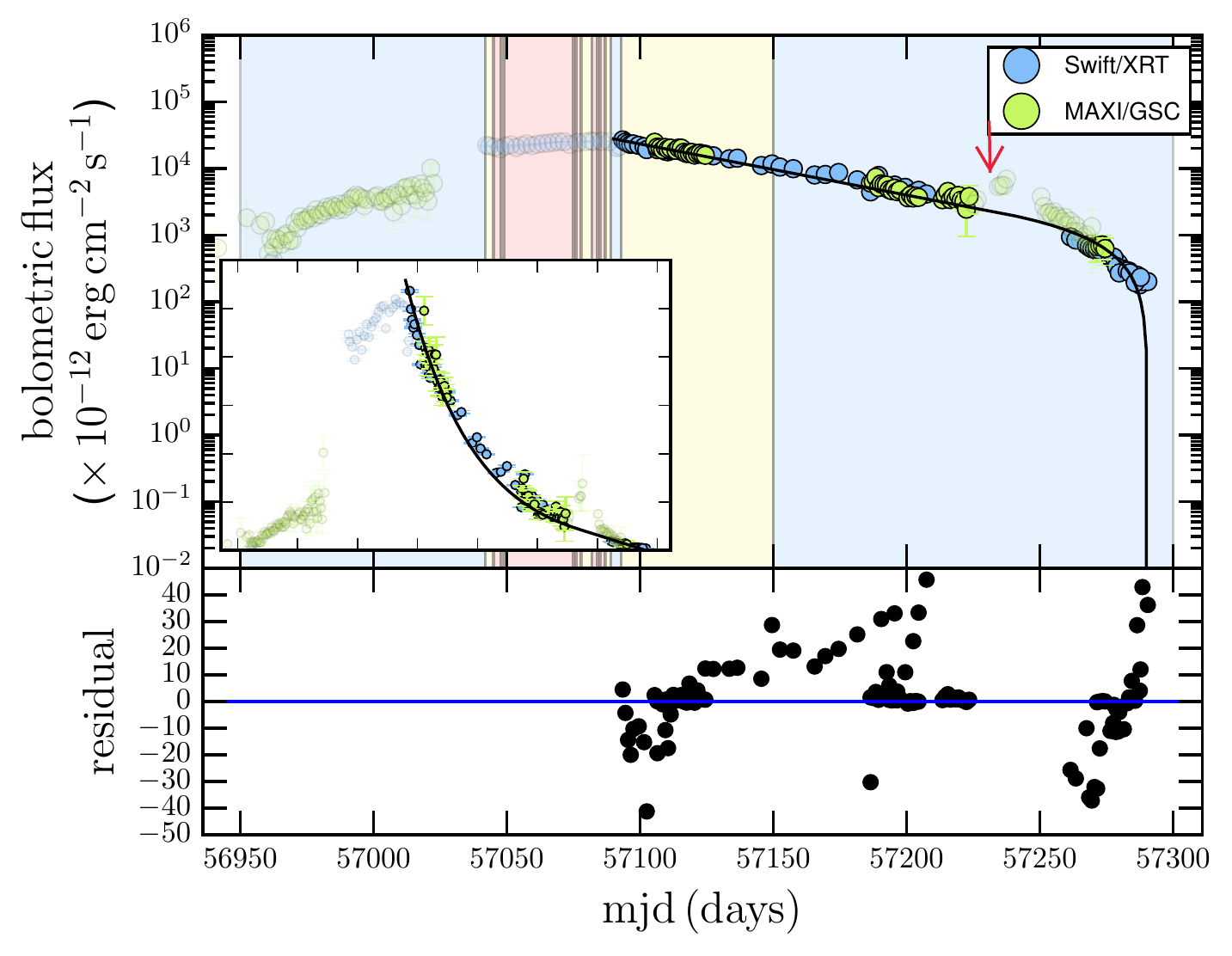}}\hfill
\subfloat[2012/2013 outburst of SWIFTJ1745$-$26\label{fig:swift1745}]
  {\includegraphics[width=.3\linewidth]{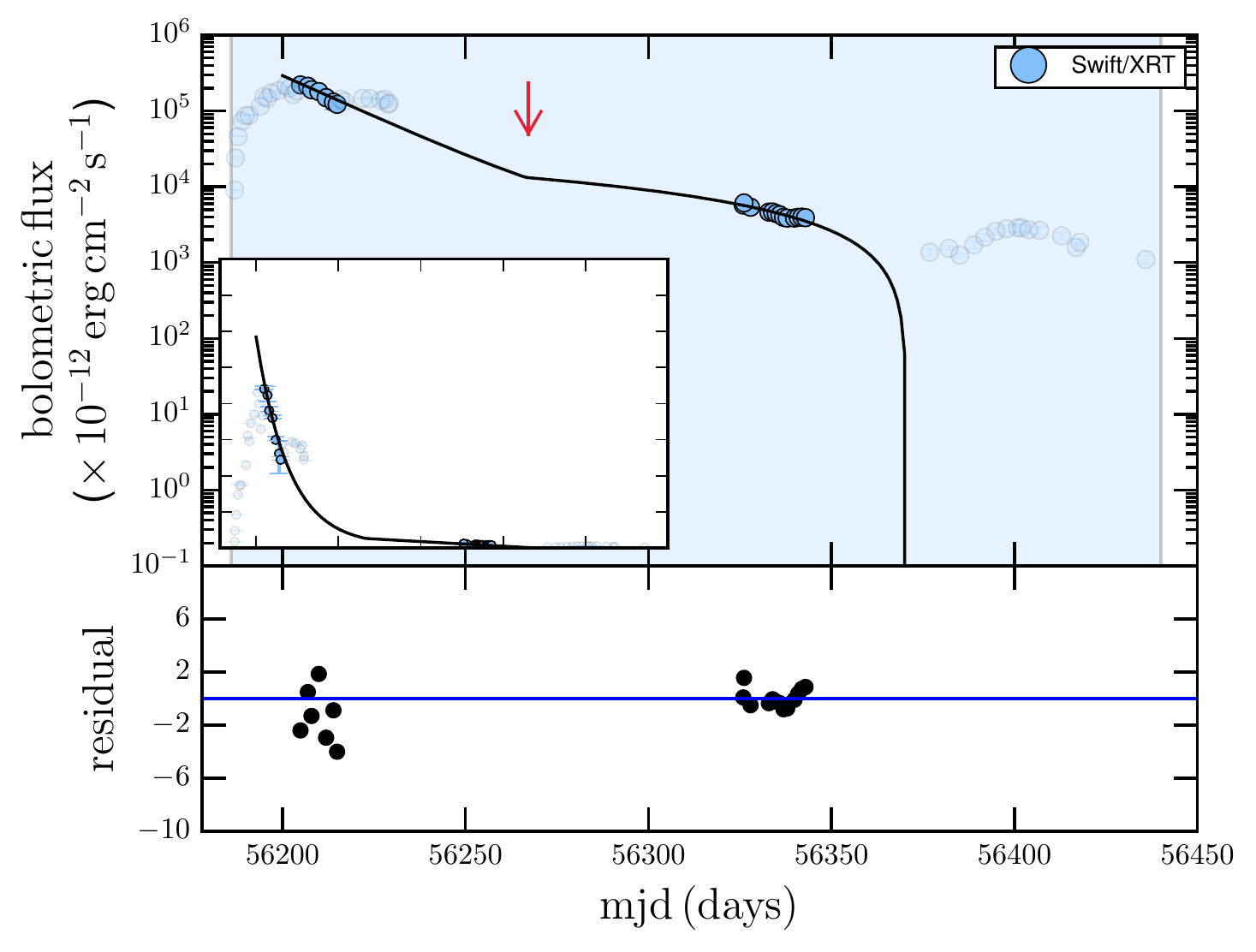}}\hfill

\subfloat[2011/2012 outburst of MAXIJ1836$-$194\label{fig:maxi1836}]
  {\includegraphics[width=.3\linewidth]{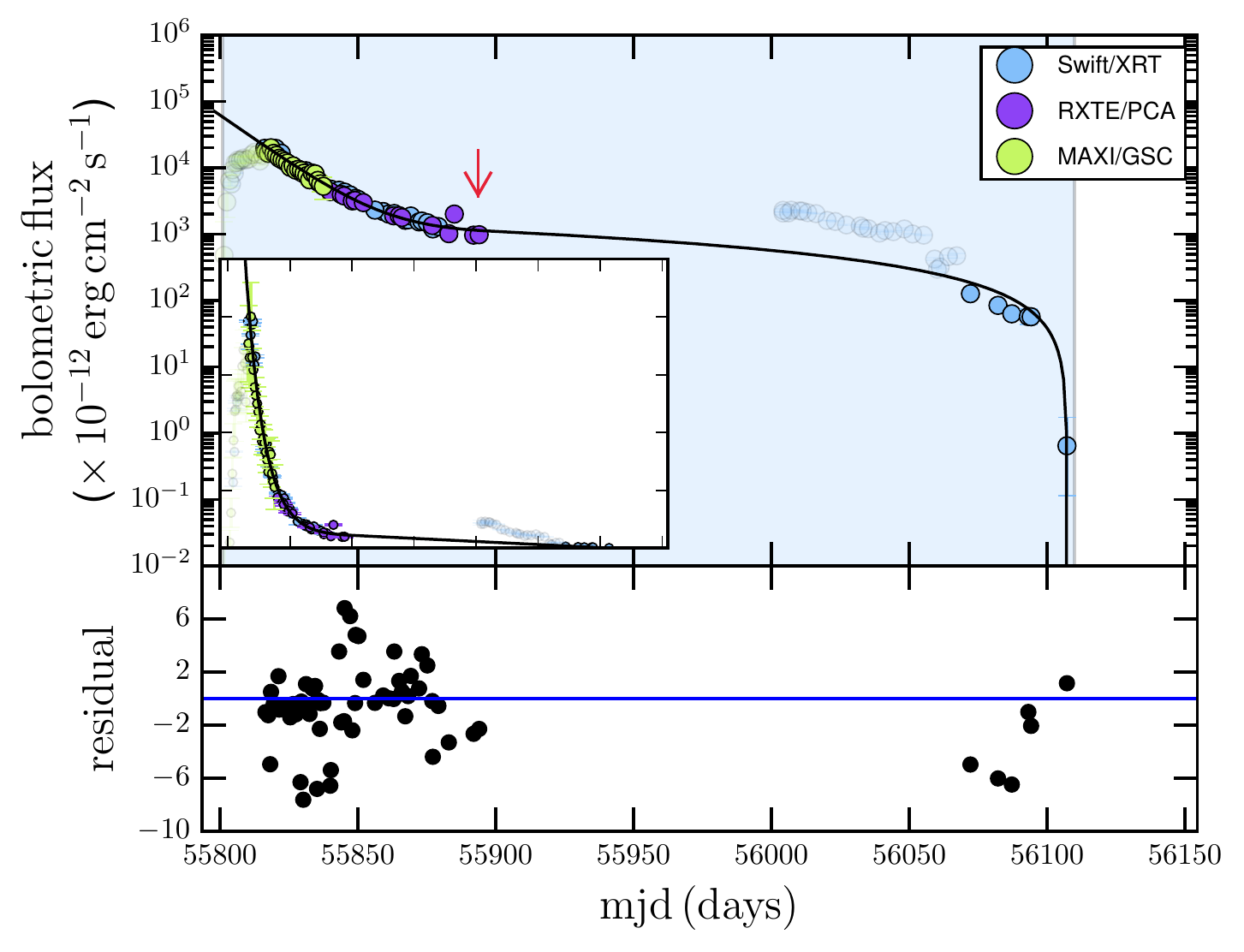}}\qquad
\subfloat[1999/2000 outburst of XTEJ1859+226\label{fig:xte1859}]
  {\includegraphics[width=.3\linewidth]{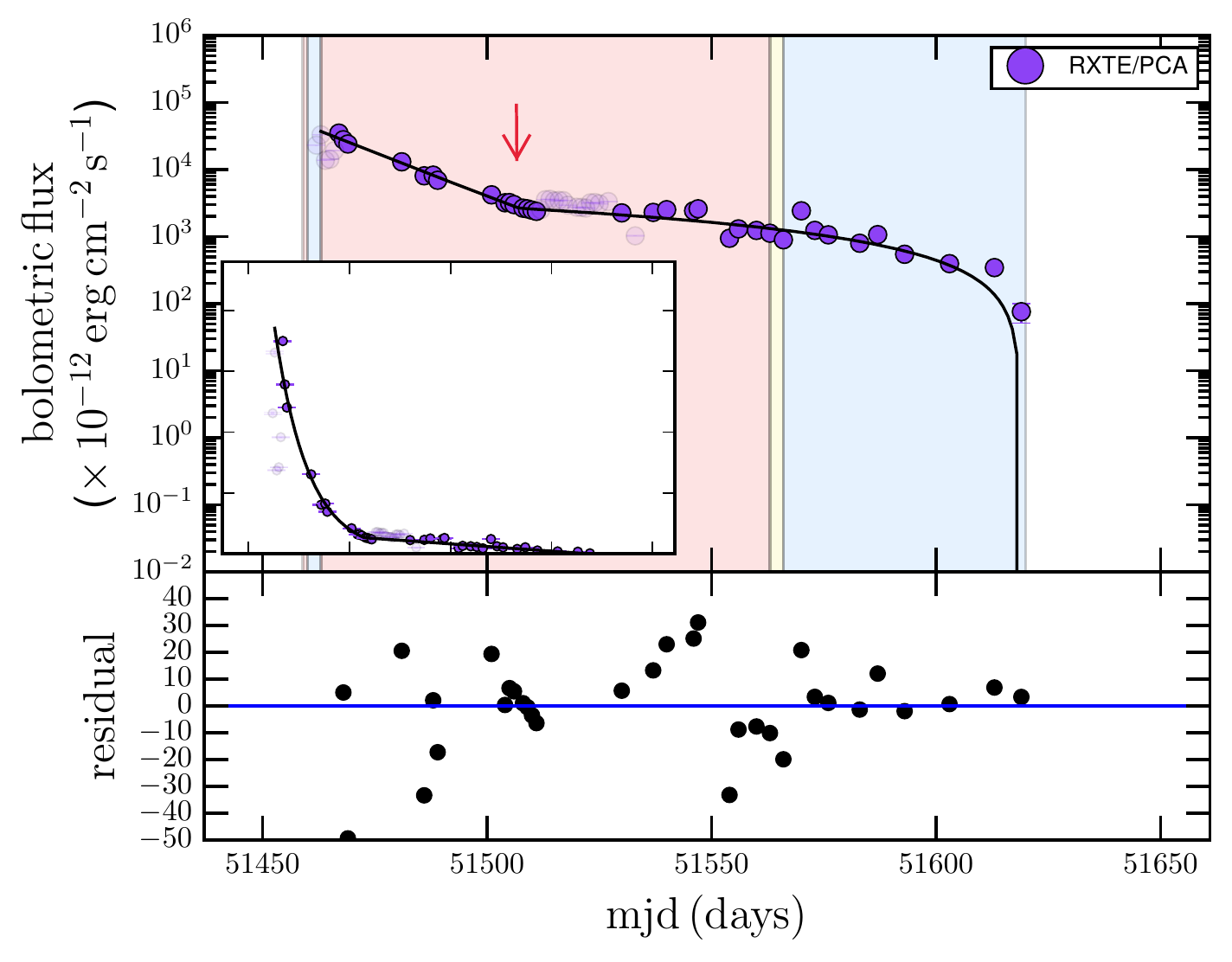}}\hfill
 \contcaption{X-ray outburst light-curves for our BH-LMXB sample. Error bars are individual instrument statistical uncertainties only. The inset axes shows the data on a linear scale. Red arrows indicate where the transition between viscous and irradiation-controlled decay stages occurs (where applicable). Background shaded colours show the accretion state(s) of the source, computed from the WATCHDOG project \citep{tetarenkob2015}, throughout the outburst: blue = hard, yellow = intermediate, red = soft. The best fit analytical model is represented by the solid black line and residuals are presented in the lower panel of each figure. Coloured circular markers represent data from individual X-ray instruments: XTE/PCA (purple), Swift/XRT (blue), MAXI/GSC (green), Chandra/ACIS-S and Chandra/HRC-S (pink), and XMM-Newton/EPIC (orange). Translucent data markers indicate portions of the outburst not included in the fit (e.g. the rise of the outburst, flares and re-brightening events).}
\end{figure*}

\begin{figure*}
\subfloat
  {\includegraphics[width=.5\linewidth,height=.35\linewidth]{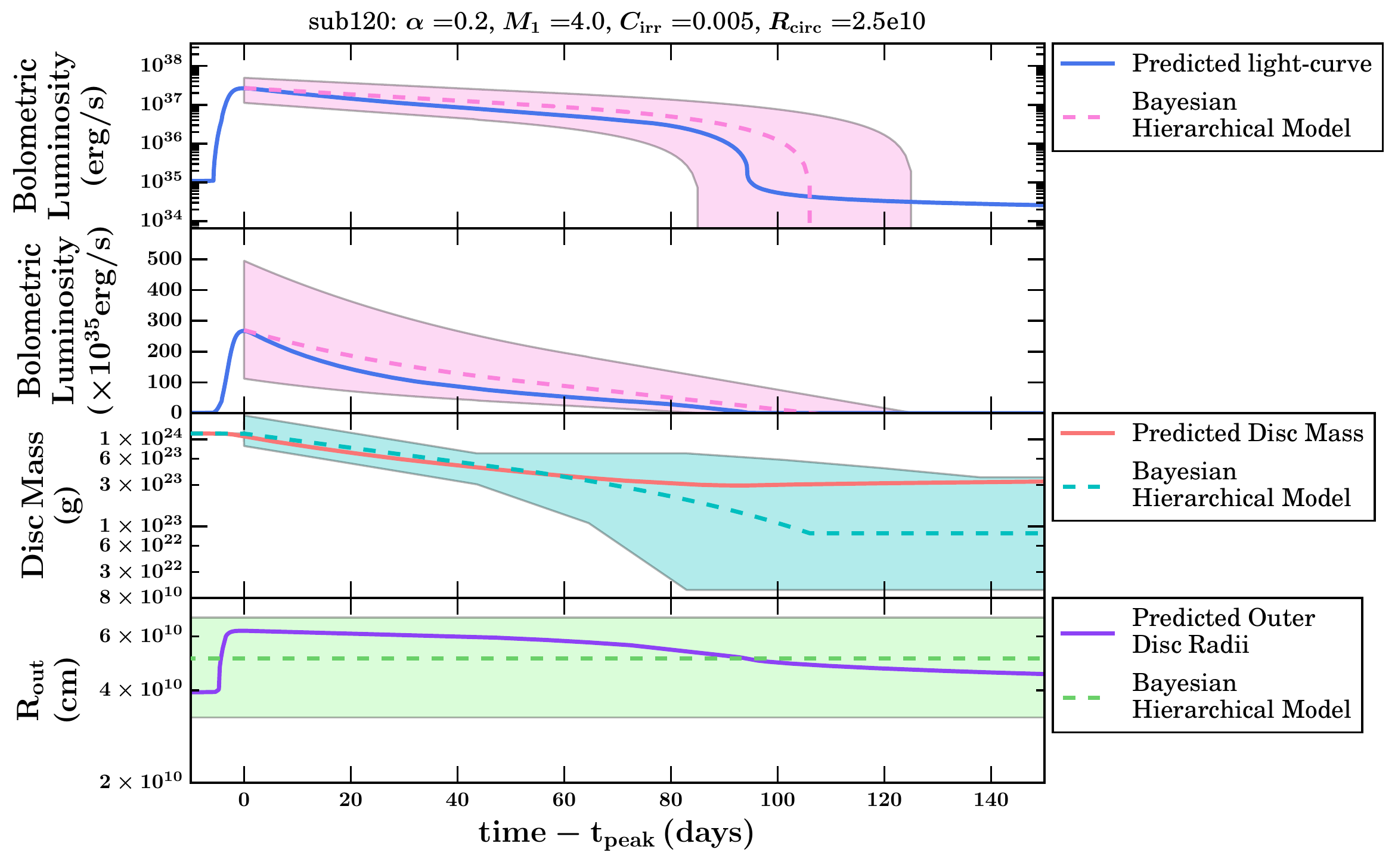}}\hfill
\subfloat
  {\includegraphics[width=.5\linewidth,height=.35\linewidth]{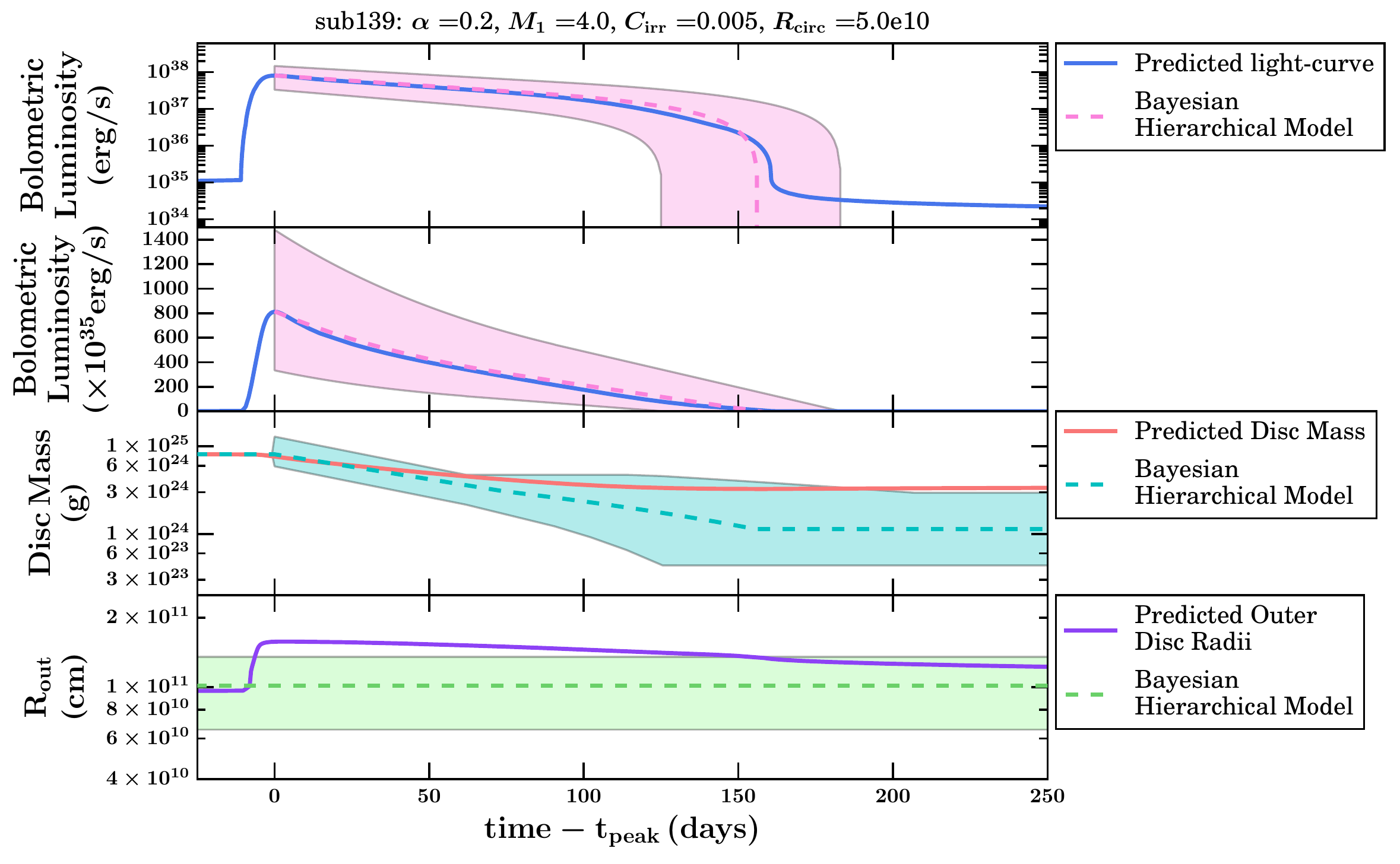}}
  
   {\includegraphics[width=.5\linewidth,height=.35\linewidth]{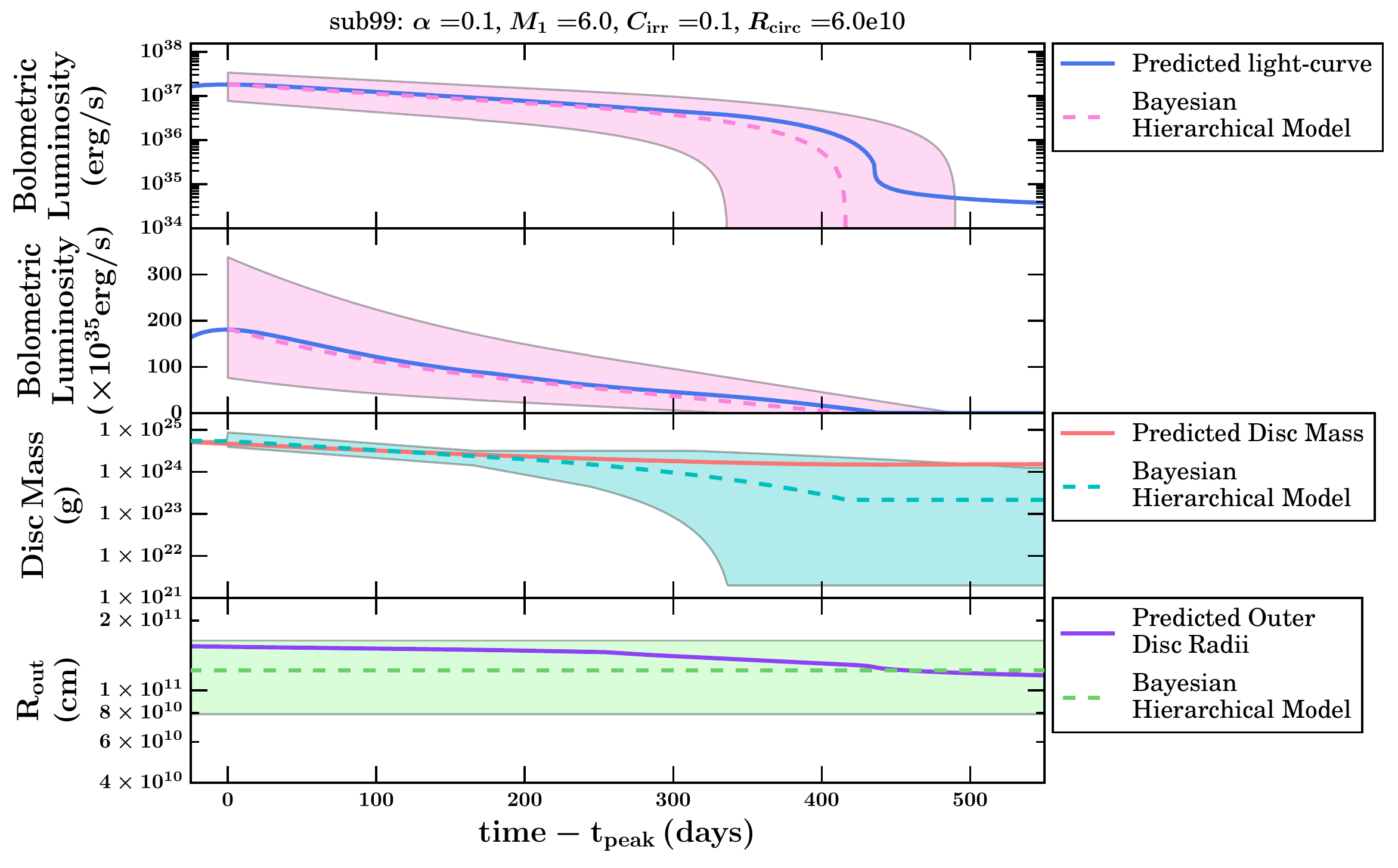}}\hfill
\subfloat
  {\includegraphics[width=.5\linewidth,height=.35\linewidth]{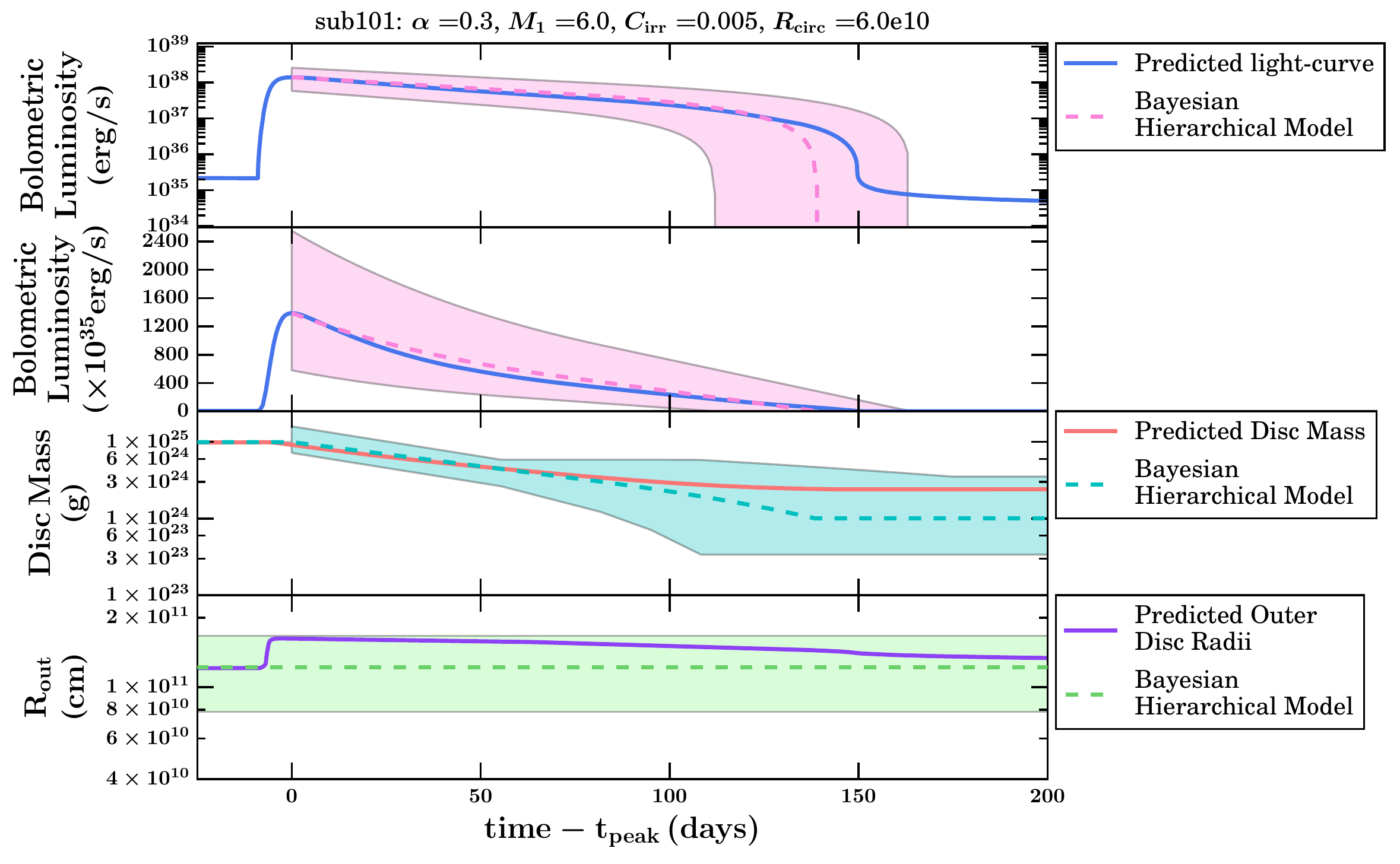}}
  
     {\includegraphics[width=.5\linewidth,height=.35\linewidth]{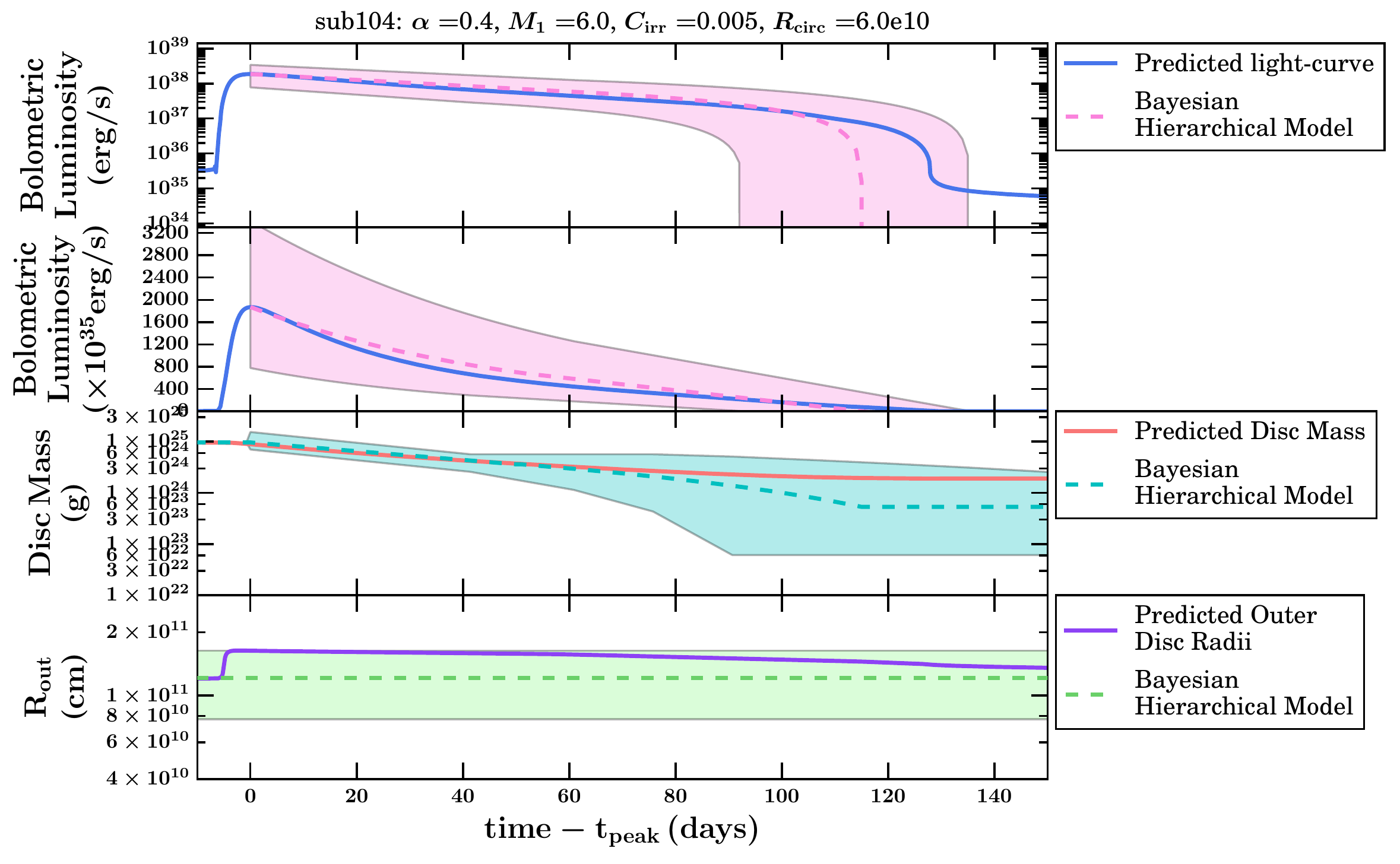}}\hfill
\subfloat
  {\includegraphics[width=.5\linewidth,height=.35\linewidth]{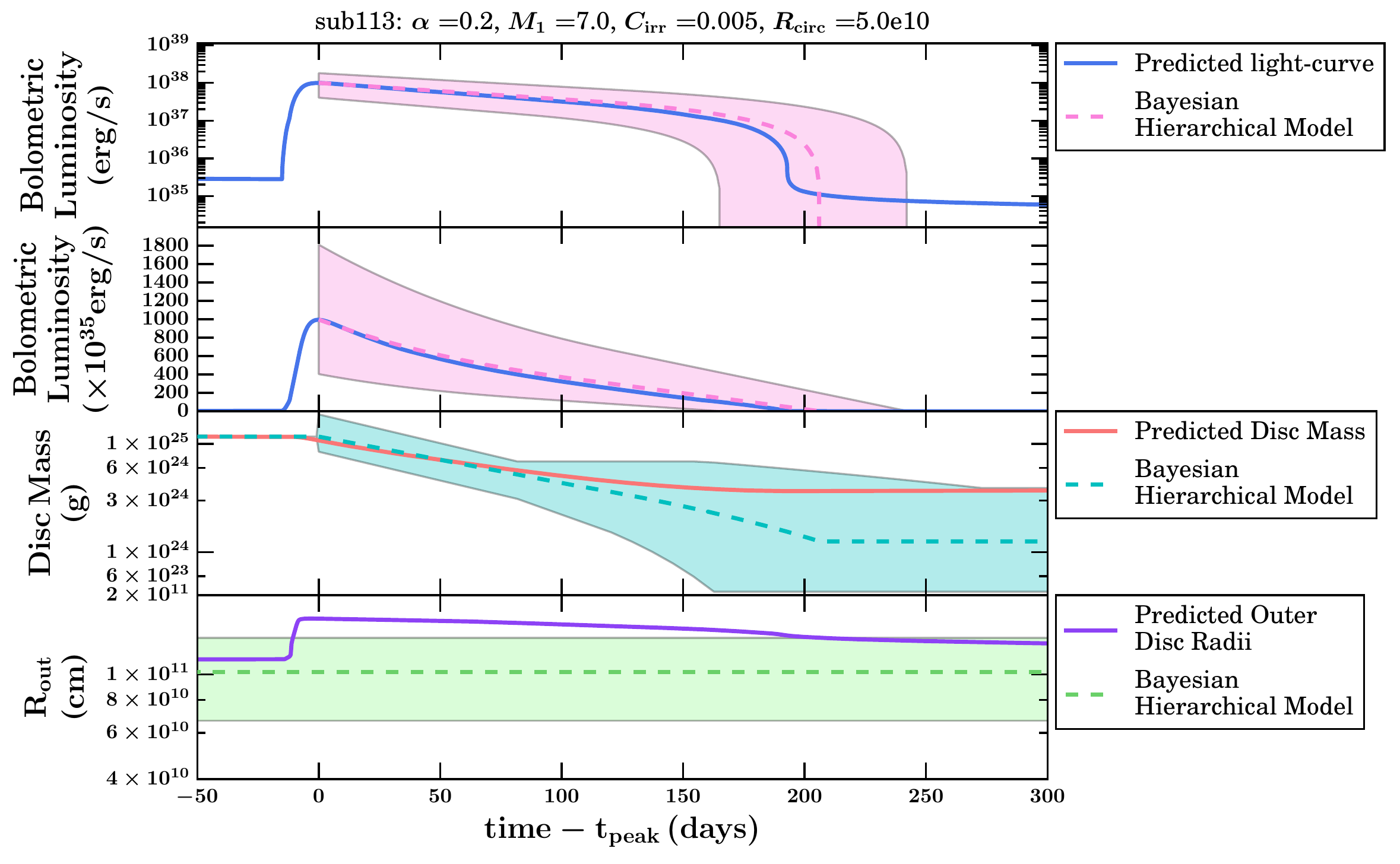}}
  \caption{Example outburst correlation plots comparing the predicted (from the numerical code) and Bayesian estimates (from our methodology) of the: light-curve decay profiles in log (top panel) and (second panel) linear space, (third panel) mass in the hot disc over time, and (bottom panel) outer disc radius, for varying $M_1$, $\alpha_h$, $C_{\rm irr}$, and $R_{\rm circ}$. The solid lines represent the output of the numerical code. The dotted lines and shaded regions represent the best-fit value and 1$\sigma$ confidence intervals from our Bayesian methodology, respectively. }%
  \label{fig:lc_compare1}%
\end{figure*}

\begin{figure*}
\subfloat
  {\includegraphics[width=.5\linewidth,height=.35\linewidth]{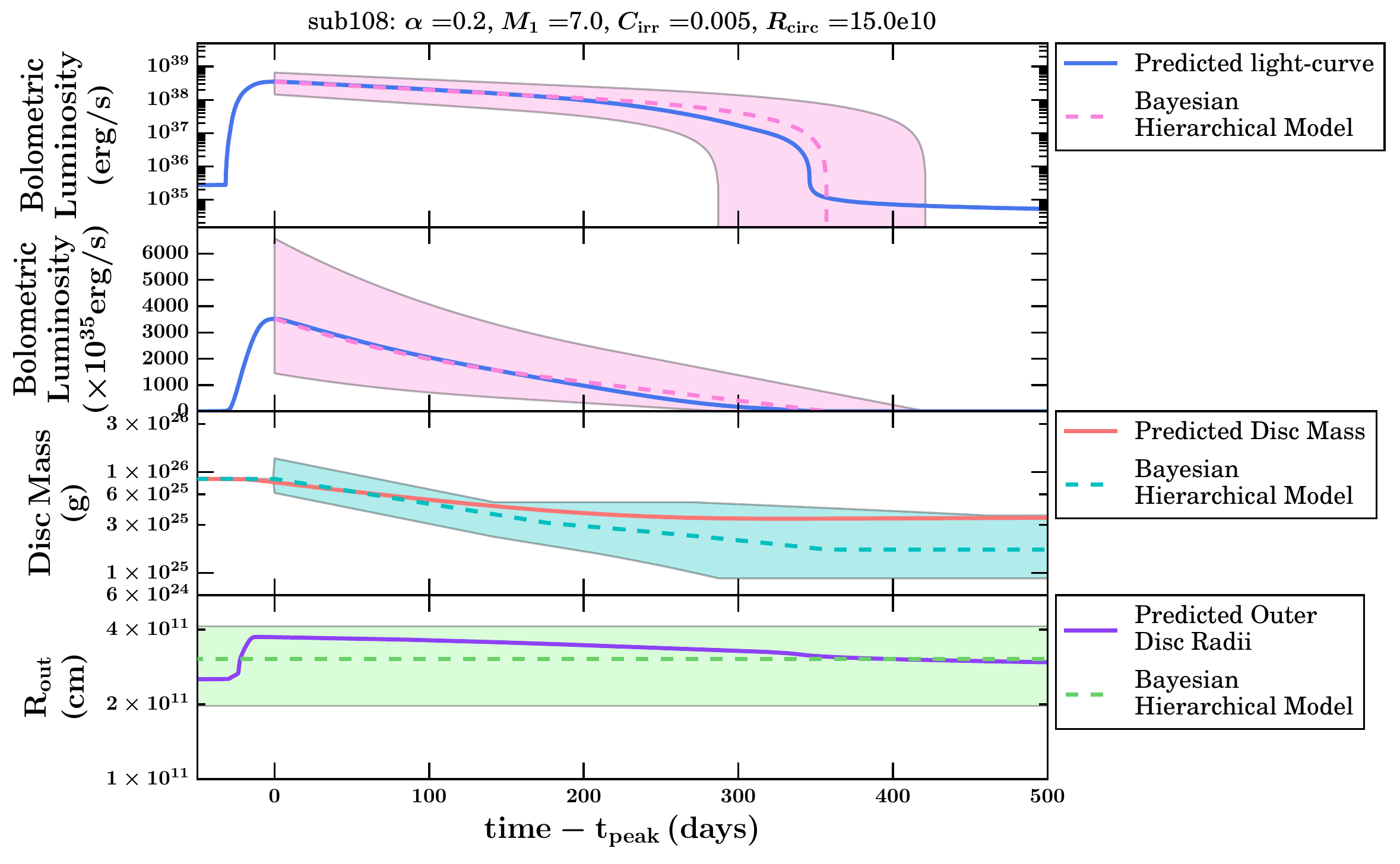}}\hfill
\subfloat
  {\includegraphics[width=.5\linewidth,height=.35\linewidth]{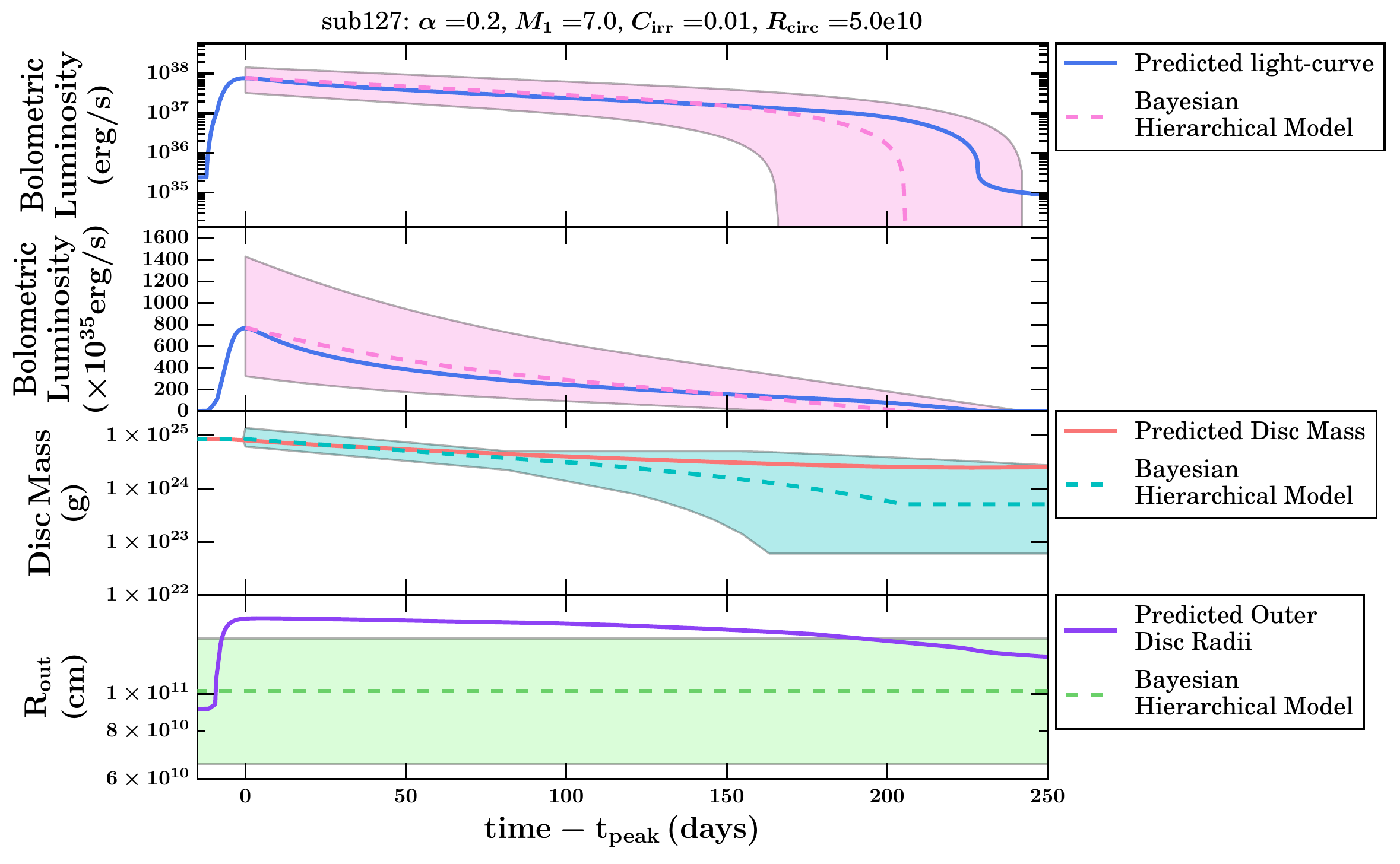}}
  
  \subfloat
  {\includegraphics[width=.5\linewidth,height=.35\linewidth]{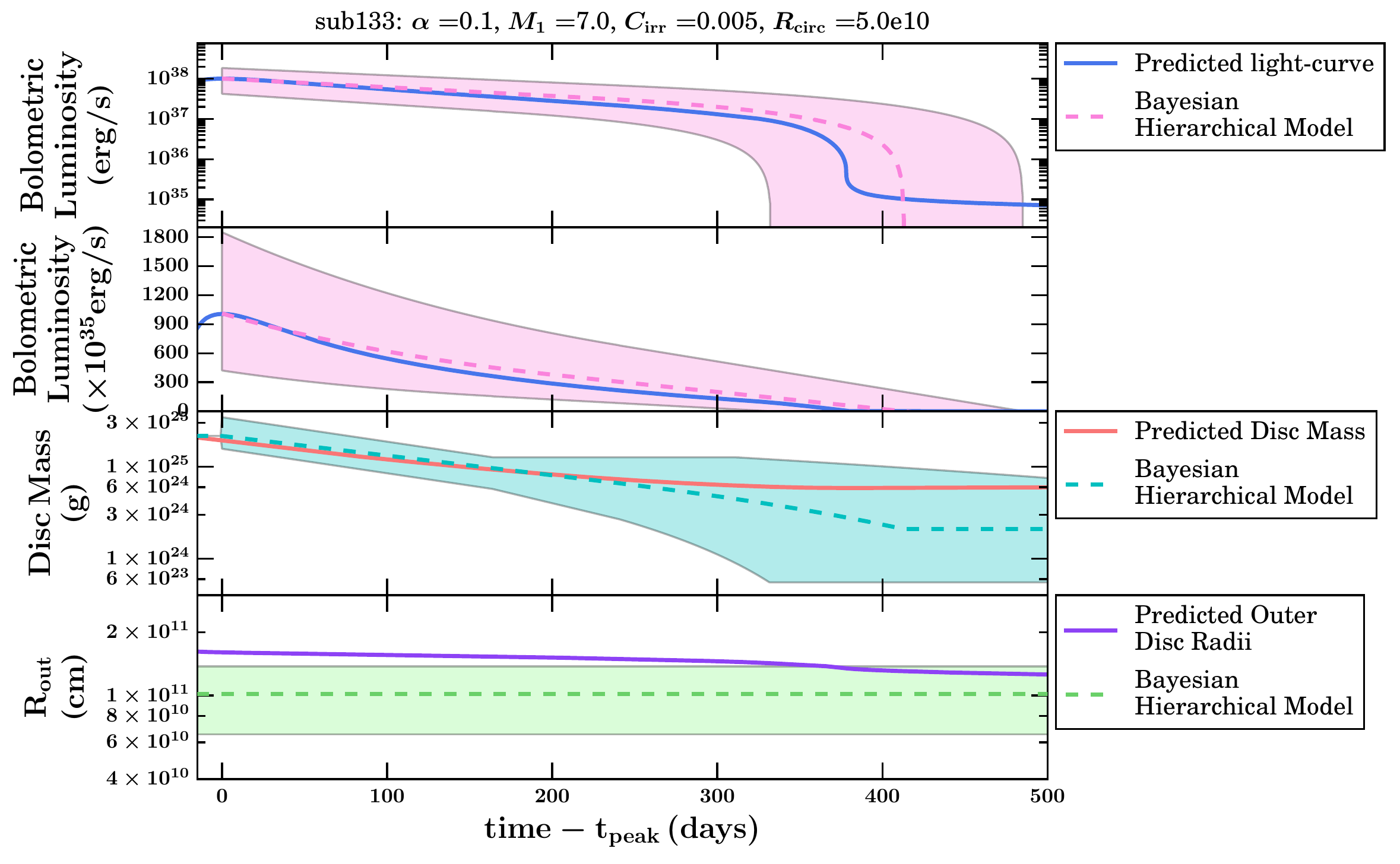}}\hfill
\subfloat
  {\includegraphics[width=.5\linewidth,height=.35\linewidth]{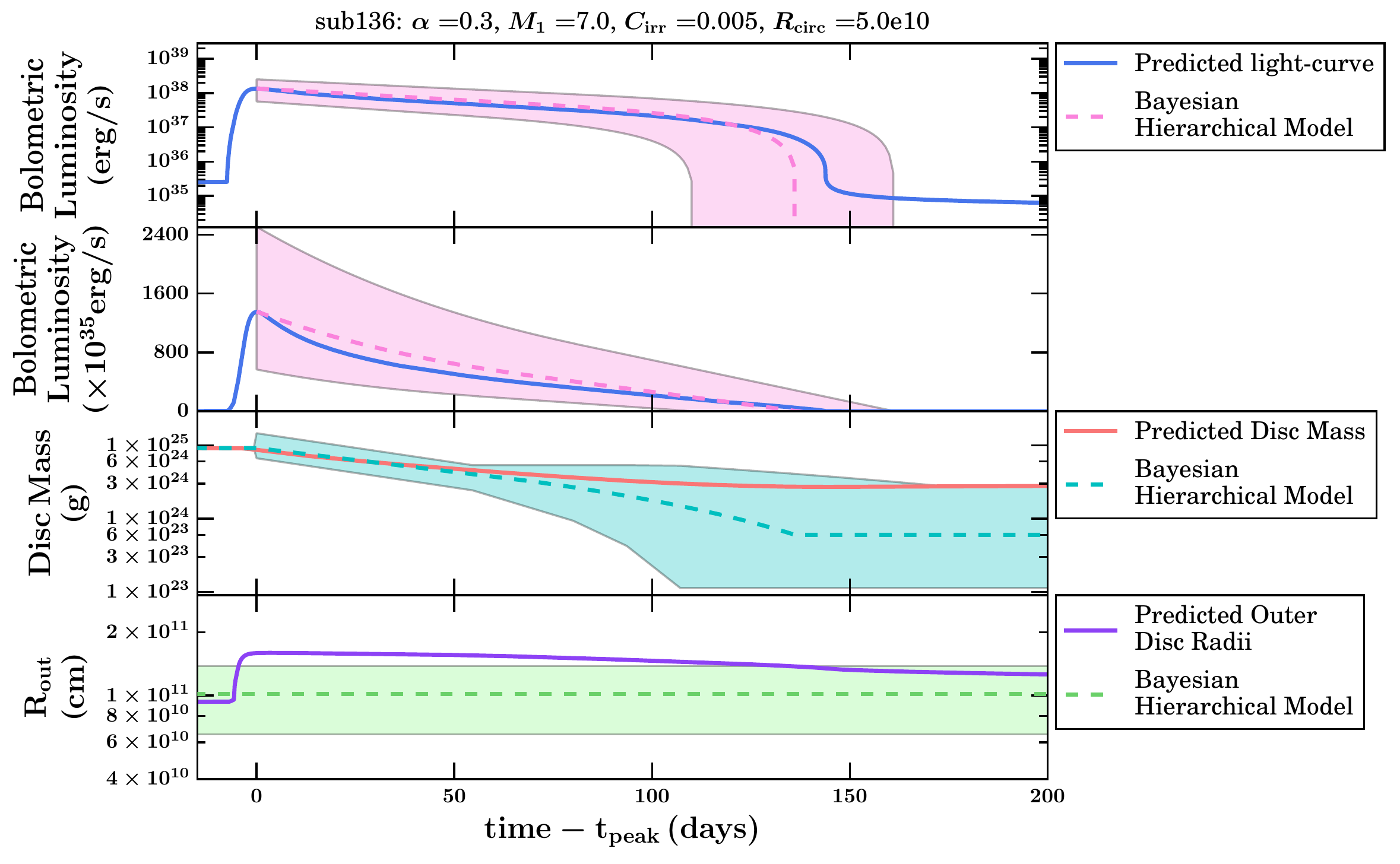}}
  
  \subfloat
  {\includegraphics[width=.5\linewidth,height=.35\linewidth]{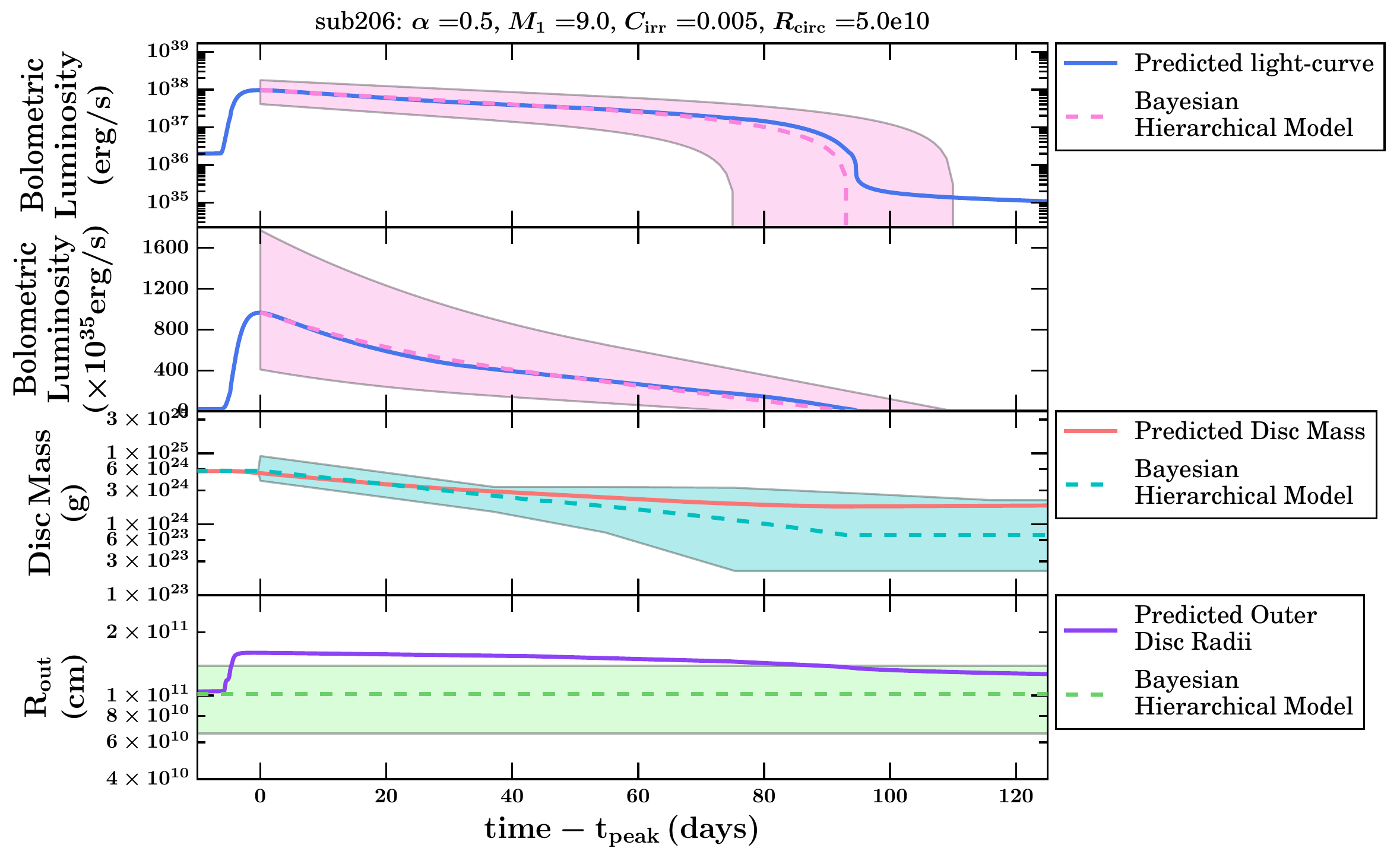}}\hfill
\subfloat
  {\includegraphics[width=.5\linewidth,height=.35\linewidth]{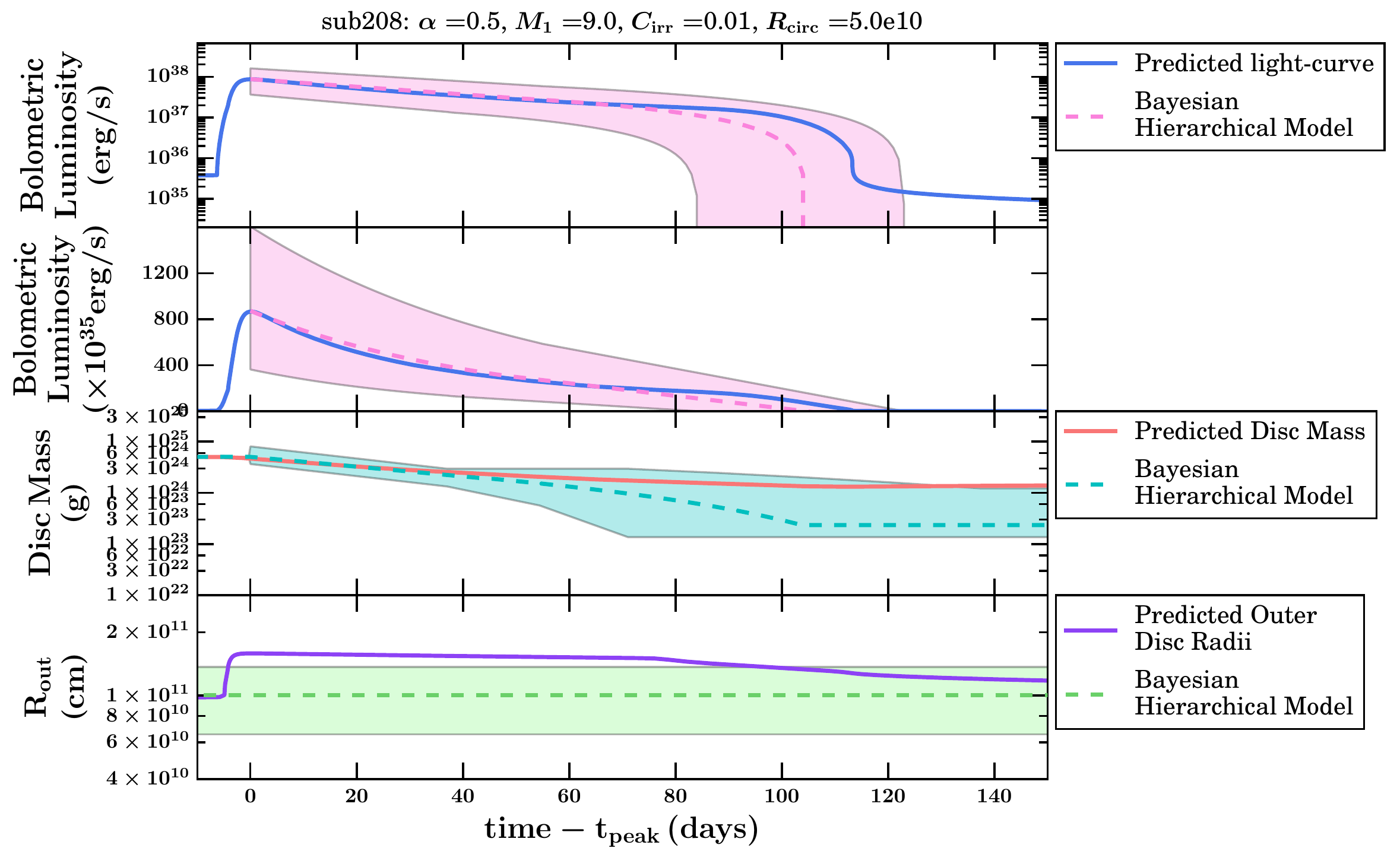}}
  \contcaption{Example outburst correlation plots comparing the predicted (from the numerical code) and Bayesian estimates (from our methodology) of the: light-curve decay profiles in log (top panel) and (second panel) linear space, (third panel) mass in the hot disc over time, and (bottom panel) outer disc radius, for varying $M_1$, $\alpha_h$, $C_{\rm irr}$, and $R_{\rm circ}$. The solid lines represent the output of the numerical code. The dotted lines and shaded regions represent the best-fit value and 1$\sigma$ confidence intervals from our Bayesian methodology, respectively.}%
  \label{fig:lc_compare2}%
\end{figure*}


\bsp	
\label{lastpage}
\end{document}